%% file: ms.tex
\newif\iftwocolumn
\twocolumntrue
\PassOptionsToPackage{hyphens}{url}
\iftwocolumn
\documentclass[format=sigconf, review=false, screen=true, dvipsnames]{acmart_arXiv}
\else
\documentclass[format=acmsmall, review=false, screen=true, dvipsnames]{acmart_arXiv}
\fi

\usepackage{amsmath}
\usepackage{fancyhdr}
\usepackage[normalem]{ulem}
\usepackage[utf8]{inputenc}
\usepackage{graphicx}
\iftwocolumn
\usepackage[keeplastbox]{flushend}
\fi
\usepackage{letltxmacro}

\usepackage{array}
\usepackage{multirow}
\usepackage{enumitem}
\usepackage{hhline}
\usepackage{dblfloatfix}
\usepackage{setspace}
\usepackage{mdframed}
\usepackage[binary-units]{siunitx}
\usepackage[us,12hr]{datetime}

\hypersetup{draft}
\setlist{leftmargin=13pt, itemsep=0ex, topsep=0pt}
\renewcommand\footnotetextcopyrightpermission[1]{}

\fancypagestyle{firststyle}
{
  \fancyhead{}
  \fancyhead[C]{\emph{Preprint; final version to appear in POMACS, 2019.}}
}

\newcommand{\changes}[0]{}
\newcommand{\changesii}[0]{}
\newcommand{\changesiii}[0]{}
\newcommand{\todo}[1][]{}
\newcommand{\changesiv}[0]{}
\newcommand{\changesv}[0]{}
\newcommand{\changesvi}[0]{}
\newcommand{\changesvii}[0]{}
\newcommand{\ch}[0]{}
\newcommand{\chii}[0]{}
\newcommand{\chiii}[0]{}
\newcommand{\chiv}[0]{}
\newcommand{\chv}[0]{}
\newcommand{\shep}[0]{}
\newcommand{\shepii}[0]{}
\newcommand{\shepiii}[0]{}
\newcommand{\shepiv}[0]{}

\newcounter{obsnumber}
\newcommand{\obs}[2]{\vspace{5pt}\begin{mdframed}[backgroundcolor=gray!10, linewidth=0.5pt]\begin{center}\emph{\textbf{OBSERVATION~\refstepcounter{obsnumber}\label{#1}\theobsnumber: } #2}\end{center}\end{mdframed}\vspace{5pt}}

\newcommand{\paratitle}[1]{\vspace{3pt}\noindent\textbf{#1.}}

\iftwocolumn
  \newcommand{\affilCMU}[0]{$^\dagger$}
  \newcommand{\affilETH}[0]{$^\S$}
  \newcommand{\affilSFU}[0]{$^\ddagger$}
\else
  \newcommand{\affilCMU}[0]{Carnegie Mellon University, USA\xspace}
  \newcommand{\affilETH}[0]{ETH Z\"{u}rich, Switzerland\xspace}
  \newcommand{\affilSFU}[0]{Simon Fraser University, Canada\xspace}
\fi

\title{Understanding the Interactions of Workloads and DRAM Types: A Comprehensive Experimental Study}

\iftwocolumn

\author{\vspace{-10pt}%
Saugata Ghose\affilCMU%
\quad\qquad%
Tianshi Li\affilCMU%
\quad\qquad%
Nastaran Hajinazar\affilSFU\affilCMU%
\\\vspace{-8pt}%
Damla Senol Cali\affilCMU%
\quad\qquad%
Onur Mutlu\affilETH\affilCMU%
}
\affiliation{%
\vspace{12pt}%\\
{\affilCMU}Carnegie Mellon University%
\quad\qquad%
{\affilSFU}Simon Fraser University%
\quad\qquad%
{\affilETH}ETH Z{\"u}rich%
\vspace{17pt}
}

\else

\author{Saugata Ghose}
\affiliation{\affilCMU}

\author{Tianshi Li}
\affiliation{\affilCMU}

\author{Nastaran Hajinazar}
\affiliation{\affilSFU \& \affilCMU}

\author{Damla Senol Cali}
\affiliation{\affilCMU}

\author{Onur Mutlu}
\affiliation{\institution{\affilETH \& \affilCMU}}

\fi

\date{}

\input{sections/abstract.tex}

\begin{document}

\fontdimen2\font=0.45ex% inter word space
\setlength{\textfloatsep}{4.0pt plus 2.0pt minus 2.0pt}
\setlength{\intextsep}{4.0pt plus 2.0pt minus 2.0pt}

\maketitle
\thispagestyle{firststyle}

\input{sections/introduction.tex}

\input{sections/background.tex}

\input{sections/methodology.tex}

\input{sections/desktop.tex}

\input{sections/multithreaded.tex}

\input{sections/server.tex}

\input{sections/mobile.tex}

\input{sections/os.tex}

\input{sections/lessons.tex}

\input{sections/related_work.tex}

\input{sections/conclusions.tex}

\section*{Acknowledgments}
\changesiii{We thank our shepherd Evgenia Smirni and the anonymous reviewers
for feedback.
We thank the SAFARI Research Group members for feedback and the
stimulating intellectual environment they provide. 
We acknowledge the generous support of our industrial partners:
Alibaba, Facebook, Google, Huawei, Intel, Microsoft, and VMware. This research was supported in part by the 
Semiconductor Research Corporation and the
National Science Foundation.}

{\renewcommand{\baselinestretch}{0.955}
\bibliographystyle{IEEEtranS}
\bibliography{references}
}

\newpage
\appendix
\section*{Appendix}
\input{sections/dramarch.tex}
\input{sections/artifactdesc.tex}
\input{sections/detail.tex}

\end{document}

%% file: sections/abstract.tex
\begin{abstract}

\changesv{It} has become increasingly difficult to
understand the complex \ch{interactions} between modern applications and 
\changes{main memory, composed of Dynamic Random Access Memory (DRAM)
chips}.
Manufacturers are now selling and proposing many different types of DRAM,
with each DRAM type catering to different needs (e.g., high throughput,
low power, high memory density).
At the same \ch{time, memory} access patterns of \changes{prevalent} and emerging \changesv{applications} 
are rapidly diverging, as these applications manipulate larger data sets
in very different ways.
As a result, the combined DRAM--workload behavior is often difficult to
\changes{\emph{intuitively}} determine today, which can hinder memory optimizations in
both hardware and software.

In this work, we identify important families of workloads, 
\changes{as well as prevalent types of DRAM chips,}
and rigorously
analyze the combined DRAM--workload behavior.
To this end, we perform a comprehensive experimental study of
the interaction between nine different DRAM types and
115~modern applications and multiprogrammed
workloads.
We draw 12~key observations from our characterization, enabled in part by
our development of new metrics that take into account contention between
memory requests due to hardware design.
Notably, we find that
(1)~newer DRAM technologies such as DDR4 and HMC often do \changes{\emph{not}}
outperform older technologies such as DDR3,
due to higher access latencies and, \changesvii{also} in the case of HMC, poor \changes{exploitation of} locality;
(2)~there is no \changes{single memory type} that can \changesv{effectively} cater to all of the 
\changesv{components of a heterogeneous system}
(e.g., GDDR5 significantly outperforms other memories for \changesv{multimedia acceleration},
while HMC significantly outperforms other memories for \changesv{network acceleration}); and
(3)~there is still a strong need to lower DRAM latency, 
\changes{but unfortunately the current design trend of commodity DRAM is toward
higher latencies to obtain other benefits.}
We hope that the trends we identify can drive optimizations in both
hardware and software design.  To aid further study, 
we \changesv{open-source} our extensively-modified simulator, as well as a 
benchmark suite \changesiii{containing our} applications.

\end{abstract}

%% file: sections/introduction.tex
% !TEX root=../dramcharacterization.tex

\section{Introduction}
\label{sec:intro}

\changes{Main memory in modern computing systems is built using Dynamic Random 
Access Memory (DRAM) technology.}
\changesv{The} performance of DRAM is an
increasingly critical factor in overall system and application performance,
\changesv{due to the increasing memory demands of modern and emerging
applications}.
As modern DRAM designers strive to improve performance \changes{and energy efficiency}, 
they must deal with three major issues.
First, DRAM consists of
\changesv{capacitive cells}, and 
\changesvii{the latency to access these DRAM cells\ch{~\cite{ddr3}}}
is \changesv{two or more} orders of magnitude greater than \changesvii{the execution latency of} a CPU \changesv{\emph{add}}
instruction\changesv{~\cite{scale.imw13}}.
Second, while the impact of long access latency can potentially be overcome by increasing \changes{data}
\emph{throughput}, \changes{DRAM chip} throughput is also constrained \changes{because
conventional} DRAM modules are discrete devices that \changesv{reside} off-chip from the CPU,
and are connected to the CPU via a narrow, pin-limited bus. For example, Double
Data Rate (e.g., DDR3, DDR4) memories exchange data with the CPU using a 64-bit 
bus.
\changes{DRAM data} throughput can be increased by increasing the DRAM bus frequency and/or
the bus pin count, but both of these options \ch{incur significant} cost in terms of
\changesv{energy and/or \changesvii{DRAM chip} area}.
Third, DRAM power consumption is not \changesv{reducing} as the memory density increases.
Today, DRAM consumes as much as half of the total power consumption of a system~\cite{lefurgy.computer03, ware.hpca10, david.icac11, 
holzle.book09, malladi.isca12, yoon.isca12, paul.isca15}.
As a result, the amount of DRAM that can be added to a system is now
constrained by its \changesv{power} consumption.

\changesv{In addition to the major DRAM design issues that need to be overcome,
memory} systems must now serve an increasingly diverse set of
applications, \changes{sometimes concurrently}.  For example, workloads designed for high-performance and cloud
computing environments process very large amounts of data\changesvii{, and} do \emph{not}
always exhibit high \changesvii{temporal} or spatial locality.  
In contrast, network processors exhibit very bursty memory access patterns
with low temporal locality.
\changesvii{As a result, it} is becoming increasingly difficult for a single \changesv{design}
point in the memory design space (i.e., one type of DRAM interface
and chip) to perform well \changesv{for all of} such a diverse set of
applications.
In response to these key challenges, \changesvii{DRAM manufacturers}
have been developing a number of different \changesvii{DRAM 
types} over the last decade,
such as Wide I/O~\cite{wideio} and Wide I/O 2~\cite{wideio2}, High-Bandwidth Memory (HBM)~\cite{AMD.hbm, lee.taco16, hbm}, and the Hybrid Memory
Cube (HMC)~\cite{jeddeloh2012hybrid, pawlowski.hc11, hmc.2.1, rosenfeld.tr12}.

With the increasingly-diversifying application behavior and the wide array of available \changes{DRAM types},
it has become very difficult to identify the \changesv{best} DRAM type for a given workload,
let alone for a system that is running a number of \changesvii{different} workloads.
Much of this difficulty lies in the complex interaction between memory access latency,
bandwidth, \changesvii{parallelism,} energy consumption, and \changes{application} memory access patterns.
Importantly, changes made by manufacturers in new \changes{DRAM} types
can significantly affect the behavior of an application in ways that are
often difficult to intuitively \changes{and easily} determine.
In response, prior work has introduced a number of detailed
memory simulators (e.g., \changesv{\cite{kim.cal15, rosenfeld-cal2011, dong.tcad12}}) to
model the performance of different DRAM types, but
\changesvii{end users must} set up and simulate each workload that
they care about, for each individual DRAM type.
\textbf{Our goal} in this work is to 
\emph{comprehensively} study the strengths and
weaknesses of each DRAM type based on the \changesv{memory demands}
of each \changesvii{of a diverse range of workloads}. 

Prior studies of memory behavior (e.g., 
\chii{\cite{cuppu.isca99, cuppu.tc01, zhu.hpca05, zheng.tc10, gomony.date12, parsec, 
henning.computer00, he.pact08, singh.icpe19, murphy.tc07, agaram.ismm06, 
charney.ibmjrd97, mccalpin.tcca95, kim.cal15, rosenfeld-cal2011, li.memsys18,
radulovic.memsys15, suresh.cluster14, giridhar.sc13, ahn.sc09, li.sc17}}) \chiii{usually} focus on a
single type of workload (e.g., desktop/scientific applications), and often examine 
only a single memory type (e.g., DDR3).
We instead aim to provide a much more \emph{comprehensive experimental study} of
the \changes{application and} memory landscape today.  Such a comprehensive study has been difficult to
perform in the past, and \emph{cannot} be conducted on real systems,
because a given \shep{CPU} chip does \changes{\emph{not}} support more than a single type
of DRAM.  
\shep{As a result, there is no way to isolate only the changes due to using
one memory type in place of another memory type on real hardware, 
\changesv{since doing so requires the use of} a different CPU to test the new memory type.}
Comprehensive simulation-based studies are also difficult,
due to the extensive time required to implement each DRAM type, to
port a large number of applications to the simulation platform,
and to capture \changes{both application-level and intricate processor-level}
interactions that impact memory access patterns.
To overcome these hurdles, we extensively modify a state-of-the-art,
\changesvii{flexible and extensible} memory simulator,
\changes{Ramulator~\cite{kim.cal15},}
to (1)~model new DRAM types that \changes{have recently appeared} on the market; and 
(2)~\changesvii{efficiently} capture processor-level interactions \changes{(e.g., instruction dependencies,
cache contention, data sharing)}
\changesvii{(see Appendix~\ref{sec:artdesc:sim}).}

Using our modified simulator, we perform \ch{a} \changesvii{comprehensive} experimental study
of the combined behavior of \changes{prevalent} and emerging applications with a large number of
contemporary DRAM types (which we refer to as the \emph{combined DRAM--workload
behavior}).
We study the design and behavior of nine different commercial DRAM 
types: \changes{DDR3~\cite{ddr3}, DDR4~\cite{ddr4}, LPDDR3~\cite{lpddr3}, LPDDR4~\cite{lpddr4}, GDDR5~\cite{gddr5}, Wide I/O~\cite{wideio}, Wide I/O 2~\cite{wideio2}, HBM~\cite{AMD.hbm},
and HMC~\cite{hmc.2.1}}.  We characterize each DRAM type using 87~applications and
28~multiprogrammed workloads (115 in total) from six diverse application
families: \changesv{desktop/scientific, server/cloud, multimedia acceleration,
network acceleration, general-purpose GPU (GPGPU),}
and common OS routines.  We perform a rigorous experimental
characterization of system performance and DRAM energy consumption,
and introduce new metrics to capture the sophisticated interactions
between \changesv{memory access patterns} and the 
\changesv{underlying hardware}.

Our characterization yields twelve key observations \chii{(highlighted in boxes) 
and many other findings (embedded in the text)} about the combined
DRAM--workload behavior \changesv{(as we describe in detail in
Sections \ref{sec:desktop}--\ref{sec:os})}.  
We highlight our five most significant experimental observations \changesv{here}:
\begin{enumerate}[leftmargin=12pt]
    \item \ch{\emph{\chii{The newer, higher bandwidth} \chiii{DDR4} \changes{rarely} outperforms DDR3
        on the applications we evaluate.}
        Compared to DDR3, DDR4 doubles the number of banks in a DRAM chip,
        in order to enable more bank-level parallelism \chii{and higher memory
        bandwidth}.  However, as a result of
        architectural changes to provide \chii{higher bandwidth and \chiii{bank-level}
        parallelism}, the access latency of DDR4
        is 11--14\% higher than that of DDR3.  We find that most of our applications
        do not exploit enough bank-level parallelism to overcome the
        increased access latency.}
    
    \item \ch{\emph{The \changes{high-bandwidth} HMC does not outperform DDR3 for most single-threaded
        and \changesvii{several} multithreaded applications.}}  This is because HMC's design trade-offs fundamentally 
	limit opportunities for exploiting spatial locality (due to its \changes{97\% smaller row width than DDR3}),
        \changes{and \chii{the} \ch{aforementioned} applications are unable to exploit the additional bank-level parallelism provided
        by HMC.  For example, \ch{single-threaded} desktop and scientific applications actually perform 
        5.8\% \emph{worse} with HMC than with DDR3, on average, even though HMC offers
        87.4\% more memory bandwidth.}
        \changesv{HMC provides significant performance improvements over other DRAM \changesvii{types
        in} cases where application spatial locality is low (or is destroyed) 
        \ch{and \chiii{bank-level} parallelism is high, such as for} highly-memory-intensive
        multiprogrammed workloads.}

    \item \ch{\emph{While low-power DRAM \changes{types \changesvii{(i.e., LPDDR3, LPDDR4, Wide I/O,
       Wide I/O 2)} typically perform} worse than
       standard-power DRAM for most memory-intensive applications, some 
       low-power DRAM types perform well when \changes{bandwidth} demand is very high.}}  For example, 
       on average, LPDDR4 performs only 7.0\% worse than DDR3 for multiprogrammed
       desktop workloads, while consuming 68.2\% less energy.
       Similarly, we find that Wide I/O 2, another low-power DRAM type,
       actually performs 2.3\% \emph{better} than DDR3 on average for
       multimedia applications, \changes{as Wide I/O 2 \ch{provides more
       opportunities for parallelism}
       while maintaining low memory access latencies}.

    \item \ch{\emph{The \changesvii{best DRAM type} for a \changesv{heterogeneous system} depends heavily on the
        predominant function(s) performed by the \changesv{system}.}}
        \changesv{We study three types of applications for heterogeneous systems:
        multimedia acceleration, network acceleration, and
        GPGPU applications.}
        \changesv{First, multimedia acceleration benefits}
        most from high-throughput memories \changes{that exploit a high amount of}
        \emph{spatial locality}, \changes{running} up to 21.6\% faster with GDDR5 and 
        14.7\% faster with HBM than \changesvii{with} DDR3, but only 5.0\% faster with HMC
        (due to \changesvii{HMC's} limited ability to exploit spatial locality).
        \changesv{Second, network acceleration memory requests} are highly bursty and do 
        \changes{\emph{not} exhibit significant} spatial locality,
        \changesv{making network acceleration a good fit for} HMC's very high bank-level
        parallelism (with a mean \changes{performance} increase of 88.4\% over DDR3).
        \changesv{Third,} GPGPU applications exhibit a wide range of memory intensity, but memory-intensive
	GPGPU applications
        \changesvii{typically take} advantage of spatial locality due to memory 
        coalescing~\cite{bakhoda.ispass09, chatterjee.sc14},
	making HBM \changes{(26.9\% higher on average over DDR3)} and GDDR5 \changes{(39.7\%)}
        more effective for GPGPU applications than \changesvii{other} \changes{DRAM}
	types such as DDR3 \changesvii{and HMC.}

    \item \ch{\emph{Several common OS routines (e.g., file I/O, process forking) 
        perform better with memories such as DDR3 and GDDR5, which have lower
        access latencies than the other memory types that we study.}
        This is because the routines exhibit \changesvii{very} high spatial locality,
        and do not benefit from high amounts of bank-level parallelism.}
        Since OS routines are used across most computer systems \changes{in a widespread manner}, 
        \changesv{we believe} \chv{that} DRAM designers
        must \chv{work to reduce the access latency}.
        \changesvii{Our recommendation goes against the current trend of} increasing the latency
	in order to deliver \ch{other benefits}.
\end{enumerate}

We hope and expect that the results of our rigorous experimental characterization 
will be \changesv{informative \ch{and useful} for}
application developers, \changesvii{system} architects, and DRAM architects alike.
To foster further work in \changes{both academia and industry}, we \changesiii{\changesvii{release} the
applications and multiprogrammed workloads that we study as a new memory benchmark suite~\cite{memben.github},
along with our heavily-modified \ch{memory} simulator~\cite{ramulator.github}}.

\changes{This paper} makes the following contributions:
\begin{itemize}[leftmargin=12pt]
    \item We perform \changes{the first} comprehensive study of the interaction between modern
    DRAM types and modern workloads.  \changesvii{Our study} covers the
    interactions of 115~applications and workloads from six different application families with nine
    different DRAM types.
    We are the first, to our
    knowledge, to (1)~quantify how new DRAM types (e.g., Wide I/O, HMC, \changesv{HBM})
    compare to commonplace DDRx and LPDDRx DRAM types 
    across a wide variety of workloads, \changes{and
    (2)~\changesvii{report findings} where newer memories often perform worse than
    older ones.}

    \item To our knowledge, \changesvii{this paper is} the first to perform a detailed
    comparison of the memory access behavior between \changesvii{desktop/scientific} applications,
    server/cloud \changesv{applications}, \changesv{heterogeneous system applications}, GPGPU applications, 
    and OS kernel routines.  These insights can help DRAM architects, \changesvii{system
    designers, and application developers} pinpoint bottlenecks in existing systems, and can inspire new
    \changesvii{memory, system, and application} designs.
    
    \item We make several new observations about the combined behavior of various DRAM
    types and different families of workloads.
    In particular, we find that new memory types, such as DDR4 and HMC, make a number
    of underlying design trade-offs that cause them to perform worse than older DRAM
    types, such as DDR3, for a \changesv{variety} of applications.
    In order to aid the development of new memory architectures and new system designs
    based on our observations, we \changesiii{\changesvii{release} our extensively-modified memory simulator~\cite{ramulator.github}
    and a memory benchmark suite~\cite{memben.github}} consisting \changesiii{of our} applications and workloads.
\end{itemize}

%% file: sections/background.tex
% !TEX root=../dramcharacterization.tex

\section{Background \& Motivation}
\label{sec:bkgd}

In this section, we provide necessary background on basic DRAM design and operation (Section~\ref{sec:bkgd:basics}),
and on the evolution of new DRAM types (Section~\ref{sec:bkgd:arch}).

\begin{table*}[!ht]
\centering
\caption{Key properties of the nine DRAM types evaluated in this work.\vspace{-10pt}}
\label{tbl:dram}
{
\footnotesize
\begin{tabular}{| c || c | c | c | c | c || c | c | c | c |}
\hline
\multirow{2}{*}{\textbf{DRAM Type}} & \multicolumn{5}{c||}{\em Standard Power} &  \multicolumn{4}{c|}{\em Low Power} \\ 
& \textbf{DDR3} & \textbf{DDR4} & \textbf{GDDR5} & \textbf{HBM} & \textbf{HMC} & \textbf{LPDDR3} & \textbf{LPDDR4} & \textbf{Wide I/O} & \textbf{Wide I/O 2} \\ 
\hhline{|=#=|=|=|=|=#=|=|=|=|}
\textbf{Data Rate} (MT/s) & 2133 & 3200 & 7000 & 1000 & 2500 & 2133 & 3200 & 266 & 1067 \\ \hline
\textbf{Clock Frequency} (MHz) & 1067 & 1600 & 1750 & 500 & 1250 & 1067 & 1600 & 266 & 533 \\ \hline
\textbf{Maximum Bandwidth} (GBps) & 68.3 & 102.4 & 224.0 & 128.0 & 320.0 & 68.3 & 51.2 & 17.0 & 34.1 \\ \hline
\textbf{Channels/Ranks per Channel} & 4/1 & 4/1 & 4/1 & 8/1 & 1/1 & 4/1 & 4/1 & 4/1 & 4/2 \\ \hline
\textbf{Banks per Rank} & 8 & 16 & 16 & 16 & 256 (32 vaults) & 8 & 16 & 4 & 8 \\ \hline
\textbf{Channel Width} (bits) & 64 & 64 & 64 & 128 & 32 & 64 & 64 & 128 & 64 \\ \hline% 
\textbf{Row Buffer Size} & 8KB & 8KB & 8KB & 2KB & 256B & 8KB & 4KB & 2KB & 4KB \\ \hline
\textbf{Row Hit/Miss Latencies} (ns) & 15.0/26.3 & 16.7/30.0 & 13.1/25.1 & 18.0/32.0 & 16.8/30.4 & 21.6/40.3 & 26.9/45.0 & 30.1/38.9 & 22.5/41.3 \\ \hline
\textbf{Minimum Row Conflict Latency}$^\dagger$ (ns) & 37.5 & 43.3 & 37.1 & 46.0 & 44.0 & 59.1 & 61.9 & 67.7 & 60.0 \\ \hline
\multicolumn{10}{l}{\vspace{-6pt}} \\
\multicolumn{10}{l}{\chii{$^\dagger$\emph{See Section~\ref{sec:metrics} for definition.}}} \\
\end{tabular}%
}\vspace{-5pt}%
\end{table*}

\subsection{Basic DRAM Design \& Operation}
\label{sec:bkgd:basics}

\begin{figure}[h]
  \centering
  \includegraphics[width=\columnwidth, trim=80 410 180 110, clip]{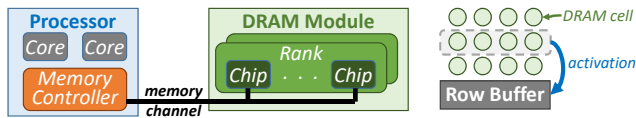}%
  \vspace{-10pt}%
  \caption{Memory hierarchy (left) and \changes{bank} structure (right).}%
  \label{fig:memsys}
  \label{fig:hierarchy}
\end{figure}

Figure~\ref{fig:memsys} (left) shows the basic overview of a DRAM-based memory system.
The memory system is organized in a hierarchical manner.
The highest level in the hierarchy is a \emph{memory channel}.  Each channel \changesvii{has
(1)~its} own bus to the host \changesvii{device} \changes{(e.g., processor)}, and 
\changesvii{(2)~}a dedicated \emph{memory controller} that interfaces
between the DRAM and the host device.  A channel
\changes{connects} to one or more \emph{dual inline memory modules} (DIMMs).
Each DIMM contains multiple DRAM \emph{chips}.  A DRAM row typically spans
across several of these chips, \changesv{and all of} the chips containing the row 
\changesv{perform operations} in lockstep with each other.  Each group of chips operating in 
lockstep is known as a \emph{rank}.  Inside each rank, there are several
\emph{banks}, where each bank is a DRAM array.  
\changesv{Each bank can operate concurrently, but the banks share a
single memory bus.  As a result, the memory controller must
schedule requests such that operations \ch{in different banks} do not interfere with
each other on the memory bus.}

A DRAM \changes{bank} typically consists of
\changes{thousands of} rows of \emph{cells}, where each cell contains a capacitor and an access 
transistor.  
To start processing a request, the controller issues a command to \emph{activate}
\changesv{the row containing the target address of the request} (i.e., open the row to perform \changes{\emph{reads} and \emph{writes}}), as shown in
Figure~\ref{fig:memsys} (right).
The \emph{row buffer} latches the opened row, at which point
the controller sends read and write commands to 
the row.  Each read/write command
operates on one \emph{column} of data at a time.  Once the read and write
operations to the row are complete, the controller issues a \emph{precharge}
command, to prepare the \changes{bank} for commands to a different row. 
For more \changes{detail} \changesv{on DRAM operation}, we refer the reader to our prior works
\changesvii{\cite{kim.isca12, lee.hpca13, lee.hpca15, chang.sigmetrics16, 
chang.sigmetrics17, lee.sigmetrics17, lee.pact15, kim.isca14, kim.hpca19,
seshadri.micro17, liu.isca12, kim.hpca18, seshadri.bookchapter20}}.

\subsection{Modern DRAM Types}
\label{sec:bkgd:arch}
\label{sec:bkgd:types}

We briefly describe several \changesv{commonly-used} and emerging DRAM types, 
all of which we evaluate in this work.  
Table~\ref{tbl:dram} summarizes the key properties of each of
these DRAM types.  We provide more detail about each
DRAM type in Appendix~\ref{sec:dramarch}.

\subsubsection{DDR3 and DDR4}
\label{sec:bkgd:types:ddr3}
\label{sec:bkgd:types:ddr4}

DDR3~\cite{ddr3} is the third generation of DDRx memory, where
a burst of data is sent on both the positive and negative edge of
the bus clock to double the data rate.  DDR3 contains eight banks of
DRAM in every rank.
DDR4~\cite{ddr4} increases the number of banks per rank,
to 16, by introducing \emph{bank groups}, a new level of hierarchy
in the \ch{DRAM subsystem}.
Due to the way in which bank groups are connected to I/O, a typical
memory access takes \emph{longer} in DDR4 than it did in DDR3,
but the bus clock frequency is significantly \changesvii{higher} in DDR4,
which \ch{enables DDR4 to have higher} bandwidth.

\subsubsection{Graphics DDR5 (GDDR5)}
\label{sec:bkgd:types:gddr5}

Similar to DDR4, GDDR5~\cite{gddr5} doubles the number of banks 
over DDR3 using bank groups.
However, unlike DDR4, GDDR5 does so by increasing the die area and energy over DDR3
instead of the memory latency.  
GDDR5 also increases memory throughput by doubling the amount of data sent in a 
single clock cycle, as compared to DDR3.

\subsubsection{High Bandwidth Memory (HBM)}
\label{sec:bkgd:types:hbm}

High Bandwidth Memory~\cite{hbm, AMD.hbm} is a 3D-stacked memory~\cite{loh2008stacked, lee.taco16}
that provides high throughput. \changesvii{designed} for devices such as GPUs.  \changes{Unlike GDDR5, which uses} faster clock frequencies to
increase throughput, HBM connects 4--8 memory channels to a \emph{single} DRAM device
to
service many more requests in parallel.

\subsubsection{Wide I/O and Wide I/O 2}
\label{sec:bkgd:types:wideio}
\label{sec:bkgd:types:wideio2}

\changes{Wide I/O~\cite{wideio} and Wide I/O 2~\cite{wideio2}} are 3D-stacked memories 
that \changesv{are designed for} low-power devices such as mobile phones.
Similar to HBM, Wide I/O and Wide I/O 2 connect multiple channels \changes{to a \emph{single}} DRAM device~\cite{kim.isscc2011},
but have fewer \changes{(2--4)} channels and contain fewer banks \changes{(8)} than \chv{HBM}
in order to lower energy consumption.

\sloppypar
\subsubsection{Hybrid Memory Cube (HMC)}
\label{sec:bkgd:types:hmc}

The Hybrid Memory Cube~\cite{jeddeloh2012hybrid, pawlowski.hc11, rosenfeld.tr12, hmc.2.1}
is a 3D-stacked memory with \changesv{more design} changes
compared to HBM and Wide I/O.  An HMC device (1)~performs request scheduling
inside the device itself, as opposed to relying on an external
memory controller for scheduling; and
(2)~partitions the DRAM array into multiple \emph{vaults}, which are small,
vertical slices of memory \changes{of which each contains} multiple banks.
\changesvii{The vault-based structure significantly increases} the amount of
bank-level parallelism inside the DRAM device \changesvii{(\ch{with 256~banks in total})},
but greatly \changesvii{reduces} the size of a row \changes{(to 256~bytes)}.
The HMC specification~\cite{hmc.2.1} provides an alternate \changes{mode}, which we call \emph{HMC-Alt},
that \changes{\changesvii{uses a different address mapping than the default mode to} maximize the limited spatial}
locality available in the smaller DRAM rows.

\subsubsection{LPDDR3 and LPDDR4}
\label{sec:bkgd:types:lpddr3}
\label{sec:bkgd:types:lpddr4}

LPDDR3~\cite{lpddr3} and LPDDR4~\cite{lpddr4} are low-power variants of DDR3
and DDR4, respectively.
These DRAM types lower power consumption by using techniques such as
a lower core voltage, \changes{two voltage domains on a single chip, 
temperature-controlled self refresh, deep power-down modes, 
reduced chip width, and fewer (1--2) chips per DRAM module~\cite{micronlp}
\changesv{than their standard-power counterparts}.}
These trade-offs \changes{\emph{increase}} the memory access latency, and limit
the total \changes{capacity of the low-power DRAM chip}.

\subsection{Motivation}
\label{sec:bkgd:motivation}

As DRAM scaling \changesv{is unable to keep pace with}
processor scaling, there is a growing
need to improve DRAM performance.
Today, conventional DDRx DRAM types suffer from three major bottlenecks.
First, prior works have shown that the
underlying design used by DDR3 and DDR4 remains largely the same as earlier
generations of DDR memory, and as a result, the DRAM access latency has not
changed significantly over the last decade\ch{~\cite{chang.sigmetrics16, son.isca13,
lee.hpca13, lee.hpca15, chang.thesis17, lee.thesis16}}.
Second, \changesv{it is becoming increasingly difficult to increase the density
of the memory chip, due to a number of challenges that DRAM vendors face
when they scale up the size of the DRAM array\chiii{~\cite{scale.imw13, 
mandelman.ibmjrd02, kang.memoryforum14, liu.isca13, liu.isca12,
kim.isca14, mutlu.tcad19, mutlu.date17}}}.
Third, DDRx connects to the host processor using a narrow, pin-limited off-chip
channel, which restricts the available memory bandwidth.

\changes{As we describe in Section~\ref{sec:bkgd:types}, new DRAM types
contain a number of key changes to \changesvii{mitigate one or more of} these bottlenecks.}
Due to the non-obvious impact of \changesvii{such} changes
\changesv{on application performance and energy
consumption},
there is a need to perform careful characterization of how various  
\changesv{applications behave} under each new \changesvii{DRAM} type, and how this 
\changesvii{behavior} compares to the 
\changesv{application behavior} under conventional DDRx architectures.
\changes{Our goal in this paper is to rigorously characterize, \changesvii{analyze, and understand} the complex
interactions between \changes{several modern} DRAM types and a diverse set of modern applications,
through the use of detailed simulation models and new metrics that capture
the sources of these interactions.}

%% file: sections/methodology.tex
% !TEX root=../dramcharacterization.tex

\section{Methodology}
\label{sec:meth}

We characterize \changes{the} nine different DRAM types on
87~different \changesv{single-threaded and multithreaded}
applications~\cite{spec2006, coral, coral-2, parsec, hadoop, ycsb, redis, Apache, 
memcached, difallah.vldb04, mediabench, nxp.networkaccel, he.pact08, rodinia, lonestar,
IOZone, Netperf, seshadri.micro13},
and 28 multiprogrammed workloads,
using a heavily-modified 
version of Ramulator~\cite{kim.cal15}, a detailed and extensible \changes{open-source}
DRAM simulator.
\shep{\shepii{Many} of these applications come from \shepii{commonly-used} benchmark suites,
including SPEC CPU2006~\cite{spec2006}, CORAL~\cite{coral} and CORAL-2~\cite{coral-2}, 
PARSEC~\cite{parsec}, the Yahoo Cloud Suite~\cite{ycsb}, MediaBench II~\cite{mediabench},
Mars~\cite{he.pact08}, Rodinia~\cite{rodinia}, LonestarGPU~\cite{lonestar},
IOzone~\cite{IOZone}, and Netperf~\cite{Netperf}.}

\shep{We categorize each of our applications into one of six families:
desktop/scientific~\cite{spec2006, coral, coral-2, parsec}, server/cloud~\cite{hadoop, ycsb, redis, Apache, 
memcached, difallah.vldb04}, multimedia \changesv{acceleration}~\cite{mediabench}, \shepiii{network \changesv{acceleration}}~\cite{nxp.networkaccel}, 
GPGPU~\cite{he.pact08, rodinia, lonestar}, and OS routines~\cite{IOZone, Netperf, seshadri.micro13}.
\shepiii{Tables~\ref{tbl:workloads:desktop}--\ref{tbl:workloads:os} in 
Appendix~\ref{sec:artdesc:workloads}} provide a 
\shepii{complete} list of the 87~applications that we evaluate.
We use these 87~applications to assemble our multiprogrammed workloads,
which we list in Tables~\ref{tbl:mwdesktop} and \ref{tbl:mwcloud} in
Appendix~\ref{sec:artdesc:workloads}.}

\shep{For our desktop/scientific and multimedia applications,
we \shepii{record} per-core traces using Intel's Pin \changesvii{software}~\cite{luk2005pin}, which uses
dynamic binary instrumentation to \shepii{analyze real CPU behavior at runtime}.
These traces \ch{are}
collected using a machine containing an Intel Core i7-975K processor~\cite{corei7-975.spec} and 
running \changesv{the Ubuntu Server 14.04 operating system~\cite{ubuntu.14.04}}.
In order to accurately \shepii{record} the behavior of multithreaded desktop/scientific applications,
we make use of a modified Pintool~\cite{multithreadedpin.github}, which
accurately captures synchronization behavior across threads.  We modify this
Pintool to \shepii{generate} traces that are compatible with Ramulator, and to \shepii{record}
\shepiii{a separate} trace \changesvii{for each} thread.  In order to test the scalability of the
multithreaded applications that we study~\cite{parsec, coral, coral-2}, we
run the applications and our modified Pintool on a machine that contains
dual Intel Xeon E5-2630 v4 processors~\cite{xeone5-2630v4.spec}, providing us with the ability to
execute 40~threads concurrently.  These machines run Ubuntu Server 14.04,
and contain 128~GB of DRAM.  We \shepii{have open-sourced} our modified Pintool\changesv{~\cite{memben.github}}
along with our modified version of Ramulator\shepii{~\cite{ramulator.github}}.}

\shep{For our server/cloud applications and OS routines, we collect per-core traces using
the Bochs full system emulator~\cite{lawton2006bochs} in order to \shepii{record}
both user-mode and kernel-mode memory operations.
Though prior works often
overlook kernel-mode memory operations, recent studies reveal that many programs spend
\changesv{the majority of} of their execution time in kernel mode~\cite{peter2016arrakis, sun2014behavior}.
Unfortunately, Pin cannot capture \shepii{kernel-mode} operations, so we cannot collect
truly-representative traces using Pin.
We use Bochs~\cite{lawton2006bochs} because it emulates both
user-mode and kernel-mode operations. 
As we are constrained to using the processor
models available in Bochs, we choose the Intel Core i7-2600K~\cite{corei7-2600k.spec}, which is the closest
available to the i7-975K processor~\cite{corei7-975.spec} we \changesvii{use} with Pin.  The emulator runs the 
Ubuntu Server 16.04 operating system\changesv{~\cite{ubuntu.16.04}}.}

\shep{Our network accelerator applications \changesvii{are} collected from a commercial network
processor~\cite{nxp.networkaccel}.  We add support in Ramulator to emulate the
injection rate of requests from the network, by limiting the total number of
requests that are in flight at any given time.  For each workload, we evaluate
four different rates: 5~in-flight requests, 10~in-flight requests, 20~in-flight requests,
and 50~in-flight requests.}

\shep{For GPGPU applications, we integrate Ramulator into GPGPU-Sim~\cite{bakhoda.ispass09},
and collect statistics in Ramulator as the integrated simulator executes.
We collect all results using the NVIDIA GeForce GTX 480~\cite{gtx480.spec} configuration.
We \shepii{have open-sourced} our integrated version of GPGPU-Sim and Ramulator\shepiii{~\cite{gpgpusimramulator.github}}.}

\shep{\shepii{All of the traces that we record include} the delays
incurred by each CPU instruction during execution, and \shepii{we} replay these traces with
our core and cache \changesii{models} in Ramulator.}
We make several modifications to Ramulator to improve the
fidelity of our experiments \changesii{for all applications}.  
\shep{We describe our modifications in Appendix~\ref{sec:artdesc:sim}.
With our modifications,}
Ramulator \changes{provides} near-identical results \shepii{(with an average
error of only 6.1\%; see Appendix~\ref{sec:artdesc:sim})} to a simulator with a
detailed, \changes{rigorously-validated out-of-order processor core model~\cite{gem5}}, while being 
\chv{significantly faster (9.8$\times$ on average for SPEC CPU2006 benchmarks; see Appendix~\ref{sec:artdesc:sim})}.
We \changesiii{have open-sourced} our modified version of Ramulator\shepii{~\cite{ramulator.github}}
\changesiii{and a benchmark suite consisting of our application traces~\cite{memben.github}}.

Table~\ref{tbl:config} shows the system configuration parameters
\changesii{used for all of our experiments}.
For all of the DRAM types, we \changesv{model a \SI{4}{\giga\byte} capacity,
\changesvii{distributed across channels and ranks as listed in Table~\ref{tbl:dram},}
and} use the widely-used FR-FCFS memory
scheduler~\cite{rixner.isca00, zuravleff.patent97}, with 32-entry read and write queues.
\changesii{For all DRAM types except HMC, we}
use cache line interleaving\ch{~\cite{kim.isca12, jeong.hpca12, 
kim.hpca10, zhang.micro00, rokicki.tr96}} for the physical address, 
where consecutive cache lines are interleaved across multiple channels 
\changesvii{(and then across multiple banks within a channel)} to
maximize the amount of memory-level parallelism.
Cache line interleaving is used by processors such as the
Intel Core~\cite{coregen7.datasheet}, Intel Xeon~\cite{xeone5.datasheet, lenovo.xeon.memconfig}, and IBM POWER9~\cite{power9.datasheet} series.
\changesii{The HMC specification~\cite{hmc.2.1} explicitly specifies two fixed 
interleavings for the physical address.  The first interleaving, which is the
default for HMC, interleaves consecutive cache lines across multiple
vaults, and then across multiple banks.
The second interleaving, which we use for HMC-Alt \changesv{(see
Section~\ref{sec:bkgd:types:hmc})}, interleaves consecutive
cache lines \changesvii{only} across multiple vaults.}
For each \changesii{DRAM type} currently in production, we select the fastest frequency variant of
the \changesii{DRAM type} on the market today \changesv{(see Table~\ref{tbl:dram}
for key DRAM properties)}, 
as we can \shepii{find} reliable latency and power information for these products.
As timing parameters for HMC have yet to be publicly released,
we use the information provided in prior 
work~\cite{kim2013memory, jeddeloh2012hybrid} to model the latencies.

\begin{table}[h]
  \centering
  \caption{Evaluated system configuration.}
  \label{tbl:config}
  \vspace{-10pt}
  \footnotesize
    \setlength{\tabcolsep}{.8em}
    \begin{tabular}{cp{6cm}}
        \firsthline
        \multirow{1}{*}[-1.2em]{\bf Processor} & x86-64 ISA, 128-entry instruction window, 4-wide issue\newline \ch{single-threaded}/multiprogrammed: 4~cores, \SI{4.0}{\giga\hertz}\newline multithreaded: 20~cores, 2 threads per core, \SI{2.2}{\giga\hertz} \\
        \hline
        \multirow{1}{*}[-1.2em]{\bf Caches} & per-core L1: \SI{64}{\kilo\byte}, 4-way set associative\newline per-core L2: \SI{256}{\kilo\byte}, 4-way set associative\newline shared L3: \SI{2}{\mega\byte} for every core, 8-way set associative \\
        \hline
        {\bf Memory} &  32/32-entry read/write request queues,
            FR-FCFS~\cite{rixner.isca00,zuravleff.patent97}, \\
        {\bf Controller} & open-page policy, cache line interleaving\ch{~\cite{kim.isca12, jeong.hpca12, kim.hpca10, zhang.micro00, rokicki.tr96}} \\
        \lasthline
    \end{tabular}
\end{table}

\changesiii{We integrate DRAMPower~\cite{drampower}, an open-source DRAM power profiling tool,
into Ramulator such that it can perform power profiling while Ramulator executes.
To isolate the effects of DRAM behavior, we focus on the power consumed by DRAM
instead of total system power.  We \changesiv{report power numbers only for the}
DRAM \changesiv{types for which} vendors have publicly released power
consumption specifications\changesiii{~\cite{micron.ddr3.2gb.datasheet, micron.ddr4.4gb.datasheet,
micron.lpddr3.8gb.datasheet, micron.lpddr4.8gb.datasheet, hynix.gddr5.2gb.datasheet}}, 
to ensure the accuracy of the results that we present.}

\section{Characterization Metrics}
\label{sec:metrics}

\paratitle{Performance Metrics}
\changes{We measure \ch{single-threaded} application performance
using \emph{instructions per cycle} (IPC).
For multithreaded applications, we show \emph{parallel speedup}
(i.e., the \ch{single-threaded execution time divided by
the parallel execution time}), \chii{which accounts} for
synchronization overheads.}
For multiprogrammed
workloads, we use \changes{\emph{weighted speedup}}~\cite{snavely.asplos00}, which represents the
job throughput~\cite{eyerman.ieeemicro08}.
We verify that trends for other metrics (e.g., harmonic speedup\changesv{~\cite{luo.ispass01}},
\changesvii{which represents the inverse of the job turnaround time}) are similar.
To quantify the memory intensity of an application, we use the number of
\changes{\emph{misses per kilo-instruction}} (MPKI) issued by the last-level cache for that application
to DRAM. 

Our network accelerator workloads \ch{are} collected from a commercial network
processor~\cite{nxp.networkaccel}, \ch{which has a microarchitecture different from}
a traditional processor. 
\changesii{We} present performance results for the network accelerator in terms of 
\changes{\changesii{sustained memory} \emph{bandwidth}}.

\paratitle{\changesiii{Parallelism} Metrics}
\changes{Prior works have used either
\emph{memory-level parallelism} (MLP)\changesv{~\cite{glew.waci98, chou.isca04, tuck.micro06, qureshi.isca06, mutlu.isca05}}
or \emph{bank-level parallelism} (BLP)]\changesvii{~\cite{mutlu.isca08, tang.micro16, lee.micro09,
kim.micro10}}
to quantify the amount of parallelism across memory requests.
Unfortunately, neither metric fully represents the actual
parallelism exploited in DRAM.
MLP measures the average number of outstanding memory requests for
an application, but this does not capture the amount of parallelism offered by
the underlying hardware.
BLP measures the average number of memory requests that are actively being
serviced for a \changesv{\emph{single} thread during a given time interval}.
\changesv{While BLP can be used to compare the bank parallelism used by
one thread within an interval to the usage of another thread, it 
does not capture the average bank parallelism exploited by \emph{all}
concurrently-executing threads \changesvii{and applications} across the \emph{entire} execution,
which can provide insight \changesvii{into} whether the additional banks \changesvii{present in}
by many of the DRAM types (compared to DDR3) are being utilized.}}

\changes{We define a new metric, called \emph{bank parallelism utilization} (BPU), which quantifies the average number of
banks in main memory that are being used concurrently.  To measure 
BPU,} we sample the number of active banks for every cycle that the
DRAM is processing a request, and report the average utilization of banks:
\begin{equation}
\text{BPU} = \frac{\sum_i \text{\# active banks in cycle~}i}{\text{\# cycles memory is active}}
\end{equation}
\changes{A \emph{larger} BPU indicates that applications 
are making \emph{better} use of the bank parallelism available
in a particular DRAM type.
Unlike MLP and BLP, BPU fully accounts for
(1)~whether requests from \emph{any} thread \changesvii{contend} with each other for the same bank, and
(2)~how much parallelism is offered by the memory device.
As we see in our analysis \changesv{(Sections \ref{sec:desktop}--\ref{sec:os})}, 
BPU helps explain why some memory-intensive applications
do not benefit from high-bandwidth memories such as HMC, while other
memory-intensive applications do benefit.}

\paratitle{Contention Metrics}
\changes{An important measure of spatial and temporal locality in memory is the 
\changesiv{\emph{row buffer hit rate}, also known as \emph{row buffer locality}}.
To quantify the row hit rate, prior works count the number of row buffer hits and
the number of row buffer misses, which they define as any request that does not
hit in the currently-open row.  Unfortunately, this categorization does \emph{not}
distinguish between misses where a bank does not have \emph{any} row open, and misses
where a bank is currently processing a request to a different row (i.e., a \emph{row
buffer conflict}).  This distinction is important, as a row buffer
conflict typically takes \emph{longer} to service than a row buffer miss,
as a conflict must wait to issue a precharge operation, and may also need to
wait for an earlier request to the bank to complete. \changesv{A row
buffer conflict takes \ch{at least} as much as} \emph{double} the row miss latency, when the 
conflicting request arrives just after a request with a row miss starts
accessing the DRAM.}
\changesvii{Table~\ref{tbl:dram} lists the \ch{\emph{minimum}} row buffer conflict latency
for each DRAM type, assuming that no prior memory 
request has already issued the precharge operation for the conflicting row.}
\ch{Note that if there is more than one pending memory request that needs to access
the conflicting row, the row buffer conflict latency could be even higher.}

\changes{To accurately capture row buffer locality, we introduce a new characterization 
\changesvii{methodology} where
we break down memory requests into:}
(1)~\emph{row buffer hits}; 
(2)~\emph{row buffer misses}, which only include misses for a DRAM request
where the bank does not have \emph{any} row open;
and
(3)~\emph{row buffer conflicts}, which \changesv{consist of} misses where
another row is currently open in the bank and must be closed (i.e., precharged)
first.
\changes{Row buffer conflicts provide us with important information about how the
amount of parallelism exposed by a DRAM type can limit opportunities to
concurrently serve multiple memory requests, which in turn hurts performance.}

%% file: sections/desktop.tex
% !TEX root=../dramcharacterization.tex

\section{Single-Threaded/Multiprogrammed Desktop and Scientific Programs}
\label{sec:desktop}

We first study the memory utilization, performance, and DRAM energy consumption
of our tested DRAM types on \ch{single-threaded} desktop and scientific applications 
from the SPEC 2006 benchmark 
suite~\cite{spec2006}, and on multiprogrammed bundles of these applications.

\subsection{Workload Characteristics}
\label{sec:desktop:workload}

Using the DDR3 memory type, we study the memory intensity of each workload.
The workloads encompass a wide
range of intensity, with some CPU-bound applications (e.g., 
\emph{gamess}, \emph{calculix}) issuing memory requests only infrequently, 
and other memory-bound applications (e.g., \emph{mcf}) issuing over 
15 \chv{last-level cache (i.e., L3)} misses
per \chv{kilo-instruction} (MPKI).  The workloads also exhibit a large
range of row buffer locality, with \changesvii{row buffer} hit rates falling
anywhere between 2.4--53.1\% (see Appendix~\ref{sec:detail:desktopchar}).

\changes{We study the relationship between the performance (IPC) and 
memory intensity (MPKI) of the desktop and scientific applications
(see Appendix~\ref{sec:detail:desktopchar} for details and plots).
In general, we observe that the IPC decreases as the MPKI increases, but there are
two notable exceptions: \emph{namd} and \emph{gobmk}.}
To understand these exceptions, we study the amount of \changesv{bank}
\changes{parallelism that an application is able to exploit by
using the BPU metric we introduced in
Section~\ref{sec:metrics}  (see Appendix~\ref{sec:detail:desktopchar}
for BPU values for all applications). In our configuration,
DDR3 has 32~banks spread across four memory channels.}
For \changes{most} applications with low memory intensity 
(i.e., MPKI $<$ 4.0), the \changes{BPU for DDR3} is very low 
\changes{(ranging between 1.19 and 2.01)}
due to the low likelihood of
having many concurrent memory requests.  \changes{The two exceptions are
\emph{namd} and \emph{gobmk}, which}
have \changes{BPU} values of 4.03 and 2.91, respectively.
The higher \changes{BPU values} at low memory intensity imply that these applications
exhibit more \emph{bursty} memory behavior, issuing requests in clusters.
\changes{Thus, they could benefit more when a DRAM type offers a greater amount 
of bank parallelism
(\ch{compared to a DRAM type that offers} \changesv{reduced latency}).}

\changes{From the perspective of \changes{\emph{memory}}, we find that 
there is \changes{\emph{no}} discernible difference between applications with
predominantly integer computation and applications with predominantly floating point
\changesv{computation} (see Appendix~\ref{sec:detail:desktopchar}).
As a result, we do not distinguish between the two in this section.}

\subsection{Single-Thread Performance}
\label{sec:desktop:single}

\begin{figure}[h]
  \centering
  \includegraphics[width=0.9\columnwidth, trim=65 168 60 155, clip]{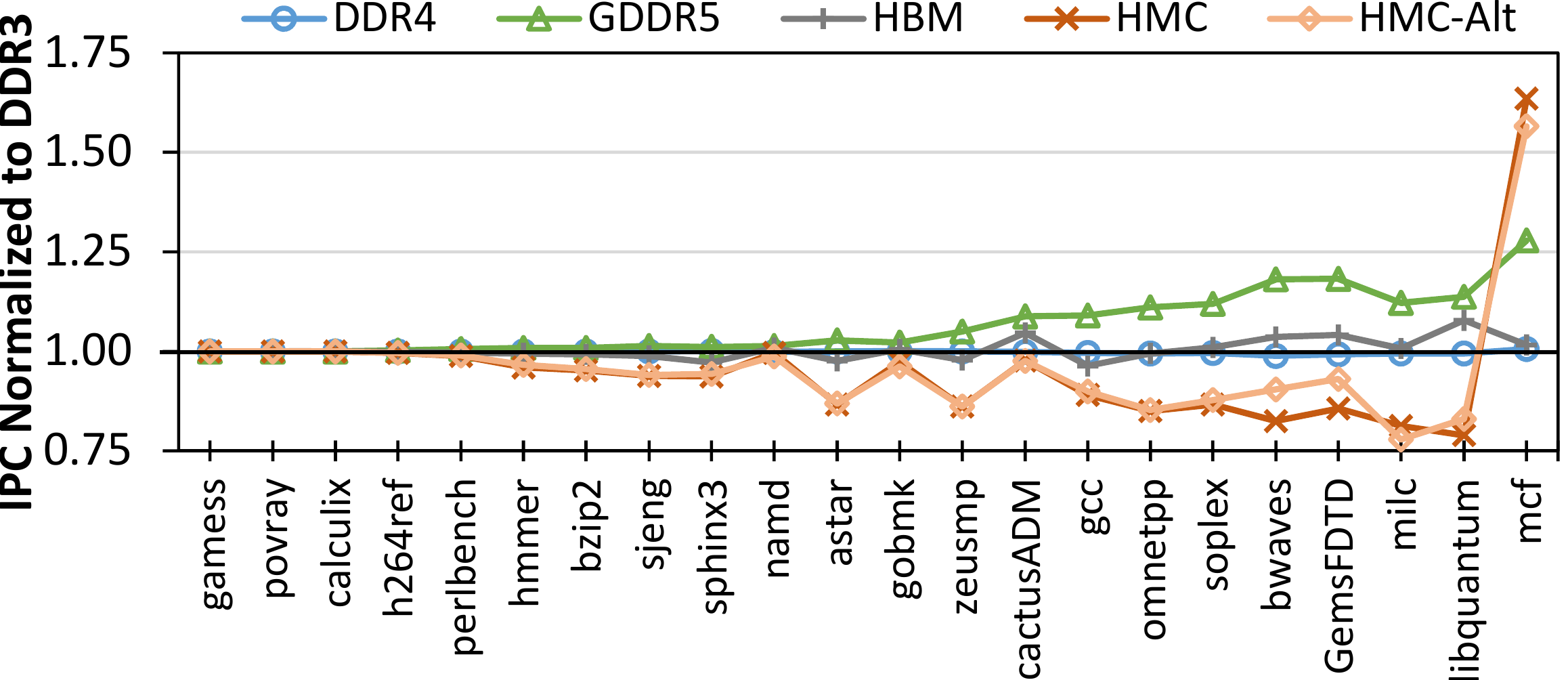}%
  \vspace{5pt}\\
  \includegraphics[width=0.9\columnwidth, trim=65 168 60 155, clip]{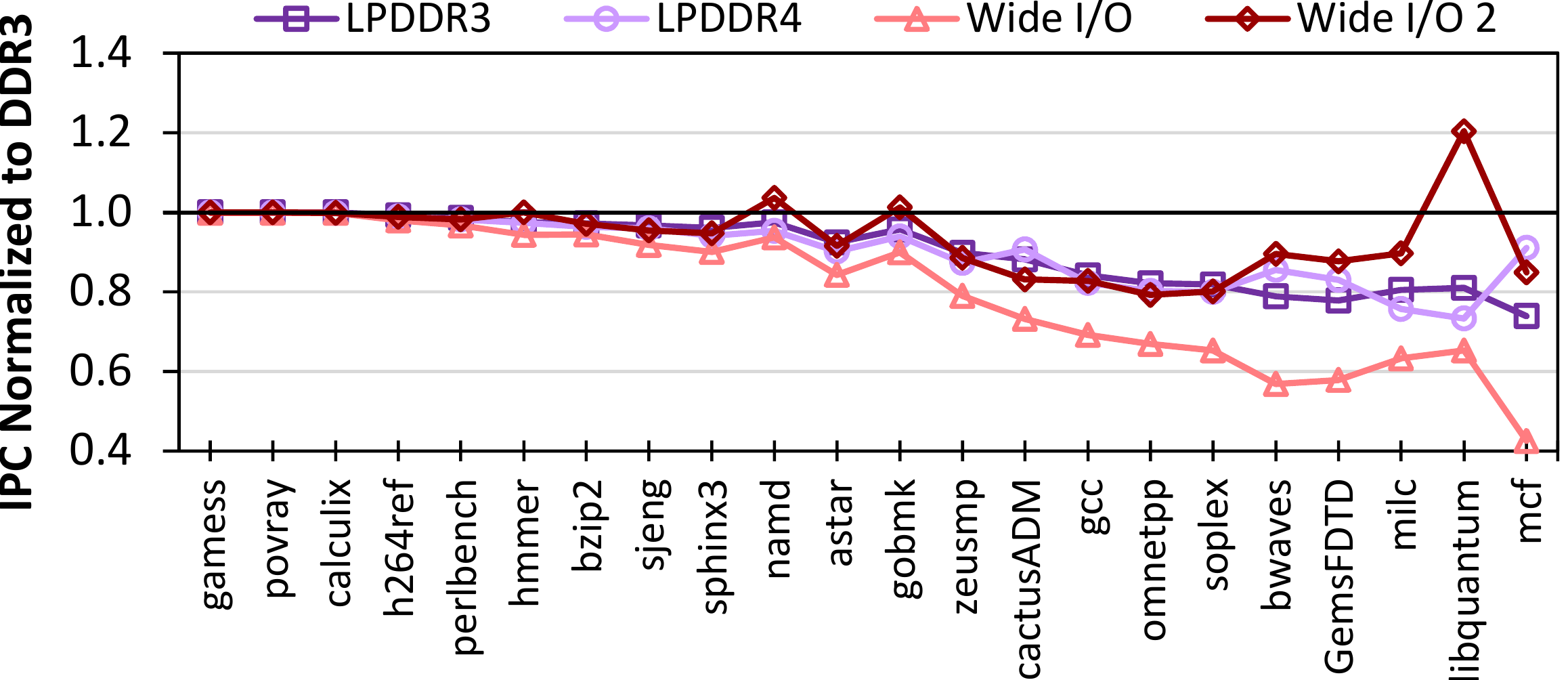}%
  \vspace{-10pt}%
  \caption{Performance of desktop and scientific applications for standard-power (top) and
  low-power (bottom) DRAM types, normalized to DDR3.}
  \label{fig:desktop:singleperf}
  \label{fig:desktop:singlelp}
\end{figure}

Figure~\ref{fig:desktop:singleperf} (top) shows the performance of the desktop
workloads under each of our standard-power DRAM types, normalized to the performance of
each workload when using a DDR3-2133 memory.  
\changes{Along the x-axis, the applications are sorted by MPKI, from least to
greatest.}
We make two observations from these experiments.

\obs{obs:ddr4}{DDR4 does not perform better than DDR3 for the vast majority of our desktop/scientific applications.}

\begin{figure}[h]
  \centering
  \includegraphics[width=0.9\columnwidth, trim=65 185 60 174, clip]{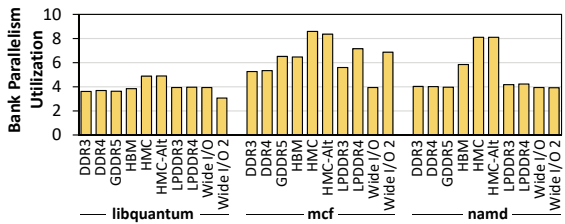}%
  \vspace{-10pt}%
  \caption{BPU for representative desktop/scientific applications.}
  \label{fig:desktop:bpu}
\end{figure}

\changes{\changesvii{Even} though DDR4 \changesv{has 50\% higher bandwidth than DDR3} and
contains double the number of banks 
\changesvii{(64 in our four-channel DDR4 configuration vs.\ 32 in our four-channel DDR3 configuration)}, 
DDR4 performs 0.2\% \emph{worse} 
than DDR3, on average across all of our desktop and scientific applications,
\changesvii{as we see in Figure~\ref{fig:desktop:singleperf} (top)}.
The best \changesv{performance with DDR4 is for \emph{mcf}, with an} improvement
of only 0.5\% \ch{over DDR3}.
We find that both major advantages of DDR4 over DDR3 (i.e., greater bandwidth,
more banks) are not useful to our applications.
Figure~\ref{fig:desktop:bpu} shows the BPU for three representative workloads
(\emph{libquantum}, \emph{mcf}, and \emph{namd}).
Across all of our applications, we find that there is not enough BPU to
take advantage of the 32 DDR3 banks, let alone the 64 DDR4 banks.
\emph{mcf} has the highest BPU, at 5.33 in DDR4, still not enough to
benefit from the additional banks.
Instead, desktop and scientific applications are sensitive to the
\emph{memory latency}.
\changesvii{Applications} are \emph{hurt} by the increased access latency in DDR4
\changesvii{(\ch{11/14\% higher in DDR4 for a row hit/miss} than in DDR3)},
which is a result of the bank group organization (which does not exist in DDR3;
see Section~\ref{sec:bkgd:types}).}

\obs{obs:hmc}{HMC performs significantly worse than DDR3 \changes{when} a workload 
\changesv{can exploit row buffer locality \\and is not highly
memory intensive}.}

\changes{From Figure~\ref{fig:desktop:singleperf} (top), we observe that
few standard-power DRAM types can improve performance across \ch{\emph{all}}
desktop and scientific applications over DDR3.  Notably, we find that
HMC actually results in significant \emph{slowdowns} over DDR3 for
most of our \ch{single-threaded} applications.  Averaged across all workloads, HMC performs 5.8\%
worse than DDR3.
To understand why, we examine the row buffer locality of our applications
\changesv{when running with different memory types}.
Recall from Section~\ref{sec:bkgd:types} that HMC reduces row buffer locality
in exchange for a much greater number of banks (256 in HMC vs.\ 32 in DDR3)
and much greater bandwidth (4.68$\times$ the bandwidth of DDR3).
We already see in Figure~\ref{fig:desktop:bpu} that, with the exception of
\emph{mcf}, HMC cannot provide significant BPU increases for our 
\ch{single-threaded} applications,
indicating that the applications cannot take advantage of 
\changesvii{the increased bank count and \chii{higher} bandwidth}.}

\begin{figure}[h]
  \centering
  \includegraphics[width=0.9\columnwidth, trim=65 185 60 185, clip]{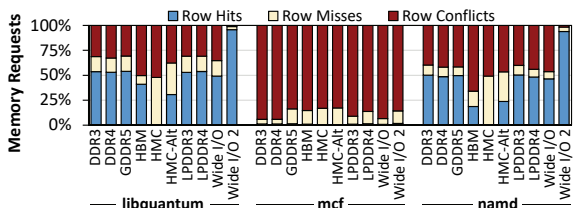}%
  \vspace{-10pt}%
  \caption{Breakdown of row buffer locality for representative \ch{single-threaded} desktop/scientific applications.}
  \label{fig:desktop:rbl}
\end{figure}

\changes{Figure~\ref{fig:desktop:rbl} shows the row buffer locality
(see Section~\ref{sec:metrics}) for our three representative applications.
As we observe from the figure, HMC eliminates \changesvii{nearly} all of the row hits
that other memories attain \changesv{in \emph{libquantum} and \emph{namd}}.  
This is a result of the row size in
HMC, which is 97\% smaller than the row size in DDR3.
This causes many more row misses to occur, without significantly
affecting the number of row conflicts.
As a result, the average \changesvii{memory request} latency \changesvii{(across all applications)} 
in HMC is 25.6\% higher than \ch{that} \changesvii{in DDR3.}
\changesvii{The only application with a lower average memory request latency in HMC}
is \emph{mcf}, 
\changesvii{because the majority of \chii{its} memory requests in \emph{all} \ch{DRAM} types
are row conflicts (see middle graph in Figure~\ref{fig:desktop:rbl})}.
\ch{Thus, due to its low spatial locality and high BPU,} \emph{mcf}
is the only application that sees a
significant speedup with HMC (\changesv{63.4\% over} DDR3).}

\changes{Unlike HMC, GDDR5 successfully improves the performance of
\ch{\emph{all}} of our desktop and scientific applications with higher memory intensity.
This is because GDDR5 delivers higher bandwidth 
\ch{at a lower latency than DDR3} \changesvii{(see Table~\ref{tbl:dram}),}
which translates into an average performance improvement of
6.4\%.
In particular, for applications with high memory intensity (i.e., MPKI $>$ 15.0),
GDDR5 has an average speedup of 16.1\%, as these applications benefit
most from a combination of \changesvii{higher memory} bandwidth and 
\changesvii{lower} memory request latencies.}

Figure~\ref{fig:desktop:singlelp} (bottom) shows the performance of the
desktop and scientific applications \changes{when we use low-power
or mobile} DRAM types.
In general, we note that as the memory intensity (i.e., MPKI) of an application
increases, its performance with low-power memory decreases 
\changes{compared to DDR3}.
In particular, LPDDR3 and LPDDR4 perform worse because
they take longer to complete a memory request,
increasing the latency for a row miss over DDR3 \changesvii{and DDR4}
by 53.2\% and 50.0\%, respectively \changesvii{(see Table~\ref{tbl:dram})}.  
Wide I/O DRAM performs significantly worse than \changes{the} other DRAM types, 
as \changesvii{(1)~}its much lower clock
frequency \changes{greatly} restricts its overall throughput,
\changesvii{and (2)~its row hit latency is \ch{longer}}.  Wide I/O 2
offers significantly higher row buffer locality \changesvii{and lower hit latency} than Wide I/O. 
As a result, applications such as \emph{namd} and \emph{libquantum}
perform well under Wide I/O 2.

\changesv{\emph{We conclude that even though \ch{single-threaded} desktop and scientific applications display a wide
range of memory access behavior, they generally need 
\changes{DRAM types that offer (1)~low access latency and
(2)~high row buffer locality.}}}

\subsection{Multiprogrammed Workload Performance}
\label{sec:desktop:mw}

We combine the \ch{single-threaded applications} into 20 four-core multiprogrammed workloads
\changes{(see Table~\ref{tbl:mwdesktop} in Appendix~\ref{sec:artdesc:workloads} 
for workload details)},
to study how the memory access behavior changes.  
\changes{Figure~\ref{fig:desktop:mw} shows the performance of the workloads
\chv{(sorted by MPKI)} with each DRAM type.
\changesvii{We \ch{draw out} three findings from the figure.}

\begin{figure}[h]
  \centering
  \includegraphics[width=0.9\columnwidth, trim=65 128 60 135, clip]{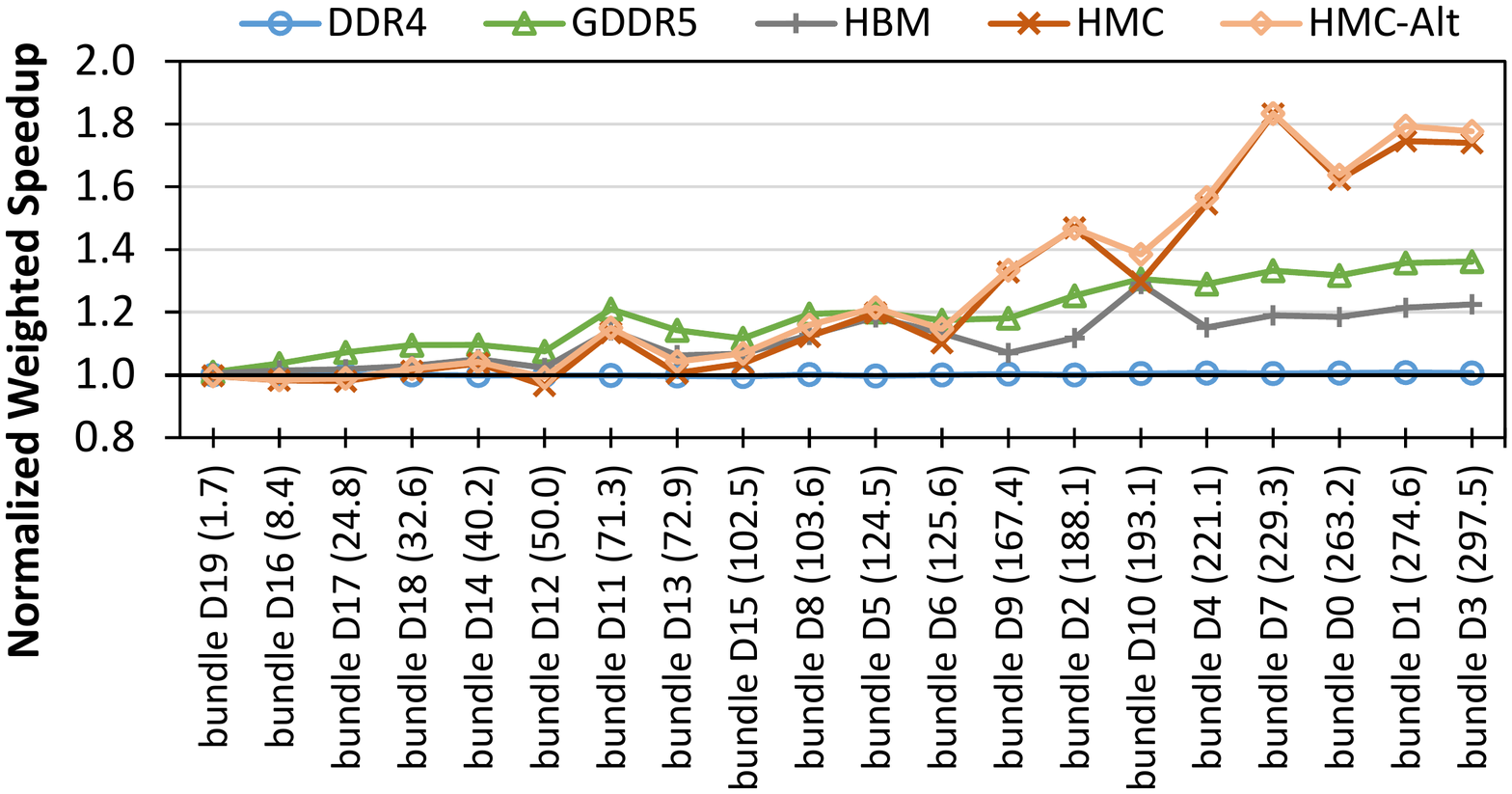}%
  \vspace{5pt}\\
  \includegraphics[width=0.9\columnwidth, trim=65 165 60 155, clip]{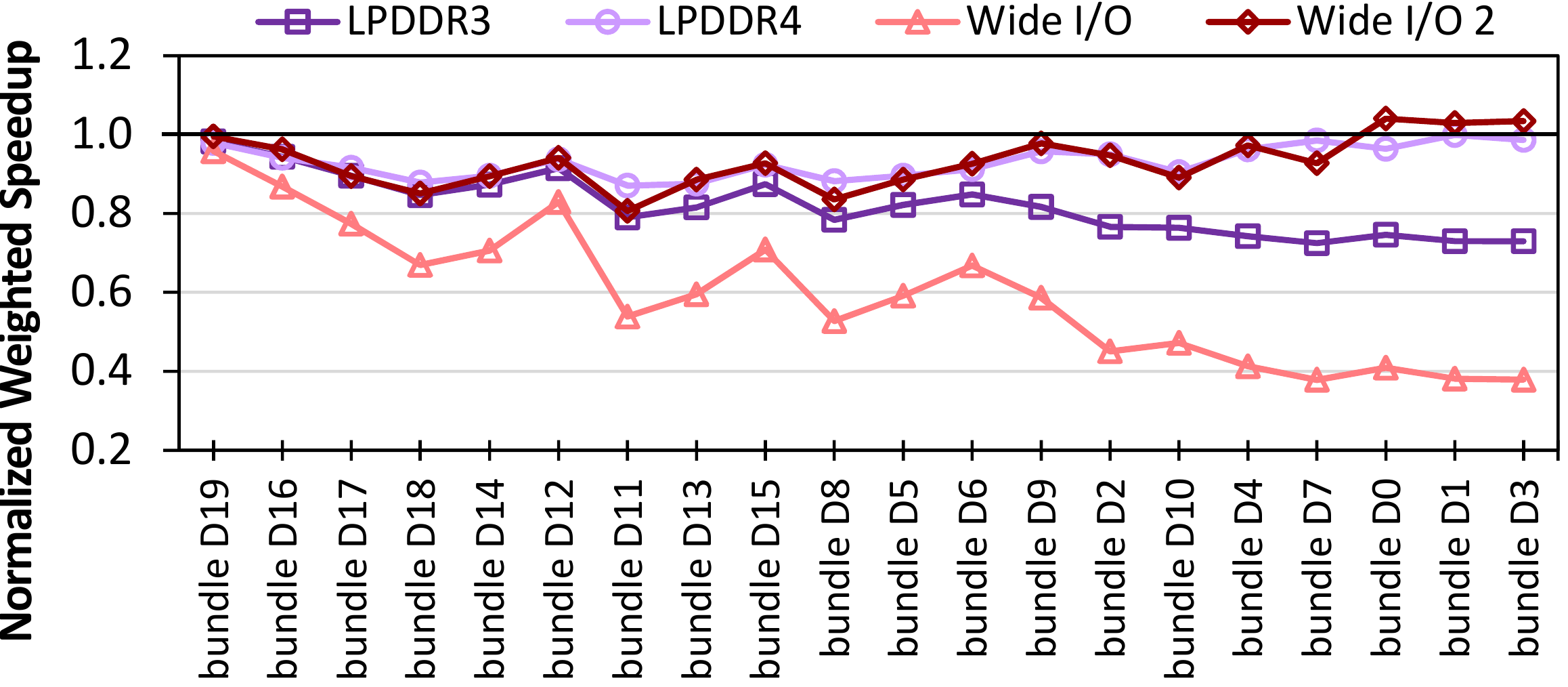}%
  \vspace{-10pt}%
  \caption{Performance of multiprogrammed desktop and scientific workloads for standard-power 
  (top) and low-power (bottom)
  DRAM types, normalized to DDR3.
  MPKI listed in parentheses.}
  \label{fig:desktop:mw}
\end{figure}

\obs{obs:mwhmc}{Multiprogrammed workloads with high aggregate memory
intensity benefit significantly from HMC, \\due to
\changes{a combination of high BPU and} poor row buffer locality.}

\changesvii{First, for} multiprogrammed workloads, HMC performs
better than the other DRAM types despite its significantly smaller row \changesvii{buffer size}.
On average, HMC improves \changesvii{system performance (as measured by
weighted speedup) by 17.0\% over DDR3}.  Note that while some workloads do
very well under HMC (\ch{with the greatest performance improvement being
\chii{83.1\%} for \emph{bundle D7}}),
\ch{many} workloads with lower memory intensity (i.e., MPKI $<$ 70) still perform 
\chii{slightly worse than they do under}
DDR3 (with the greatest \changesvii{performance loss} being \ch{3.4\% for
\emph{bundle D12}}).}
\changesv{We find two \changesvii{major} reasons for HMC's high performance with
multiprogrammed workloads: \changesvii{poor row buffer locality and high BPU}.}

\begin{figure}[h]
  \centering
  \includegraphics[width=0.9\columnwidth, trim=65 185 60 185, clip]{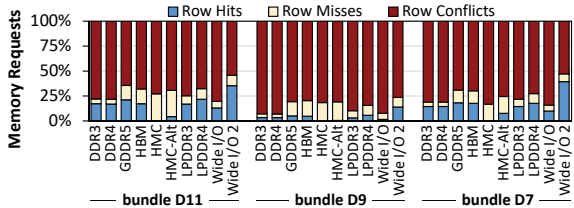}%
  \vspace{-10pt}%
  \caption{Breakdown of row buffer locality for representative multiprogrammed desktop/scientific workloads.}
  \label{fig:desktop:mw-rbl}
\end{figure}

\changesvii{The} row buffer locality of the multiprogrammed workloads is much
\emph{lower} than \changesvii{that of} the \ch{single-threaded} applications.
\changes{Figure~\ref{fig:desktop:mw-rbl} shows row buffer locality for three representative
workloads.  For \emph{bundle~D9}, which has an
MPKI of 167.4, the row buffer hit rate never exceeds 5.6\% on any DRAM type.
We observe that for all of our workloads, the vast majority of memory accesses are 
row conflicts.
\changesvii{This is because each application in a multiprogrammed workload accesses
a different address space, and these competing applications frequently interfere
with each other when they conflict in banks or channels within the shared DRAM,
\ch{as also observed in prior works~\cite{kim.isca12, hassan.hpca16, hassan.isca19,
lee.hpca13}}.}}

\begin{figure}[h]
  \centering
  \includegraphics[width=0.9\columnwidth, trim=65 185 60 174, clip]{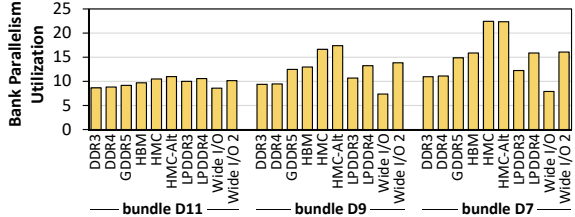}%
  \vspace{-10pt}%
  \caption{BPU for representative multiprogrammed desktop/scientific workloads.}
  \label{fig:desktop:mw-bpu}
\end{figure}

\changesvii{With HMC, we find that the BPU of highly-memory-intensive workloads is 
significantly higher than the BPU with DDR3.}
Figure~\ref{fig:desktop:mw-bpu} shows the BPU for the \ch{three} representative workloads.
\changesv{\emph{Bundle~D11}, which has an MPKI of 71.3, does} not issue enough
parallel memory requests, limiting its BPU.
For \emph{bundle~D7}, which has a much higher MPKI of 229.3, concurrent memory
requests are distributed across the memory address space, as three out of the four applications
in the workload (\emph{libquantum}, \emph{mcf}, and \emph{milc}) are memory
intensive \changesvii{(i.e., MPKI $\geq$ 4.0 for \ch{single-threaded} applications)}.  
As a result, with HMC, the workload achieves 2.05$\times$ the BPU that
it does with DDR3.

\changes{\changesvii{Second, unlike HMC, which does not perform well for \ch{most}
non-memory-intensive multiprogrammed workloads,} GDDR5 \ch{improves performance} for 
\emph{all} 20 of our multiprogrammed workloads.  This is because
GDDR5 provides a balanced combination of low memory latencies,
\changesvii{high} bank parallelism, and high bandwidth.
However, GDDR5's balance \changesvii{across these metrics}
is not enough to maximize the performance
of our highly-memory-intensive workloads, \ch{which require very high bandwidth,}
and thus \ch{GDDR5's} average \changesvii{performance}
improvement over \changesv{DDR3
\changesvii{on multiprogrammed workloads}, 13.0\%, is lower than that of HMC (17.0\%)}.}

\changesvii{Third, some low-power DRAM types can provide energy savings
(see Section~\ref{sec:desktop:energy}) 
for multiprogrammed workloads \ch{\emph{without}} sacrificing performance.}
\changes{From Figure~\ref{fig:desktop:mw} (bottom), we observe that
LPDDR4 and Wide I/O 2 perform competitively with DDR3 for
highly-memory-intensive workloads.  This is because both DRAM types
provide \chiii{higher amounts of parallelism and bandwidth than DDR3,}
and the \chii{highly-memory-intensive}
\ch{applications make significant use of the available parallelism
and memory bandwidth, which \chii{lowers application} execution time.
As a \chiv{result,}
such applications are not
significantly impacted by the increased memory access latency
\chii{in LPDDR4 and Wide I/O 2}.}}

\changesv{\emph{We conclude that for multiprogrammed workloads,
DRAM types that provide high bank parallelism and bandwidth
can significantly improve performance when a workload
exhibits (1)~high memory intensity, (2)~high BPU, and
(3)~poor row buffer locality.}}

\subsection{DRAM Energy Consumption}
\label{sec:desktop:energy}

\begin{figure}[h]
  \centering
  \includegraphics[width=0.9\columnwidth, trim=70 120 60 270, clip]{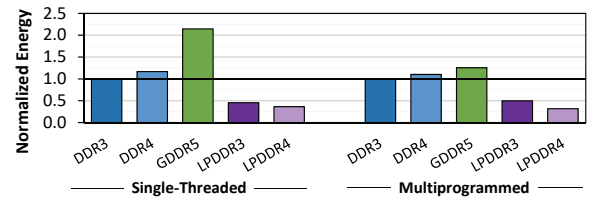}%
  \vspace{-10pt}
  \caption{Mean DRAM energy consumption for \ch{single-threaded} (left) and multiprogrammed (right) desktop and scientific applications,
    normalized to DDR3.}
  \label{fig:desktop:energy}
\end{figure}

\changes{We 
characterize the energy consumption of our desktop \changesv{and scientific} workloads
for the DRAM types that we have accurate power models for (i.e.,
datasheet \ch{values for power consumption that are} provided by vendors for actual off-the-shelf parts).}
Figure~\ref{fig:desktop:energy} shows the average DRAM energy consumption 
by DDR3, DDR4, GDDR5, LPDDR3, and LPDDR4 for our
\ch{single-threaded applications and multiprogrammed workloads}, normalized to the energy consumption of DDR3.
\changes{We make two \ch{new observations} from the figure.}

\obs{obs:lpddrpower}{\ch{LPDDR3/4 reduce} DRAM energy consumption by
\ch{as much as} 54--68\% over DDR3/4, but \changesvii{LPDDR3/4 provide worse
performance for \ch{single-threaded} applications, with \chii{their performance 
loss increasing}}
as the memory intensity increases.}

\changes{\ch{For} all of our desktop/scientific workloads,
LPDDR3/4 consume significantly less energy than DDR3/4 due to the numerous
low-power features incorporated in their design (see Section~\ref{sec:bkgd:types}).
In particular, as we discuss in Appendix~\ref{sec:detail:desktoppwr}, standby power is the
\changesv{single} largest source of power consumption for these workloads, and
LPDDR3/4 incorporate a number of optimizations to reduce standby power.
Unfortunately, 
these optimizations lead to increased memory request latencies \changesvii{(see Table~\ref{tbl:dram})}.  
This, in turn, hurts the overall performance
of \ch{single-threaded} applications, as we see in Figure~\ref{fig:desktop:singlelp} (bottom).
GDDR5 makes the opposite trade-off, with reduced memory request latencies and
\ch{thus} higher performance, but at the cost of 2.15$\times$ more energy than DDR3
for \ch{single-threaded} applications.

\obs{obs:lpddrpowermw}{For highly-memory-intensive multiprogrammed workloads,
LPDDR4 provides significant energy savings over DDR3 without sacrificing performance.}

\changes{For multiprogrammed workloads, LPDDR4 delivers a 68.2\% reduction in 
energy consumption, on average across all workloads, while losing only 7.0\%
\changesv{performance compared to} DDR3 \changesvii{(see Section~\ref{sec:desktop:mw})}.  
This is because LPDDR4 compensates for its higher
memory request latency \ch{over DDR3 by having} 
a greater number
of banks.  As we discuss in Section~\ref{sec:desktop:mw}, highly-memory-intensive
multiprogrammed workloads can achieve a high BPU, which allows them to
take advantage of the increased bank parallelism available in \changesv{LPDDR4}.
As a comparison, LPDDR3 still performs poorly with these workloads
because it \changesv{has lower} bandwidth and a lower bank count
\ch{than LPDDR4}.}
\changesvii{In contrast, GDDR5 provides higher throughput than LPDDR4,
and due to the high memory intensity of multiprogrammed workloads, 
the workloads complete much faster with GDDR5 than DDR3
(13.0\% higher performance on average; see Section~\ref{sec:desktop:mw}).
The increased performance of GDDR5 comes at the cost of consuming
only 25.6\% more energy on average than DDR3, which is a much smaller
increase than \ch{what we observe for the single-threaded} applications.}

\changesvii{\emph{We conclude that
(1)~low-power DRAM variants (LPDDR3/4) are effective at 
reducing overall DRAM energy consumption, especially for
applications that exhibit high BPU; and
(2)~the performance improvements of GDDR5 come with a 
significant energy penalty for \ch{single-threaded} applications,
but with a smaller penalty for multiprogrammed workloads.}}

%% file: sections/multithreaded.tex
% !TEX root=../dramcharacterization.tex

\section{Multithreaded Desktop and Scientific Programs}
\label{sec:mt}

Many modern applications, especially in the high-performance computing
domain, launch multiple threads on a machine to exploit the \changesv{thread-level} parallelism 
available in multicore systems.  We evaluate the following applications:
\begin{itemize}
    \item \emph{blackscholes}, \emph{canneal}, \emph{fluidanimate},
    \emph{raytrace}, \emph{bodytrack}, \emph{facesim}, \emph{freqmine},
    \emph{streamcluster}, and \emph{swaptions} from PARSEC 3.0~\cite{parsec}, and
    \item \emph{miniFE}, \emph{quicksilver}, and \emph{pennant} from
    CORAL~\cite{coral}/CORAL-2~\cite{coral-2}.
\end{itemize}

\subsection{Workload Characteristics}
\label{sec:mt:workload}

\ch{Multithreaded} workloads often work on very large datasets 
(e.g., several gigabytes in size) that 
are partitioned across the multiple threads.
A major component of multithreaded application behavior is how
the application scales with the number of threads.  This scalability is 
typically a function of (1)~how memory-bound an application is,
(2)~how much synchronization must be performed across threads, and
(3)~how \changesv{balanced} the work done by each thread is.

We provide a detailed experimental analysis of the IPC and MPKI 
of the multithreaded applications in Appendix~\ref{sec:char:mt}.
From the analysis, we find that these applications
have a narrower IPC range than the \ch{single-threaded} desktop applications.
This is often because multithreaded applications are designed to strike a
careful balance between computation and memory usage, which is
necessary to scale the algorithms to large numbers of threads.
Due to this balance, memory-intensive 
multithreaded applications have significantly higher IPCs compared to
\ch{memory-intensive single-threaded} desktop/scientific applications,
even as we scale the number of threads.
For example, the \changesvii{aggregate} MPKI of \emph{miniFE} increases from 11.5 with only
one thread to 68.1 with 32~threads, but
its IPC \changesvii{per thread} remains around 1.5 \chii{(for both one thread
and 32~threads)}.
\ch{The \chii{relatively} high IPC indicates that the application is not completely
memory-bound even when its MPKI is high}.

\subsection{Performance}
\label{sec:mt:perf}

To study performance and scalability,
we evaluate 1, 2, 4, 8, 16, and 32~thread runs of each multithreaded
application on each DRAM type.  
All performance plots show parallel speedup, normalized to
one-thread execution on DDR3, \changesvii{on the y-axis,
and the thread count \ch{(in log scale)} on the x-axis}.
For brevity, we do not
show individual results for each application.  We find that the applications
generally \changesv{fall} into one of three categories:
(1)~\emph{memory-agnostic}, where the application is able to
achieve near-linear speedup across most thread counts \changesvii{for all} DRAM
types;
(2)~\emph{throughput-bound memory-sensitive}, where the application
is highly memory-intensive, and has trouble approaching
linear speedup for most DRAM types; and
(3)~\emph{irregular memory-sensitive}, where the application is
highly memory-intensive, \changesvii{and} its irregular memory access patterns
allow it to benefit from either lower memory latency or 
higher memory throughput.

\begin{figure}[h]
  \includegraphics[width=0.47\columnwidth, trim=65 215 375 162, clip]{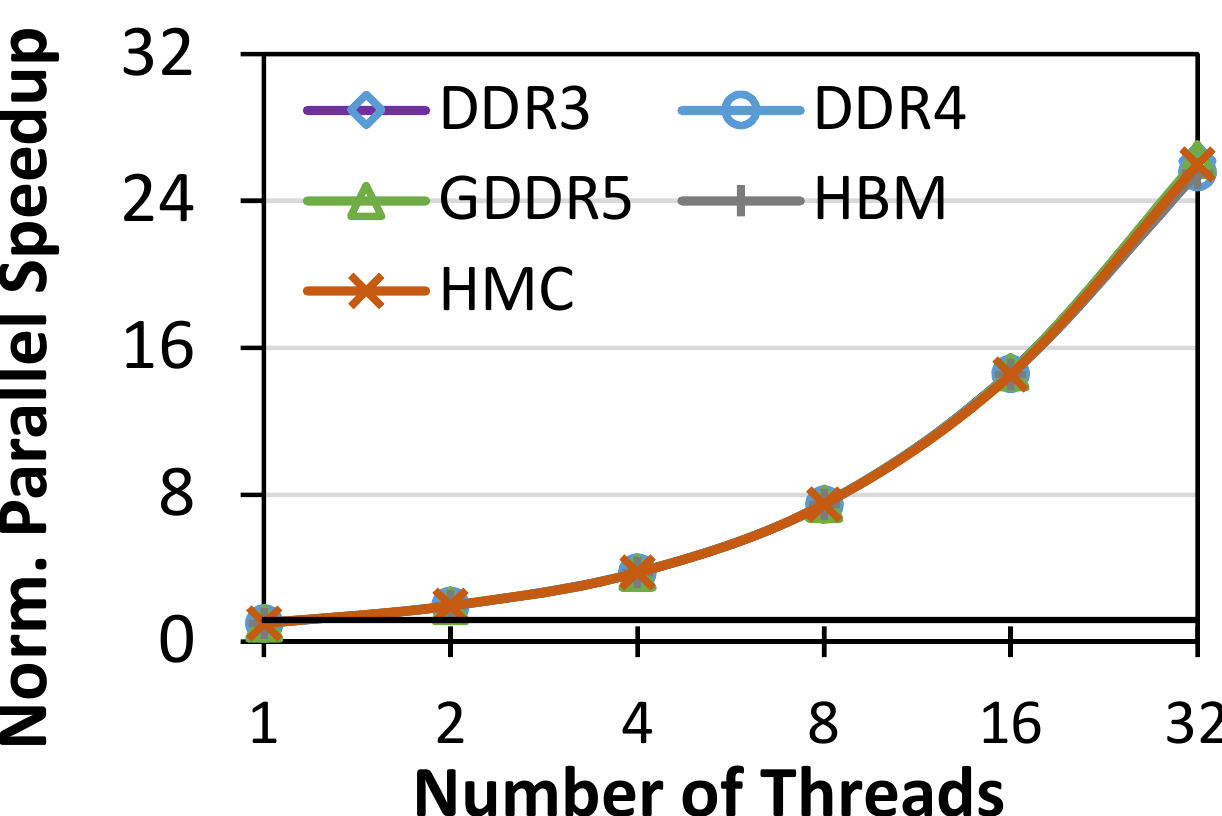}%
  \hfill%
  \includegraphics[width=0.47\columnwidth, trim=65 215 375 162, clip]{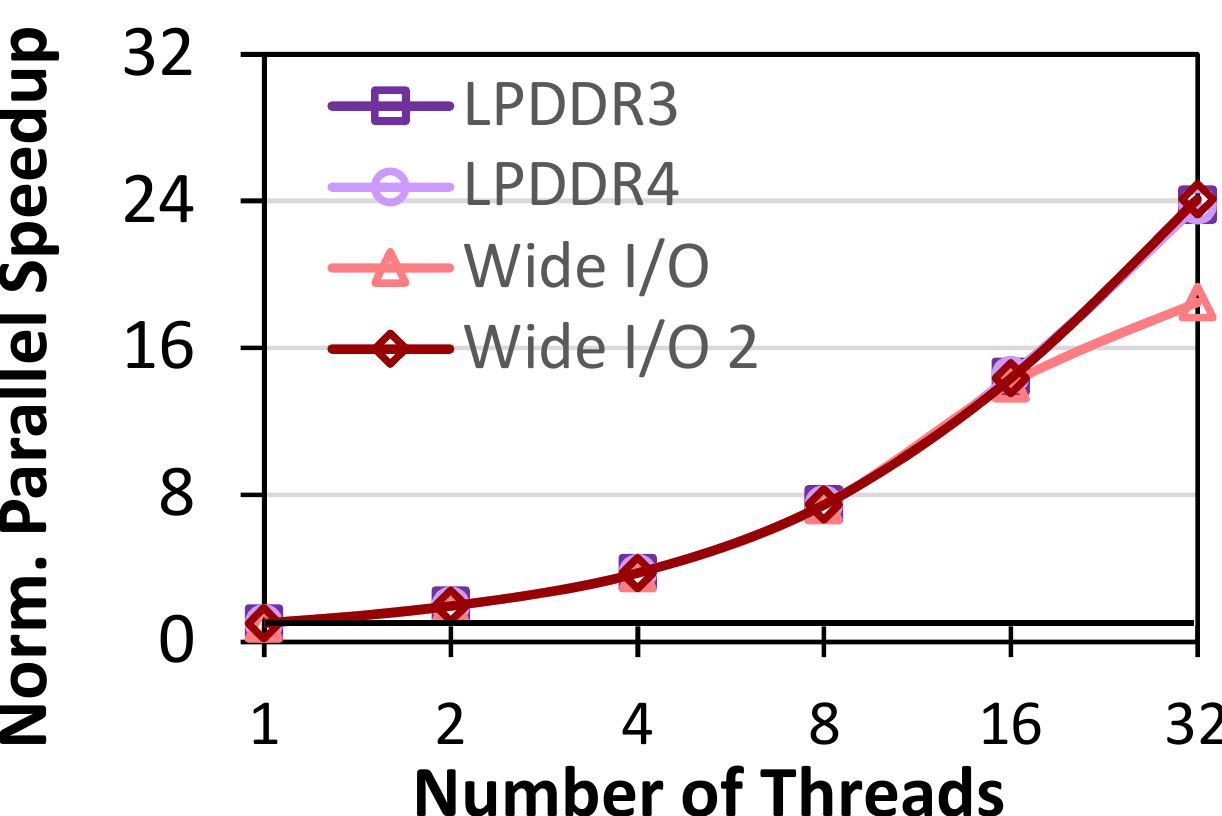}%
  \vspace{-10pt}%
  \caption{Performance of \emph{quicksilver} for standard-power (left) and
  low-power (right) DRAM types, normalized to single-thread performance with DDR3.}
  \label{fig:mt:quicksilver}
\end{figure}

\paratitle{\ch{Memory-Agnostic Applications}}
Six of our applications are \emph{memory-agnostic}: \emph{blackscholes},
\emph{raytrace}, \emph{swaptions}, \emph{quicksilver},
\emph{pennant}, and \emph{streamcluster}.
Figure~\ref{fig:mt:quicksilver} shows the performance of \emph{quicksilver}
across all thread counts, which is representative of the memory-agnostic
applications.  \ch{We draw out three findings from the figure.}

\ch{First, regardless of the DRAM type, the performance of \emph{quicksilver}
scales well with the thread count, with no tapering of
performance improvements for the standard-power DRAM types (i.e.,
DDR3/4, GDDR5, HBM, HMC). 
This is because memory-agnostic applications have relatively low MPKI values
(see Appendix~\ref{sec:char:mt}), even at high thread counts (e.g., 
\emph{quicksilver} has an MPKI of 20.9 at 32~threads).
Therefore, all of the standard-power DRAM types are able to
keep up as the thread counts increase, and the memory-agnostic applications
do not benefit significantly from one DRAM type over another.}

\begin{figure}[t]
  \centering
  \includegraphics[width=0.9\columnwidth, trim=65 185 60 185, clip]{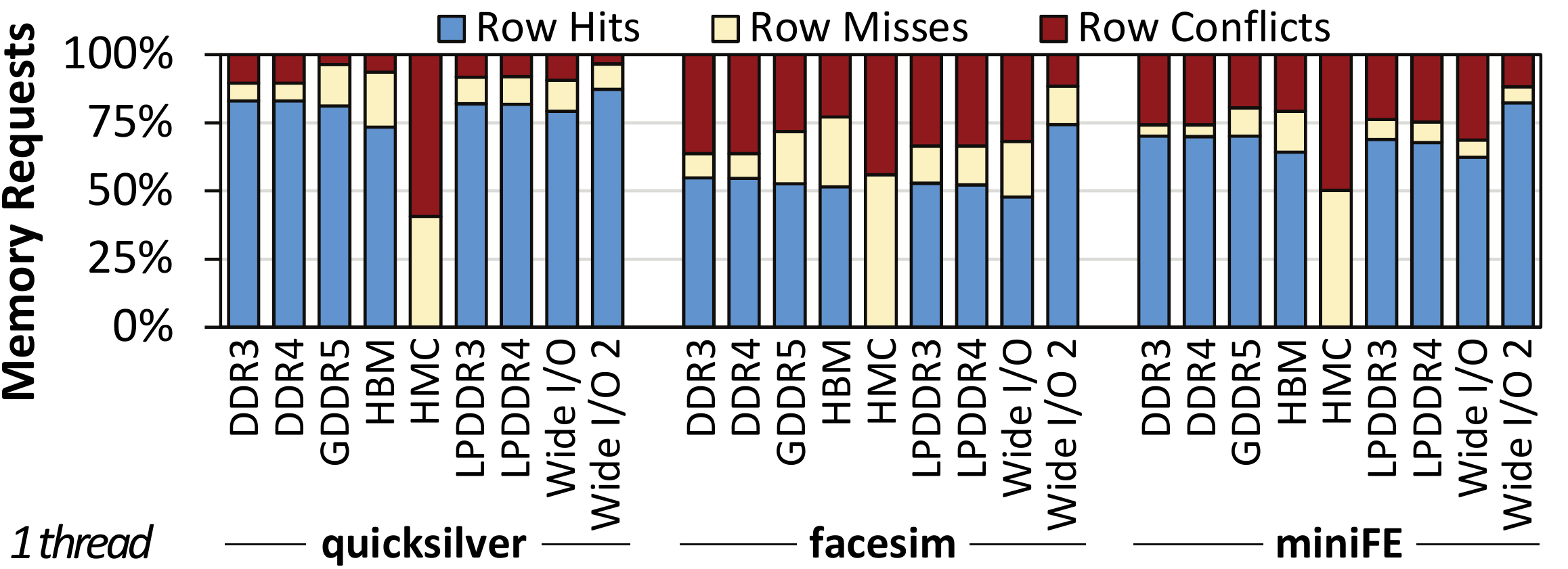}%
  \vspace{5pt}\\
  \includegraphics[width=0.9\columnwidth, trim=65 185 60 185, clip]{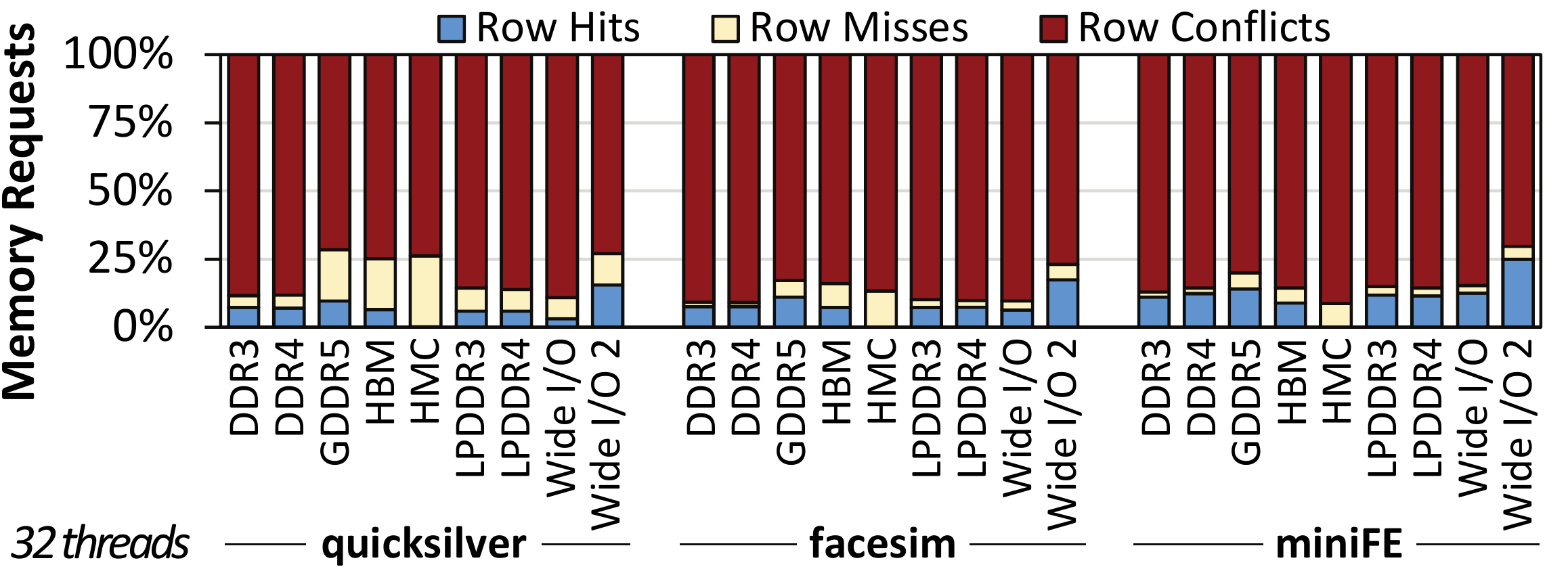}%
  \vspace{-10pt}%
  \caption{Breakdown of row buffer locality for 
  \changesvii{three representative} multithreaded applications.}
  \label{fig:mt:rbl}
\end{figure}

\ch{Second, like many of our memory-agnostic applications, \emph{quicksilver} 
does not have a fully-linear speedup at 32~threads.
This is because when the number of threads increases from 1 to 32, the row hit rate
decreases significantly (e.g., for DDR3, from 83.1\% with one thread to 7.2\% with
32~threads), as shown in Figure~\ref{fig:mt:rbl},
due to contention among the threads for shared \chiii{last-level} cache space
\chii{and \chiii{shared} DRAM banks}.
The significantly lower row hit rate results in an increase in the average memory request latency.
Two of our memory-agnostic applications (\emph{swaptions} and \emph{pennant})
maintain higher row hit rates (e.g., 46.6\% for \emph{pennant} at 32~threads; not shown)
because they have significantly lower memory intensity (i.e., MPKI $<$ 3 at 32~threads)
than our other \chii{memory-agnostic} applications, generating less contention 
at the last-level cache \chii{and DRAM banks}.
As a result, these two applications have a fully-linear speedup at 32~threads.}

\ch{Third, due to the memory-agnostic behavior of these applications,
there is no discernible difference in performance between standard-power
DRAM types, LPDDR3/4, and Wide I/O 2.  Given the minimal memory needs
of these applications, the increased latencies and reduced bandwidth of
\chiii{low-power DRAM types do} not have a significant impact on the 
applications \chiii{in most cases}.\chii{\footnote{The one exception is Wide I/O, 
\chiii{for which performance scaling begins to taper off at 32~threads.  
Wide I/O's poor scalability is a result of \chiv{its combination of} high} memory access latency 
\chiv{and} a memory bandwidth that is significantly lower than the other DRAM
types (see Table~\ref{tbl:dram}).}} Based on these observations, we believe that}
\changesvii{the LPDDR3/4 and Wide I/O 2 low-power DRAM types 
are promising to use for memory-agnostic applications, as they can lower the
DRAM power consumption with \ch{little} impact on performance.}

\begin{figure}[h]
  \includegraphics[width=0.47\columnwidth, trim=65 215 375 162, clip]{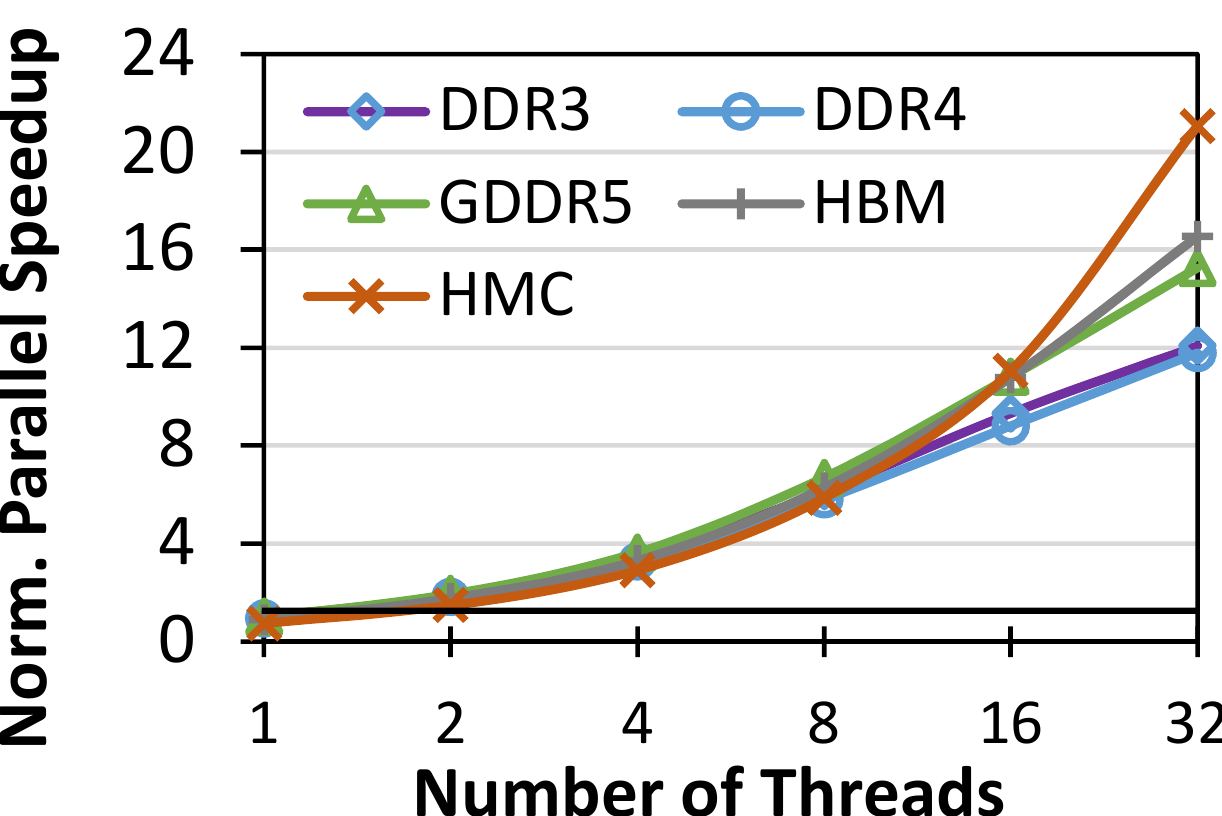}%
  \hfill%
  \includegraphics[width=0.47\columnwidth, trim=65 215 375 162, clip]{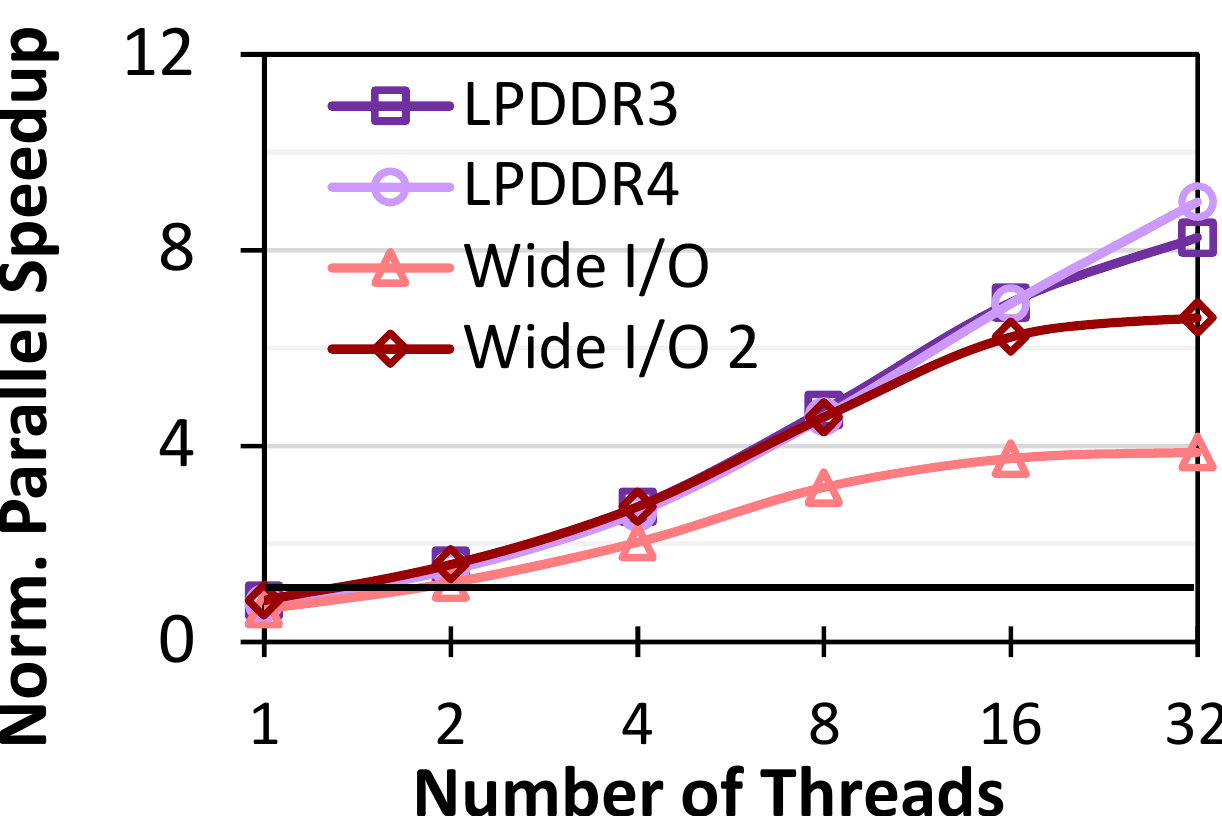}%
  \vspace{-10pt}%
  \caption{Performance of \emph{facesim} for standard-power (left) and
  low-power (right) DRAM types, normalized to single-thread performance with DDR3.}
  \label{fig:mt:facesim}
\end{figure}

\paratitle{\ch{Throughput-Bound Memory Sensitive Applications}}
Five of our applications are \emph{throughput-bound memory-sensitive}:
\emph{bodytrack}, \emph{canneal}, \emph{fluidanimate}, \emph{facesim},
and \emph{freqmine}.
Figure~\ref{fig:mt:facesim} shows the performance of \emph{facesim}
across all thread counts, which is representative of the throughput-bound
memory-sensitive applications.  These applications become highly
memory-intensive (i.e., they have very high \changesvii{aggregate} MPKI values) at high thread counts.  
\changesvii{As} more threads contend for the limited shared space in the
last-level cache, the cache hit rate drops, placing greater pressure on the memory system.
This has two effects.  First, since the memory requests \changesvii{are generated}
across multiple threads, where each thread \changesvii{operates} on its own working set
of data, there is little spatial locality among the requests that are waiting
to be serviced by DRAM at any given time.
As we see in Figure~\ref{fig:mt:rbl}, \emph{facesim} does not exploit
row buffer locality at 32~threads.
Second, because of \chiv{their high memory intensity and}
poor spatial locality, \chiv{these} applications benefit greatly from
a memory like HMC, \chiii{which delivers higher memory-level parallelism
and higher bandwidth \chiv{than DDR3} at the expense of spatial locality and latency}.
As Figure~\ref{fig:mt:facesim}
shows, \changesvii{(1)~the performance provided by the other memories cannot} 
scale at the rate \changesvii{provided by HMC}
at higher thread counts; and \changesvii{(2)~HMC outperforms even GDDR5 and HBM,
which in turn outperform other DRAM types}.

\begin{figure}[h]
  \includegraphics[width=0.47\columnwidth, trim=65 215 375 162, clip]{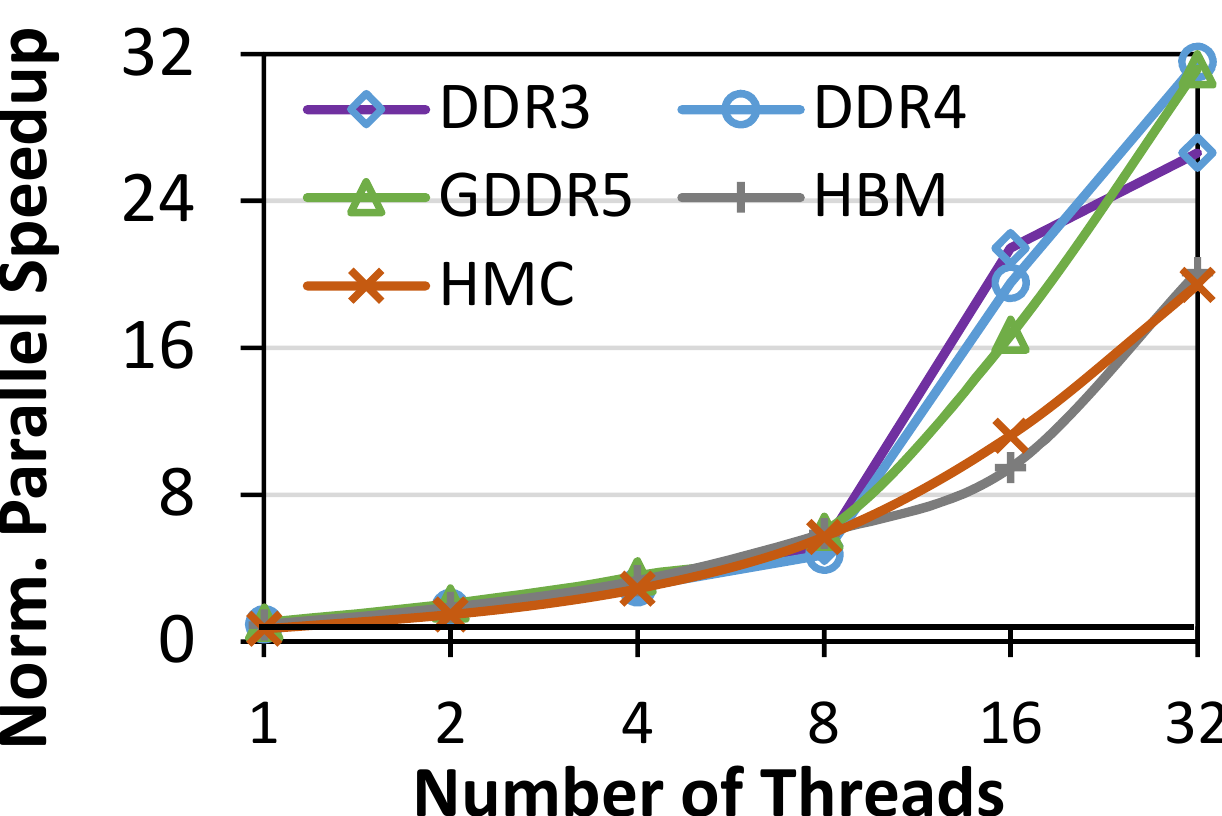}%
  \hfill%
  \includegraphics[width=0.47\columnwidth, trim=65 215 375 162, clip]{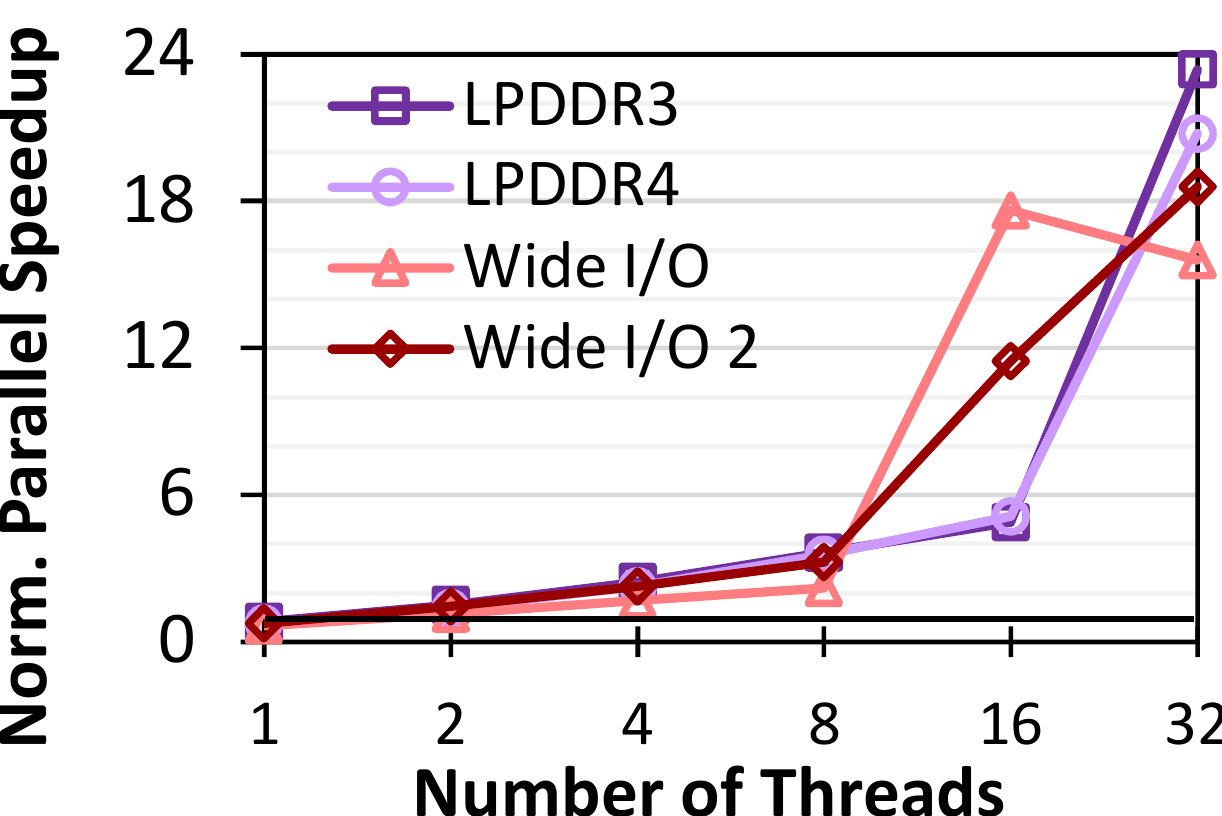}%
  \vspace{-10pt}%
  \caption{Performance of \emph{miniFE} \changesiv{with a 32 x 32 x 32 problem size}
  for standard-power (left) and
  low-power (right) DRAM types, normalized to single-thread performance with DDR3.}
  \label{fig:mt:minife}
\end{figure}

\paratitle{\ch{Irregular Memory-Sensitive Applications}}
Only one application is \emph{irregular memory-sensitive}:
\emph{miniFE}.  
Figure~\ref{fig:mt:minife} shows the performance of \emph{miniFE}
across all thread counts.
\emph{miniFE} operates on sparse matrices, which results in irregular
memory access patterns that compilers cannot easily optimize.
One result of this irregular behavior is low
\changesv{BPU} at all thread counts, corroborating similar observations
by prior work~\cite{tang.micro16} for \emph{miniFE} and 
other irregular multithreaded workloads.  
As a result, \changesv{for smaller problem sizes (e.g., 32 x 32 x 32
for \emph{miniFE}),} \changesvii{\emph{miniFE}
becomes memory-latency-bound, and behaves
much like our \ch{single-threaded} desktop applications in Section~\ref{sec:desktop}.
We \ch{draw out} two findings from Figure~\ref{fig:mt:minife}.
First, \emph{miniFE}
\changesiv{with a 32 x 32 x 32 problem size
benefits most from traditional, low-latency memories such as DDR3/4 and
GDDR5, while \ch{it fails to achieve such high benefits with}
throughput-oriented memories such as HMC and HBM.}
In fact, just as we see for memory-agnostic applications, 
many of the low-power memories
outperform HMC and HBM at all thread counts.  
\changesvii{Second,} as the
core count increases, \emph{miniFE} benefits more from high memory
throughput and high bank-level parallelism.  As a result, while the performance
\changesvii{improvement with DDR3 starts leveling} off after 16~threads,
the performance improvements with HBM and with HMC continue to scale
at 32~threads.
\ch{Unlike DDR3, DDR4 continues to scale as well, as DDR4 provides higher
throughput and more banks than DDR3.}}

\begin{figure}[h]
  \includegraphics[width=0.47\columnwidth, trim=65 215 375 162, clip]{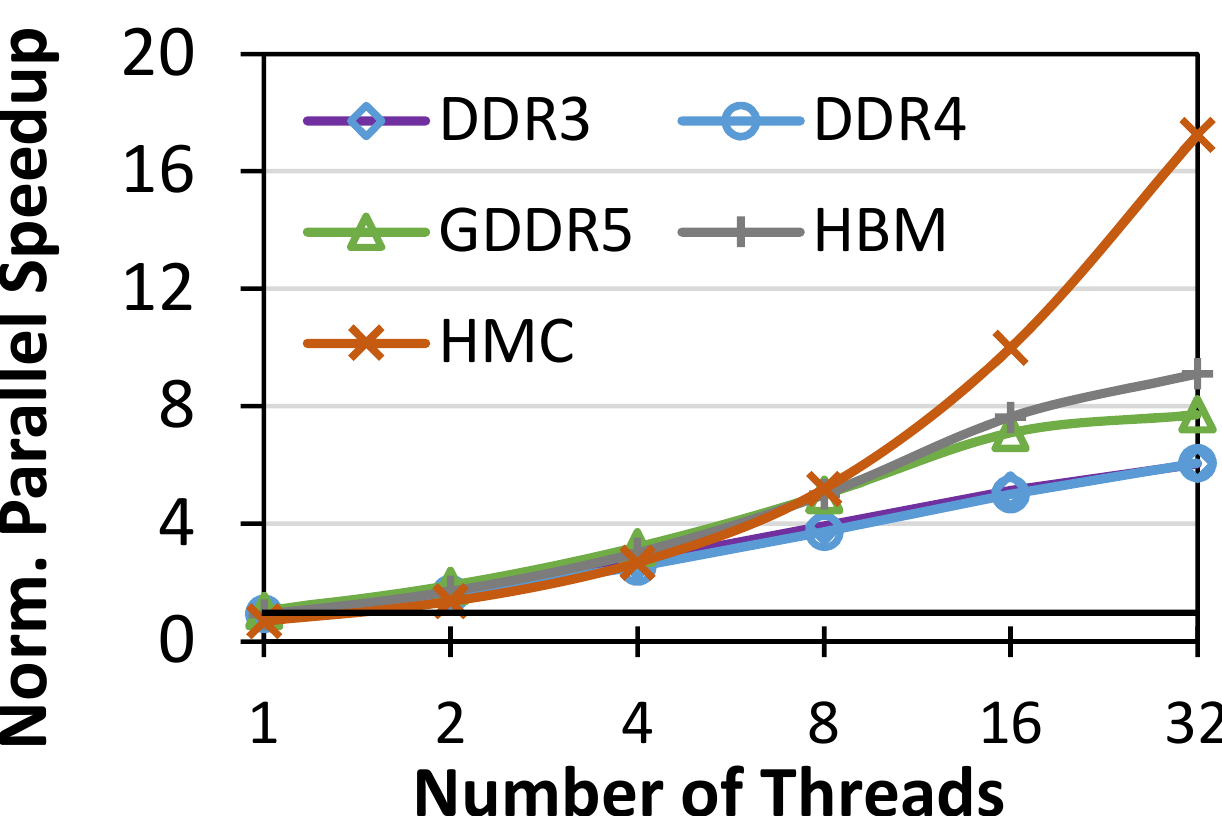}%
  \hfill%
  \includegraphics[width=0.47\columnwidth, trim=65 215 375 162, clip]{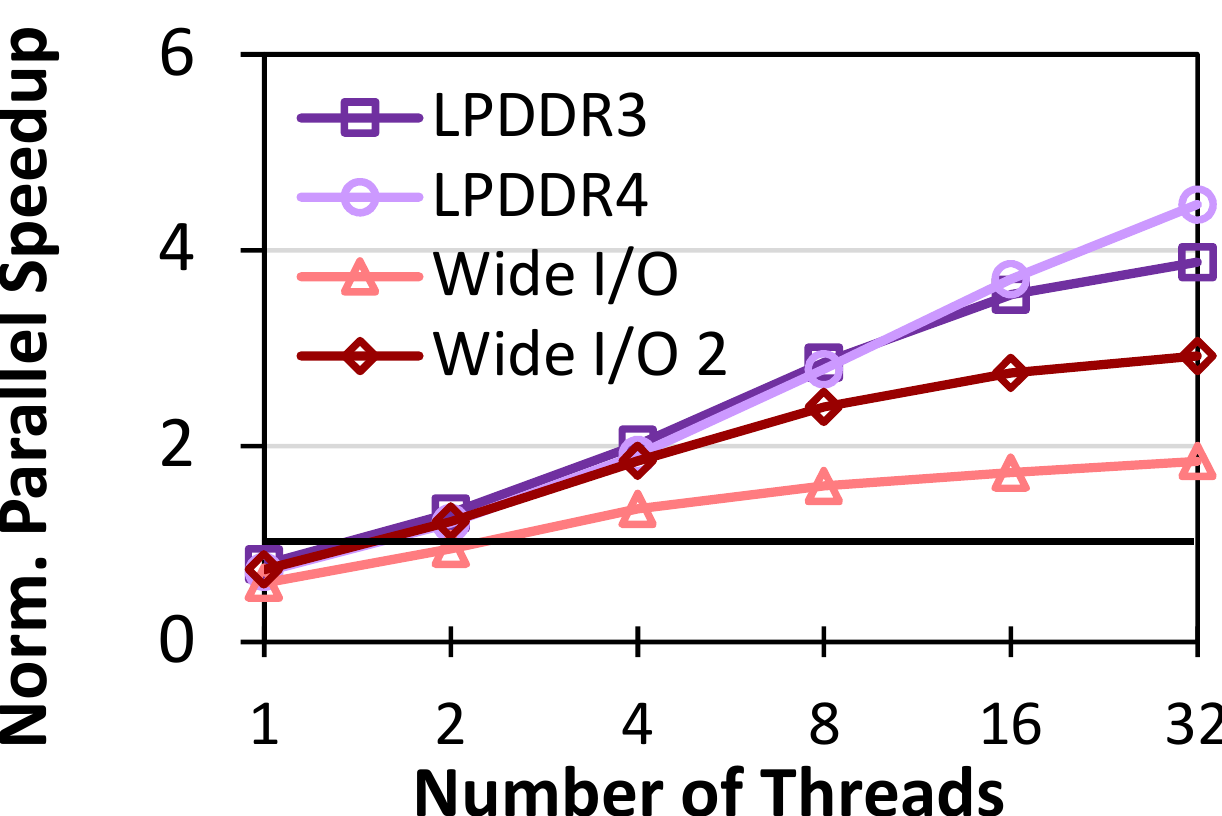}%
  \vspace{-10pt}%
  \caption{\changesiv{Performance of \emph{miniFE} with a 64 x 64 x 64 problem size
  for standard-power (left) and
  low-power (right) DRAM types, normalized to single-thread performance with DDR3.}}
  \label{fig:mt:minife64x64x64}
\end{figure}

\changesiv{The irregular behavior of \emph{miniFE} changes as the problem size
grows.  Figure~\ref{fig:mt:minife64x64x64} shows the performance of \emph{miniFE}
as we increase the problem size to 64 x 64 x 64.  We observe that as the number of threads
increases, the scaling trends look significantly different than they do for the smaller
32 x 32 x 32 problem size (Figure~\ref{fig:mt:minife}).  For reference, with a single thread
and with the DDR3 DRAM type, the 64 x 64 x 64 problem size takes 13.6x longer
than the 32 x 32 x 32 problem size.  The larger problem size puts more pressure on
the cache hierarchy and the memory subsystem \changesv{(e.g., the
memory footprint increases by 449\%; see Table~\ref{tbl:workloads:desktop} in
Appendix~\ref{sec:artdesc:workloads})},
which causes \emph{miniFE} to transition from memory-latency-bound to
memory-throughput-bound.
As a result, when the number of threads increases, lower-throughput DRAM types
such as DDR3 and DDR4 become the bottleneck to scalability, limiting
parallel speedup at 32~threads to only 6.1x.
Likewise, we observe that all of our low-power DRAM types cannot deliver
the throughput required by \emph{miniFE} at high thread counts.
In contrast, HMC can successfully take advantage of the high
throughput and high contention between threads, due to its
\ch{large number of banks and high bandwidth}.
As a result, HMC reaches a parallel speedup of 17.3x at 32~threads,
with no drop-off in its scalability as the number of threads increases
from 1 to 32.
This behavior is similar to what we observe for
throughput-bound memory-sensitive applications
(e.g., the performance of \emph{facesim} in Figure~\ref{fig:mt:facesim}).}

\obs{obs:mt}{\ch{The behavior of multithreaded applications with irregular memory access patterns}\\
\changesv{depends on the problem size.\\~
\newline  At small problem sizes, these applications
are latency-bound, \changesvii{and thus benefit from DRAM types that provide low latency}.
\newline\newline  As the problem size increases, these applications become throughput-bound, 
\changesvii{behaving} like the throughput-bound memory-sensitive applications,
\changesvii{and thus benefit from DRAM types that provide high throughput.}}}

\changesv{\emph{We conclude that the ideal DRAM type for a multithreaded application
is highly dependent on the application behavior, and that for many such applications,
such as memory-agnostic or irregular memory-sensitive applications \changesvii{with
smaller problem sizes,
low-power DRAM types such as LPDDR4 can perform competitively
with standard-power DRAM types}.}}

%% file: sections/server.tex
% !TEX root=../dramcharacterization.tex

\section{Server and Cloud Workloads}
\label{sec:server}

Server and cloud workloads are designed to accommodate
very large data sets, and can often coordinate requests between multiple 
machines across a network.  
We evaluate the following workloads with representative inputs:
\begin{itemize}
    \item the \emph{map} and \emph{reduce} tasks~\cite{mapreduce} for 
            \emph{grep}, \emph{wordcount}, and \emph{sort}, implemented using
            Hadoop~\cite{hadoop} for scalable distributed processing
            \shepiii{(we use four \changesiv{\emph{map}} threads for each application)};
    \item YCSB~\cite{ycsb} OLTP (OnLine Transaction Processing) \emph{server workloads A--E}, and
            the background process forked by workload A to write the log to disk (\emph{bgsave}), 
            executing on the Redis in-memory key-value store~\cite{redis}, 
    \item an \emph{Apache} server~\cite{Apache}, which services a series of \texttt{wget} requests from a
            remote server;
    \item \emph{Memcached}~\cite{memcached}, using a microbenchmark that inserts 
            key-value pairs into a memory cache; and
    \item the \emph{MySQL} database~\cite{difallah.vldb04}, using a microbenchmark that loads the sample 
            \emph{employeedb} database.
\end{itemize}

\subsection{Workload Characteristics}
\label{sec:server:workload}

From our analysis, we find that while server and cloud workloads tend to work 
on very large datasets (e.g., gigabytes of data), the workloads are written to
maximize the efficiency of their on-chip cache utilization.  As a result, these
applications only infrequently issue requests to DRAM, and typically exhibit low
memory intensity (i.e., MPKI $<$ 10) and high IPCs.  We show IPC plots 
\ch{for these workloads} in Appendix~\ref{sec:char:server}.

\changesiii{For each of our Hadoop applications,
we find that the four \changesiv{\emph{map}} threads exhibit near-identical behavior.
As a result, we show the characterization of only one out of the four \changesiv{\emph{map}} threads 
(map~0) in the remainder of this section.}

\subsection{Single-Thread Performance}
\label{sec:server:perf}
\label{sec:server:single}

\begin{figure}[h]
  \centering
  \includegraphics[width=0.9\columnwidth, trim=60 145 60 155, clip]{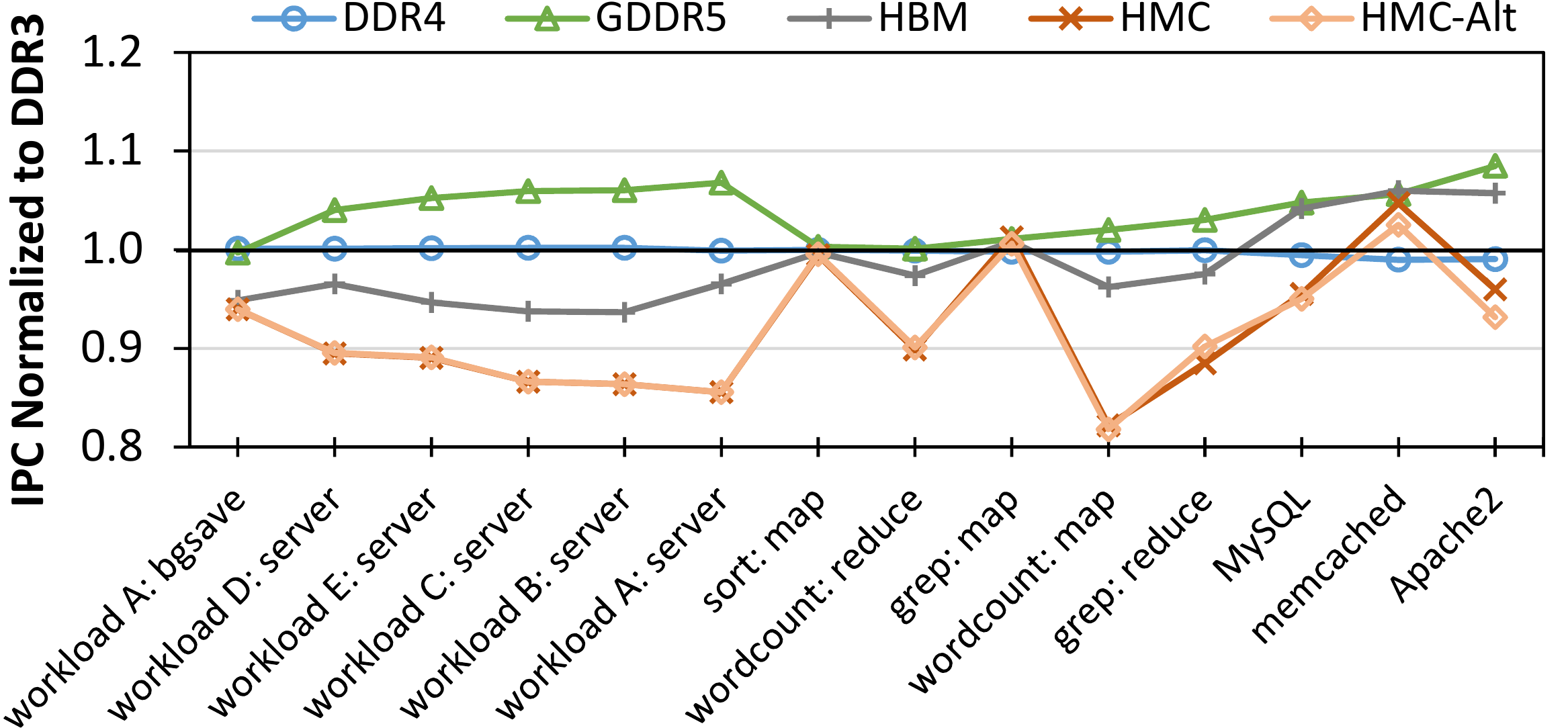}%
  \vspace{5pt}\\
  \includegraphics[width=0.9\columnwidth, trim=60 145 60 155, clip]{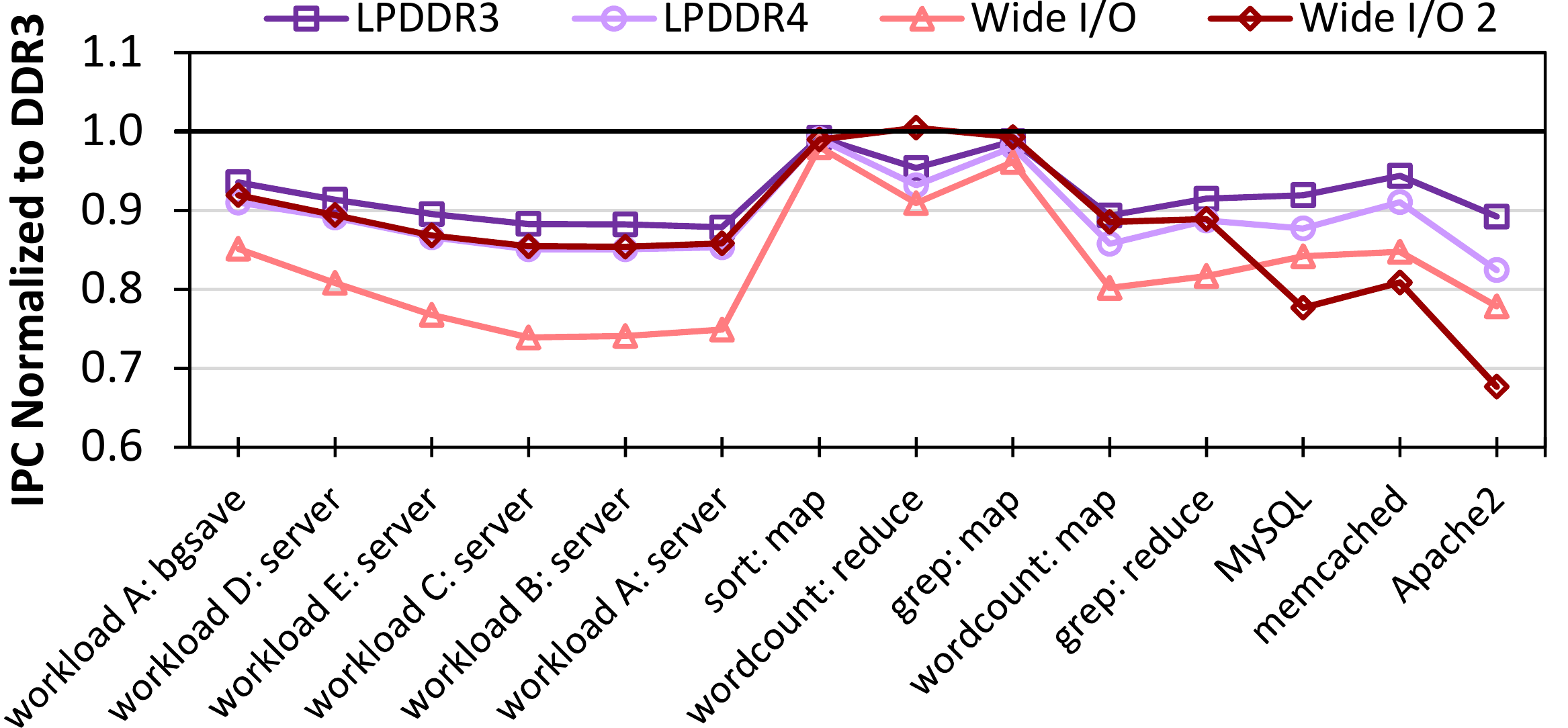}%
  \vspace{-5pt}%
  \caption{Performance of server/cloud applications for standard-power (top) and
  low-power (bottom) DRAM types, normalized to DDR3.}
  \label{fig:server:single}
  \label{fig:server:singlelp}
\end{figure}

Figure~\ref{fig:server:single} shows the performance of \ch{single-threaded applications}
for server and cloud environments when run on our evaluated DRAM types,
normalized to DDR3.  
We find that none of our workloads benefit significantly
from using HBM, HMC, or Wide I/O 2.
These DRAM types sacrifice DRAM latency to provide high throughput.
Since our workloads have low memory \ch{bandwidth needs}, they are unable to benefit
significantly from this additional throughput.

\obs{obs:serverperf}{Due to their low memory intensity and poor BPU, most of the server and
cloud workloads that we study do not benefit significantly from high-throughput
memories.}

\begin{figure}[h]
  \centering
  \includegraphics[width=0.375\columnwidth, trim=65 195 450 174, clip]{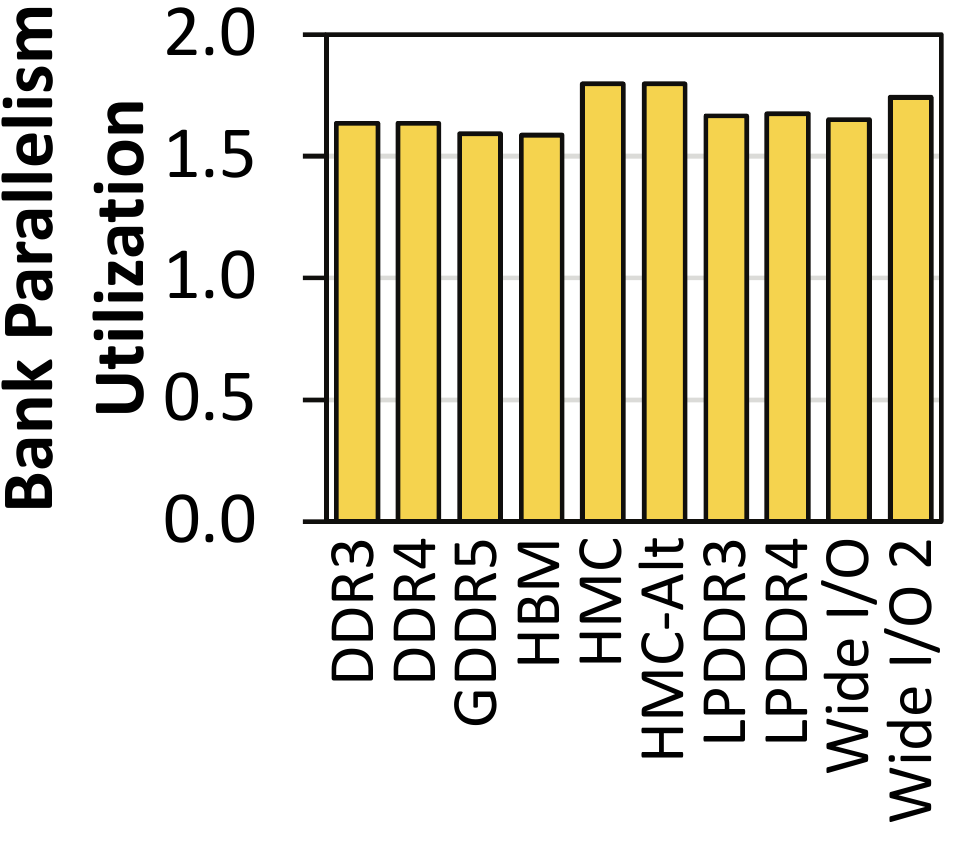}%
  \hfill%
  \includegraphics[width=0.58\columnwidth, trim=65 195 315 174, clip]{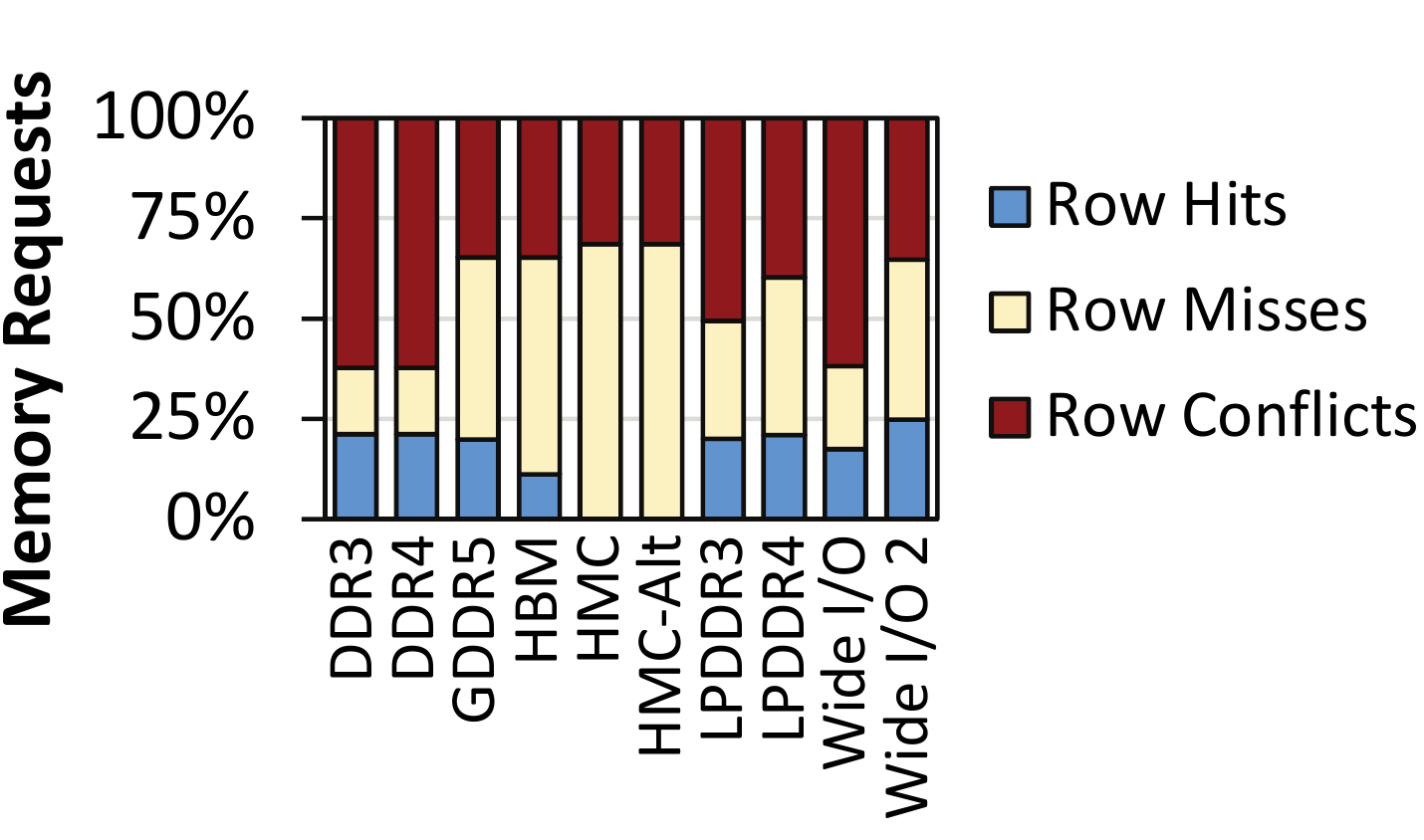}%
  \vspace{-10pt}%
  \caption{BPU (left) and row buffer locality (right) of \emph{workload A: server}.}
  \label{fig:server:bpu}
  \label{fig:server:rbl}
\end{figure}

To understand why high-throughput memories do not benefit these applications,
we focus on YCSB (the leftmost six workloads in Figure~\ref{fig:server:single}).
For these workloads, we observe that as the memory intensity
increases, HMC performs increasingly worse compared to DDR3.  
We find that the YCSB workloads exhibit low BPU values (never exceeding 1.80).
Figure~\ref{fig:server:bpu} shows the BPU (left) and row buffer locality (right)
for \emph{workload A: server}, as a representative YCSB workload.
\ch{Due to the low BPU of the workload across all DRAM types,
the high number of banks provided by HBM, HMC, and Wide I/O are wasted.}
HMC also destroys the row hits (and thus
lower access latencies) that other DRAM types provide, resulting in
a significant \emph{\ch{performance loss}} of 11.6\% over DDR3, on average across the
YCSB workloads.
HMC \changesvii{avoids} \ch{performance loss} for applications
\ch{that have high BPU, such as the map process for \emph{grep}
(with a BPU of 18.3; not shown).  However, such high} BPU values are not
typical for \chii{the server and cloud workloads we examine}.

We find \ch{two effects of the} low memory intensity and \ch{low} BPU
of server and cloud workloads.
First, these workloads are highly sensitive to memory request latency.
\ch{The limited memory-level parallelism exposes the latency 
of a memory request to} the processor pipeline\chii{~\cite{ghose.isca13, 
qureshi.isca06, mutlu.hpca03, mutlu.micro07, mutlu.ieeemicro06, kirman.hpca05,
das.micro09}}.
\ch{Second, the performance loss due to using the low-power DRAM types
is \chii{mainly due} to the increased memory access latencies, and
\chii{not} reduced throughput.
For example, as we observe in Figure~\ref{fig:server:single} (bottom),
Wide I/O's performance loss is comparable to the performance loss
with other low-power DRAM types for many of our server and cloud workloads, 
even though the available bandwidth of Wide I/O is only 25\% of the bandwidth
available with LPDDR4 (see Table~\ref{tbl:dram}).}

\ch{\emph{We conclude that \chiii{the server and cloud workloads we evaluate} are 
\chii{highly sensitive to the memory access latency, and are}
not significantly impacted by memory throughput.}}

\subsection{Multiprogrammed Performance}
\label{sec:server:mw}

\begin{figure}[h]
  \centering
  \includegraphics[width=0.9\columnwidth, trim=65 185 60 135, clip]{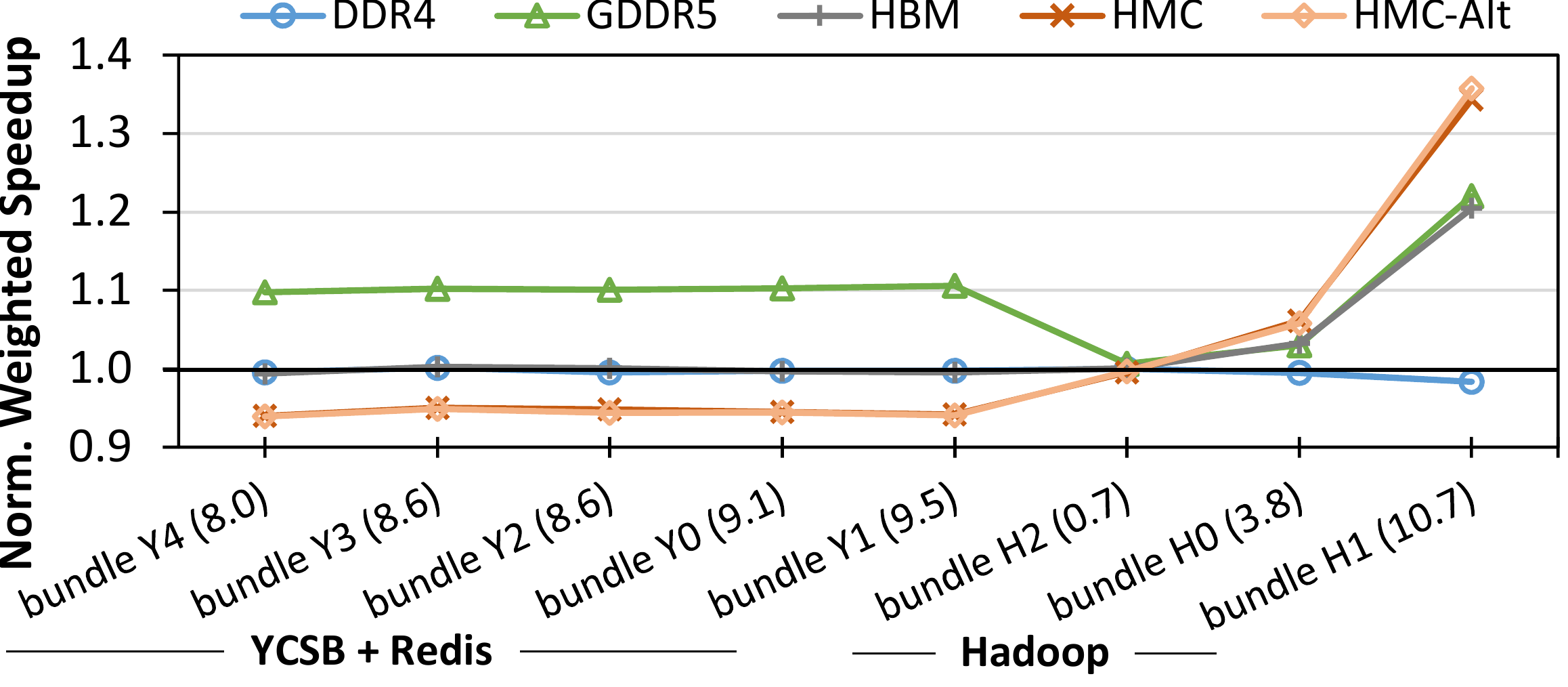}%
  \vspace{5pt}\\
  \includegraphics[width=0.9\columnwidth, trim=65 205 60 155, clip]{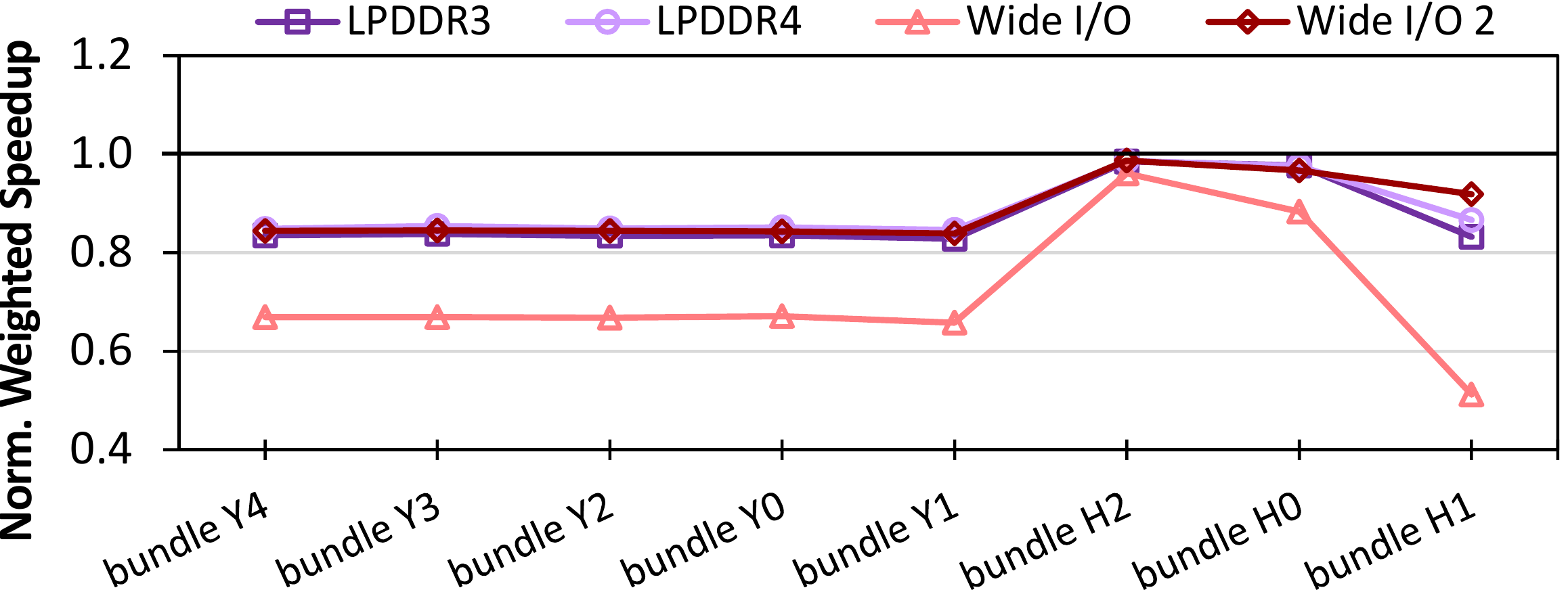}%
  \vspace{-5pt}%
  \caption{Performance of multiprogrammed server/cloud workloads for standard-power (top) and
  low-power (bottom) DRAM types, normalized to performance with DDR3.
  MPKI listed in parentheses.}
  \label{fig:server:mw}
  \label{fig:server:mwlp}
\end{figure}

\changesvii{We combine the \ch{single-threaded server and cloud applications}
into eight four-core multiprogrammed workloads (see Table~\ref{tbl:mwcloud}
in Appendix~\ref{sec:artdesc:workloads} for workload details).}
Figure~\ref{fig:server:mw} shows the performance of executing four-core 
multiprogrammed workloads for our 
\ch{YCSB workload bundles (\emph{Y0}--\emph{Y4}) and 
Hadoop workload bundles (\emph{H0}--\emph{H2})}
with each DRAM type.  
We \changesvii{identify two trends} from the figure.

First, \changesvii{we observe that the multiprogrammed YCSB workloads see little benefit from high-throughput
memories, much like the \ch{single-threaded} YCSB applications}.  
\changesvii{One exception to this is when the workloads run on GDDR5,
which provides a mean}
speedup of 10.2\% over DDR3
due to the increased memory intensity \ch{(and thus higher memory bandwidth
demand)} of the multiprogrammed workloads.
On low-power DRAM types, 
\changesvii{the multiprogrammed YCSB workloads experience larger performance
drops over DDR3 than the \ch{single-threaded} applications.}
\changesvii{For LPDDR3, LPDDR4, and Wide I/O 2, the performance drop ranges
between 14.5\% and 17.2\%.
For Wide I/O, the performance drop is even worse, with an
average drop of 33.3\%.
Because the multiprogrammed workloads are more memory-intensive than
the \ch{single-threaded} applications, the reduced throughput of low-power DRAM types
compared to DDR3 has a greater negative impact on the multiprogrammed
workloads \chii{than on the single-threaded applications}.}

\changesvii{Second, 
\ch{we observe that HMC significantly improves the performance of
the Hadoop workloads, because
the working sets of the individual applications in each workload}}
conflict with each other in the \chii{last-level CPU} cache.  \changesvii{This increases
the last-level cache miss rate, which in turn significantly increases the memory intensity
compared to the memory intensity of the \ch{single-threaded} Hadoop applications}.
\changesvii{Due to the increased memory intensity, the queuing latency
of memory requests} make up a significant fraction of the DRAM access
latency.  \ch{For example, on DDR3, queuing accounts for 77.2\% of the
total DRAM access latency for \emph{workload~H0} (not shown).  
HMC} is able to alleviate queuing
significantly for the multiprogrammed Hadoop workloads compared to DDR3
\chii{(reducing it to only 23.8\% of the total DRAM access latency)},
\ch{similar to what we saw for multiprogrammed desktop and scientific workloads
in Section~\ref{sec:desktop:mw}.
On average, with HMC, the Hadoop workloads achieve} 2.62$\times$ the BPU,
with an average performance improvement of 9.3\% \chii{over DDR3}.

\ch{\emph{We conclude that the performance of multiprogrammed server and cloud
workloads depends highly on the \chii{interference that occurs} between the applications
in the workload, and that HMC provides performance benefits when 
\chii{\chiii{such application interference results} in high memory intensity}.}}

\subsection{DRAM Energy Consumption}
\label{sec:server:energy}

\obs{obs:serverenergy}{For server and cloud workloads, LPDDR3 and LPDDR4 
greatly minimize standby power
consumption without imposing a large performance penalty.}

\begin{figure}[h]
  \centering
  \includegraphics[width=0.9\columnwidth, trim=70 120 60 270, clip]{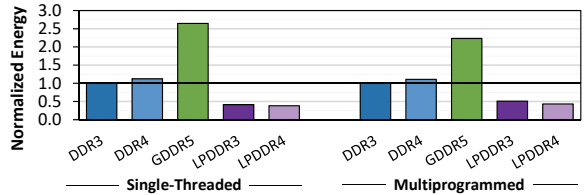}%
  \vspace{-10pt}
  \caption{Mean DRAM energy consumption for \ch{single-threaded} (left) and multiprogrammed (right) server/cloud applications, normalized to DDR3.}
  \label{fig:server:energy}
\end{figure}

Figure~\ref{fig:server:energy} shows the DRAM energy consumption for the
\ch{single-threaded} and multiprogrammed server and cloud workloads.  
\changesvii{GDDR5} 
consumes a significant amount of energy \changesvii{(2.65$\times$ the energy
consumed by DDR3 for
\ch{single-threaded} applications, and 2.23$\times$ for multiprogrammed workloads).
Given the modest performance gains over DDR3 (3.8\% for \ch{single-threaded} applications, and
9.4\% for multiprogrammed workloads), GDDR5 is much less energy efficient than
DDR3. This makes GDDR5} especially unsuitable for a
datacenter setting, where energy \changesvii{consumption and efficiency are
first-order design concerns}.  In contrast, we find
that \ch{LPDDR3/LPDDR4} \chiii{save a significant amount of DRAM energy 
\ch{(58.6\%/61.6\% on average)}, while} \chii{causing only relatively} small performance degradations
\ch{for single-threaded applications \chiv{compared} to DDR3 (8.0\%/11.0\% on
average).}
\chii{Thus, we believe} LPDDR3 and LPDDR4 \chii{can be} competitive candidates for the
server and cloud environments, as they \ch{have higher} memory energy efficiency than
DDR3.

\changesvii{\emph{We conclude that low-power DRAM types can be viable options 
to improve energy efficiency in server and cloud environments, 
while DRAM types such as GDDR5 are not as energy efficient as DDR3.}}

%% file: sections/mobile.tex
% !TEX root=../dramcharacterization.tex

\section{Heterogeneous System Workloads}
\label{sec:mobile}
\label{sec:hetero}

In this section, we study the performance and energy consumption of
workloads that are representative of \changesvi{those that run on} three major types of 
processors and accelerators in heterogeneous systems, such as
systems-on-chip (SoCs) and mobile
processors: (1)~\emph{multimedia acceleration},
which we approximate using benchmarks from the MediaBench~II suite for JPEG-2000 and
H.264 video encoding and decoding~\cite{mediabench};
(2)~\emph{network acceleration}, for which we use \ch{traces} collected from a
commercial network processor~\cite{nxp.networkaccel}; and
(3)~\emph{general-purpose GPU (GPGPU) applications}
\changesii{from the Mars~\cite{he.pact08}, Rodinia~\cite{rodinia},
and LonestarGPU~\cite{lonestar} suites}.

\subsection{Multimedia Workload \changesiii{Performance}}
\label{sec:mobile:multimedia}

Multimedia accelerators are designed to perform high-throughput parallel
operations on media content, such as video, audio, and image processing~\cite{mediabench}.
Often, the content is encoded or decoded in a \emph{streaming} manner,
where pieces of the content are accessed from memory and processed in order.
Multimedia accelerators typically
work one file at a time, and tend to exhibit high spatial locality due to the
streaming behavior of their algorithms.  The algorithms we explore are
often bound by the time required to encode or decode each piece 
\ch{of media (e.g., a video frame)}.
We find that \changesvii{JPEG processing and H.264 encoding} applications are
highly compute-bound (i.e., MPKI $<$ 5.0), and exhibit \emph{very slow} streaming
behavior \changesvi{(i.e., the requests are issued in a streaming fashion,
but exhibit low memory intensity)}.  \changesvii{In contrast, we find that} H.264 decoding 
exhibits a highly memory-bound \ch{\emph{fast}} streaming behavior, with an MPKI of 124.5.

\obs{obs:stream}{\ch{Highly-memory-intensive}
multimedia applications 
benefit from high-bandwidth DRAM types with wide rows.}

\begin{figure}[h]
  \includegraphics[width=0.47\columnwidth, trim=65 140 403 127, clip]{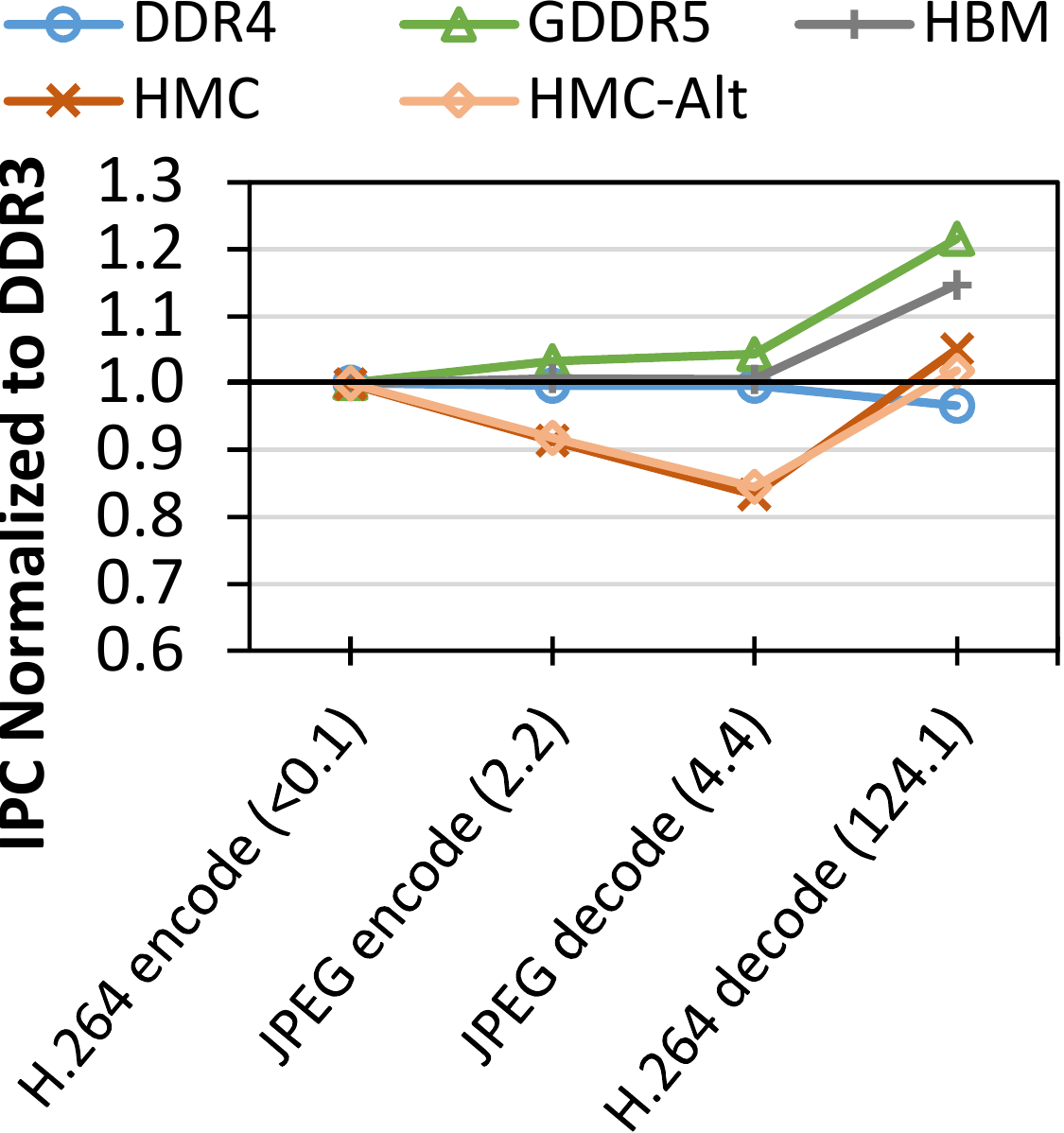}%
  \quad%
  \includegraphics[width=0.47\columnwidth, trim=65 122 403 145, clip]{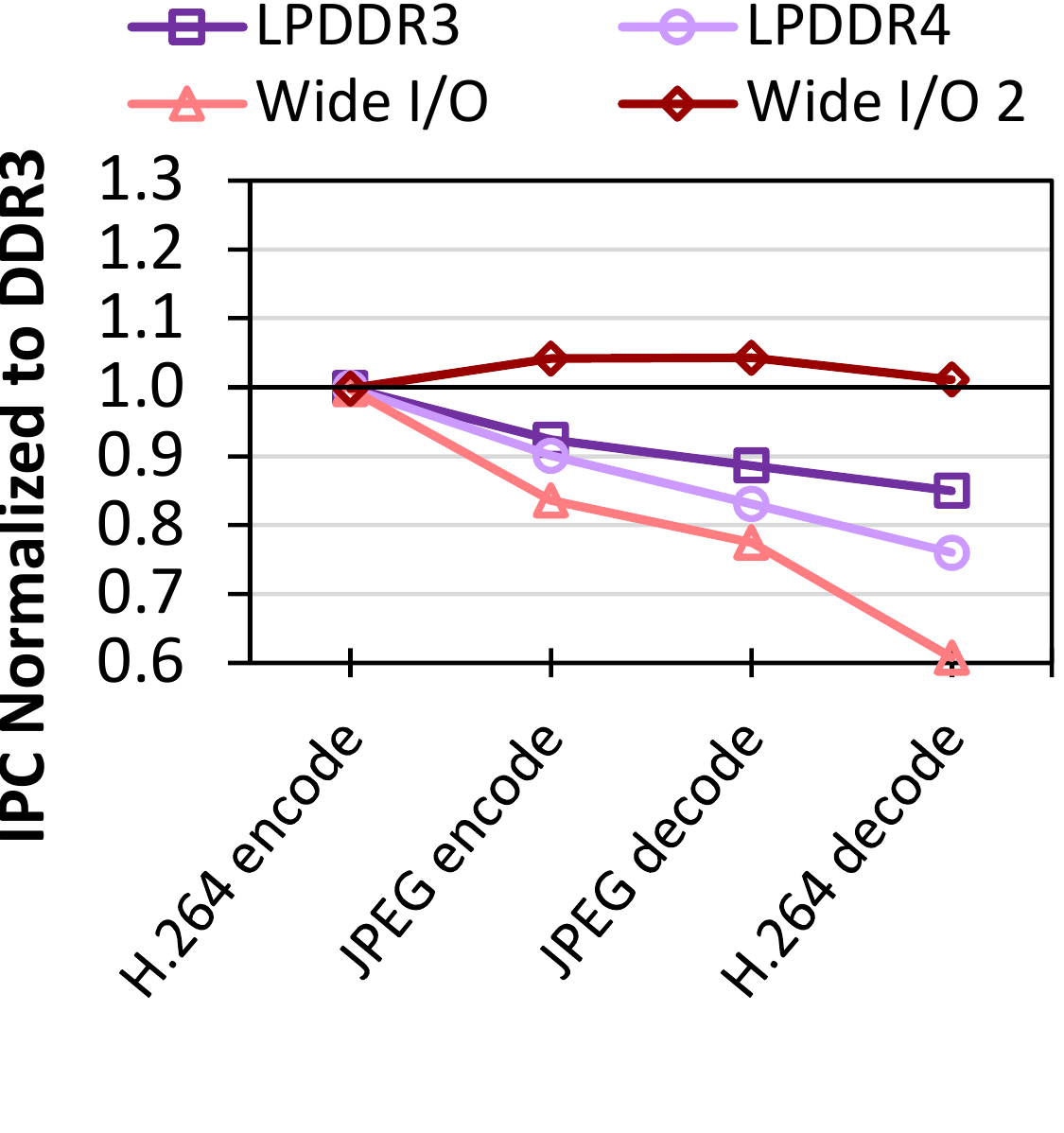}%
  \vspace{-5pt}%
  \caption{Performance of multimedia applications for standard-power (left) and
  low-power (right) DRAM types, normalized to performance with DDR3.
  MPKI listed in parentheses.}
  \label{fig:mobile:multimedia}
  \label{fig:mobile:multimedialp}
\end{figure}

Figure~\ref{fig:mobile:multimedia} shows the performance of the multimedia
applications on each DRAM type, normalized to DDR3.
\changesvi{We \changesvii{draw out} two findings from the figure.  First,}
JPEG encoding/decoding and H.264 encoding
do not benefit from any of the
high-bandwidth DRAM types, \changesvi{due to the applications'
low memory intensity.} \ch{The performance of some of these
applications is} actually hurt significantly by HMC,
due to HMC's small row size and high access latencies.
In contrast, the larger row width of Wide I/O 2 allows these
applications to experience modest speedups \changesvi{over DDR3}, by
increasing the row hit rate.

\changesvi{Second, unlike} the other \chii{multimedia} applications, H.264 decoding performs significantly better with
\changesvi{certain high-bandwidth memories:
\ch{with GDDR5/HBM, its performance improves by 21.6\%/14.7\% over DDR3.}
\changesvi{We find that GDDR5 and HBM cater well to H.264 decoding, as
the application exhibits high memory intensity with highly-localized
streaming access patterns, causing the majority of its memory requests to be
row hits.
\ch{Due} to its streaming nature, H.264 decoding still relies heavily on
DRAM types with wide rows, which can take advantage of spatial locality.
\ch{As a result, even though \chii{the BPU of H.264 decoding increases \chiii{in}
HMC by 177.2\% over DDR3 (due to the distribution of streaming requests across
multiple banks),}
the application does not see large performance improvements
with HMC.}
\ch{The} highly-localized access pattern also hurts the performance of H.264 decoding
with DDR4.  \chii{Much like with DDR3, the \chiii{application's memory requests exploit
spatial locality within a DDR4 DRAM row}, \chiv{but} make use of only a limited amount of bank parallelism.}
As a result, \changesvii{the application cannot take advantage of
the additional banks in DDR4 over DDR3, and} DDR4 \emph{slows down}
\changesvii{H.264 decoding} by 2.6\% compared to DDR3
\chii{due to its increased access latency}.}

\subsection{Network Accelerator \changesiii{Performance}}
\label{sec:mobile:network}

The network accelerators we study handle a number of data processing tasks
(e.g., processing network packets, \changesvi{issuing network} responses, storing the
data in an application buffer).
Such network accelerators can be found in dedicated network processing
\changesvi{chips, SoCs, and server chips}~\cite{nxp.networkaccel}.
Unlike
multimedia accelerators, which exhibit regular streaming access patterns, 
network accelerator memory access patterns are dependent on the rate of
incoming network traffic.  A network accelerator monitors traffic entering
from the network adapter, performs depacketization and error correction,
and transfers the data to the main memory system.  As a result of its
dependency on incoming network traffic, the network accelerator exhibits
highly bursty behavior, where it occasionally writes to DRAM, but has a high
memory intensity during each write burst.

\obs{obs:network}{Network accelerators experience very high queuing latencies
at DRAM even at low MPKI, 
and benefit greatly from a high-bandwidth DRAM \ch{with large bank
parallelism,} such as HMC.}

Figure~\ref{fig:mobile:network} shows the \changesvii{sustained bandwidth
provided by} \changesvi{different
memory types when running} the network accelerators,
normalized to DDR3.  We sweep the number of network accelerator requests that 
are allowed to be in flight at any given time, to emulate different network
injection rates.  We find that the network accelerator workloads behave \changesvii{quite}
differently than our other applications.  Thanks to the highly-bursty nature of
the memory requests, the queuing latency accounts for 62.1\%
of the total request latency, averaged across our workloads.  
\changesvii{For these workloads,} HMC's combination of high
available bandwidth and a very large number of banks allows it to 
increase the BPU by 2.28$\times$ over DDR3, averaged across all of our
workloads.  This reduces the average queuing latency by 91.9\%,
leading to an average performance improvement of 63.3\% \chii{over DDR3}.
HMC-Alt combines HMC's low queuing latencies with improved \changesvii{row} locality,
\changesvii{which better exploits the large (i.e., multi-cache-line) size
of each network packet.}
As a result, HMC-Alt performs 88.4\% better than DDR3, on average.

\begin{figure}[h]
  \includegraphics[width=0.47\columnwidth, trim=60 170 403 127, clip]{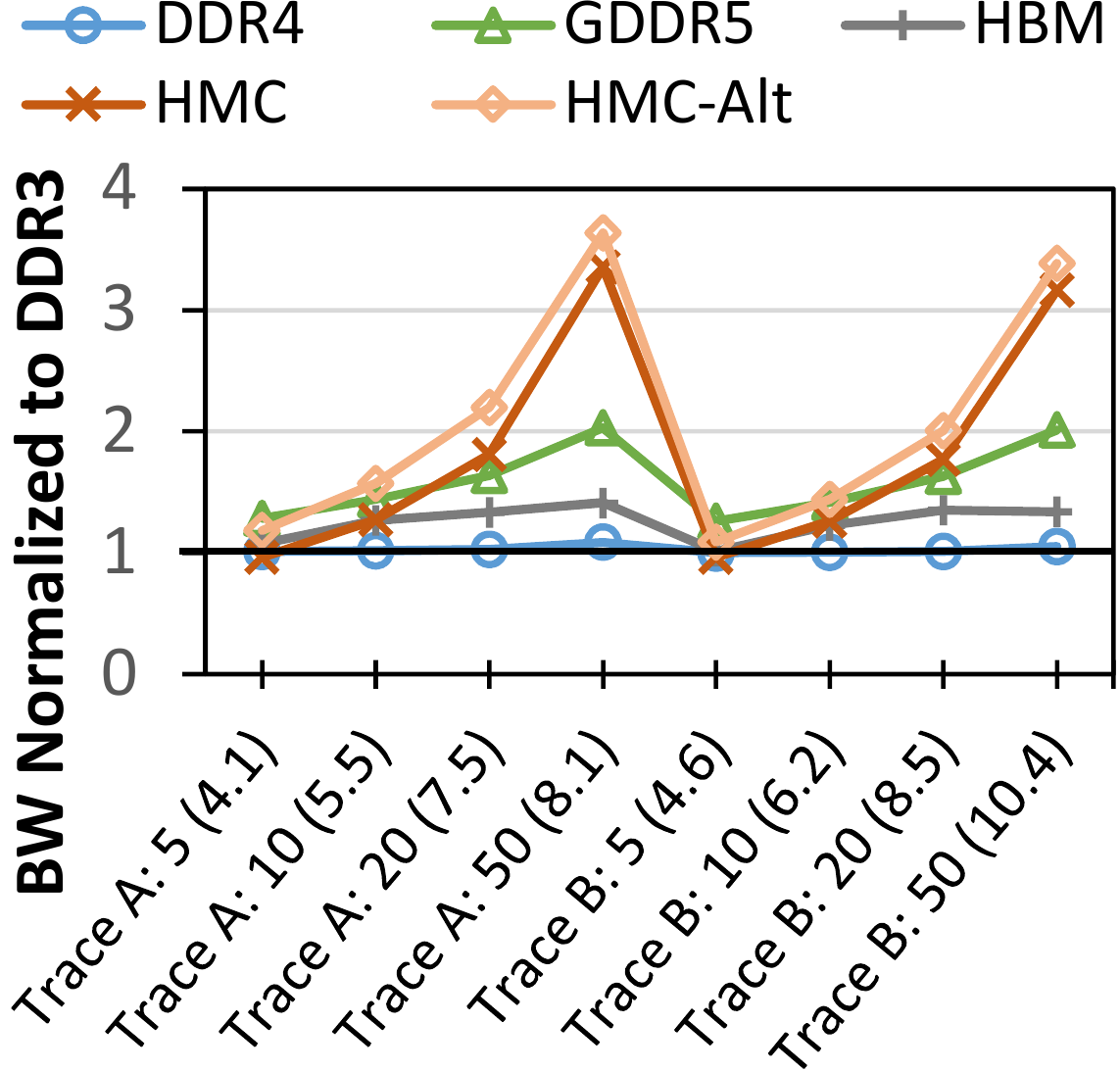}%
  \quad%
  \includegraphics[width=0.47\columnwidth, trim=60 152 403 145, clip]{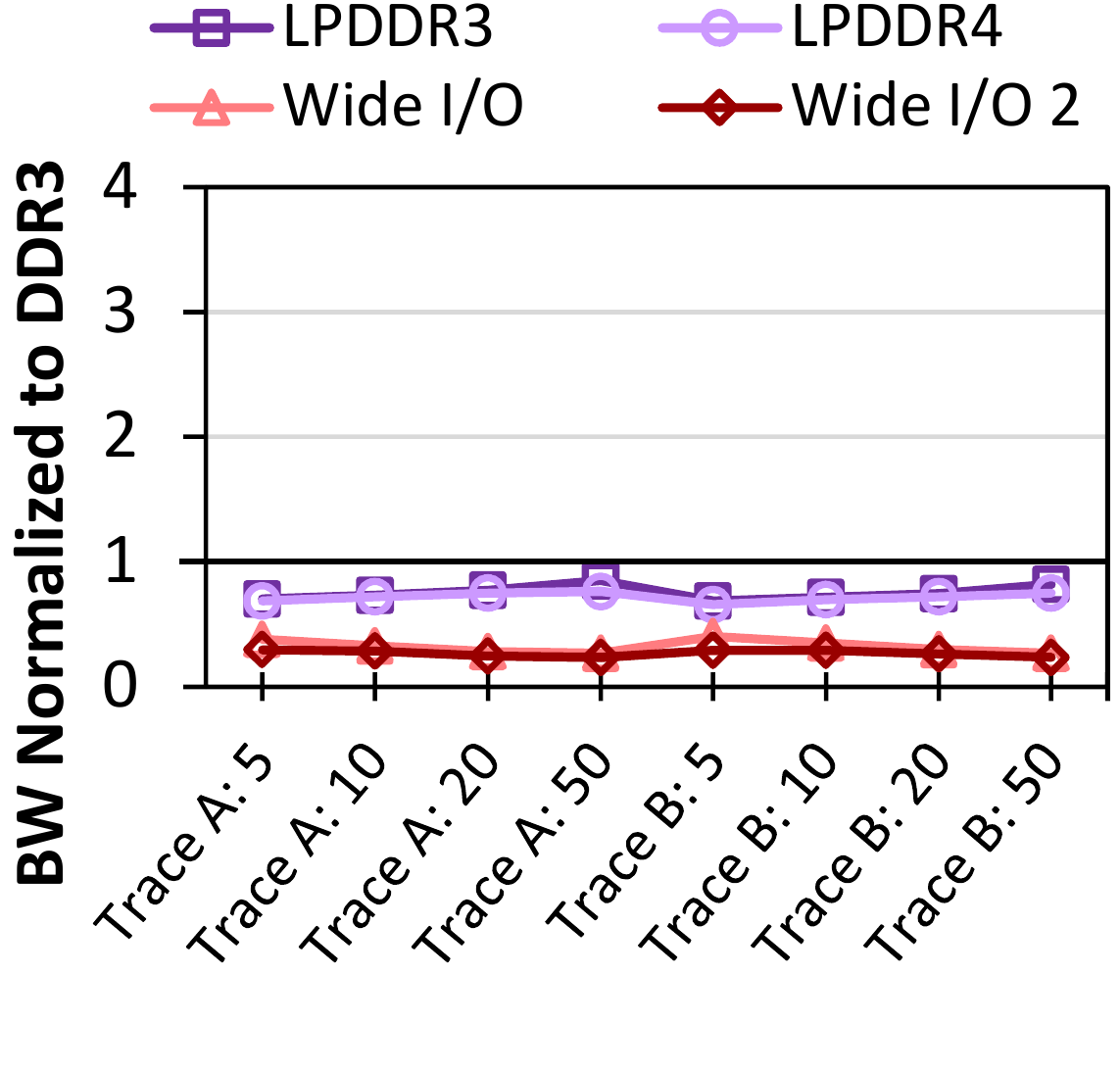}%
  \vspace{-5pt}%
  \caption{Network accelerator bandwidth (BW) for standard-power (left) and
  low-power (right) DRAM types, normalized to BW with DDR3.
  \changesvi{Maximum in-flight requests listed after the trace name, and}
  MPKI listed in parentheses.}
  \label{fig:mobile:network}
  \label{fig:mobile:networklp}
\end{figure}

\changesvi{\emph{We conclude that SoC \ch{accelerators benefit} significantly from high-bandwidth
memories (e.g., HMC, GDDR5), \ch{but the} diverse behavior of the different \changesvii{types of} accelerators 
\changesvii{(e.g., multimedia vs.\ network)} makes it
difficult to identify a single DRAM type that performs best across the 
board.}}

\subsection{GPGPU Application \changesiii{Performance}}
\label{sec:mobile:gpgpu}
\label{sec:char:gpgpu} % moved back from appendix
\label{sec:gpgpu}

We study ten applications from the Mars~\cite{he.pact08}, Rodinia~\cite{rodinia},
and LonestarGPU~\cite{lonestar} suites.  These applications have diverse memory
intensities, with \changesvii{last-level cache} MPKIs ranging from 0.005 (\emph{dmr}) to 25.3 (\emph{sp}).
Figure~\ref{fig:gpgpu} shows the performance of the applications.

\begin{figure}[h]
  \includegraphics[width=\columnwidth]{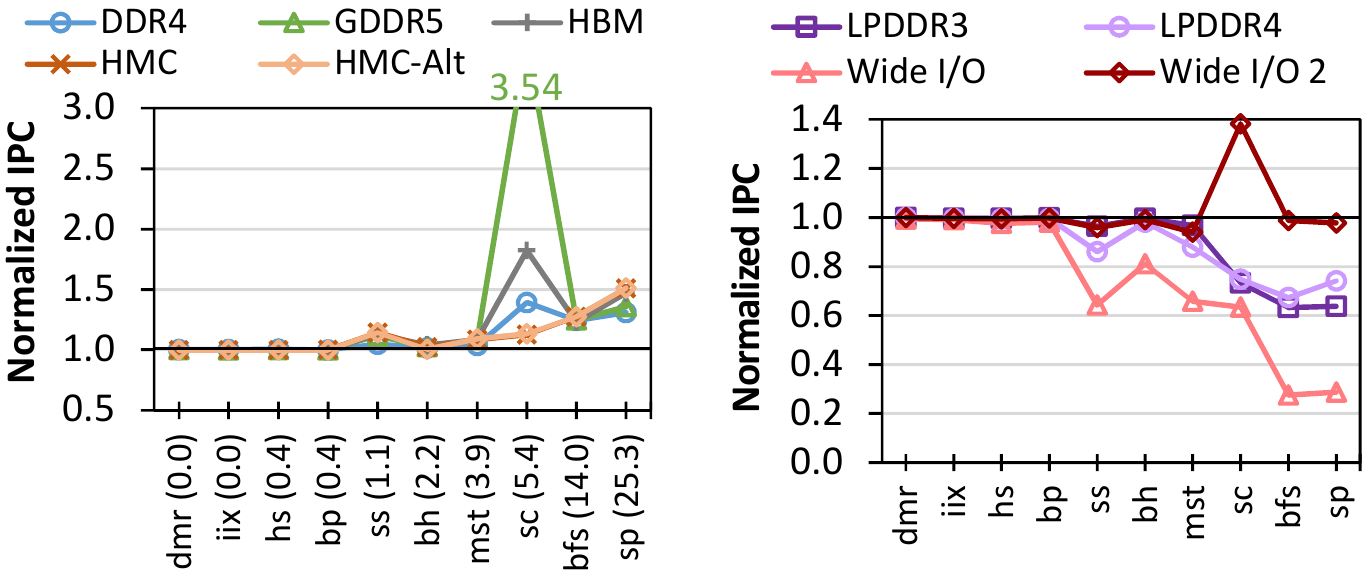}%
  \vspace{-8pt}%
  \caption{Performance of GPGPU applications for standard-power (left) and
  low-power (right) DRAM types, normalized to DDR3.
  MPKI in parentheses.}
  \label{fig:gpgpu}
\end{figure}

\changesvi{We \changesvii{draw out} two key findings from the figure.}
\changesvi{First,} we find that for our applications that are not memory intensive (MPKI $<$ 1
\changesvi{for GPGPU applications}), 
all of our DRAM types perform
near identically.  
\changesvi{Second, six} of our applications \changesvi{(\emph{ss}, \emph{bh}, \emph{mst},
\emph{sc}, \emph{bfs}, \emph{sp})} are memory intensive, and benefit significantly
from executing on a system with \chii{high-bandwidth \chiii{DRAM types} (i.e., GDDR5, HBM, or HMC)}.
On average, the IPC of memory-intensive GPGPU applications 
\changesvi{is 39.7\% higher with GDDR5, 26.9\% higher with HBM, and
18.3\% higher with both HMC and HMC-Alt, compared to DDR3.}
Unlike the other \changesvii{applications we} study, the
memory-intensive GPGPU applications also see significant performance improvements
with \changesvi{DDR4, which provides} an average performance improvement of 16.4\%
\changesvi{over DDR3}.

A large reason for the high speedups \changesvii{of the memory-intensive
GPGPU workloads} \chii{with high-bandwidth DRAM types is} 
\emph{memory coalescing}~\cite{bakhoda.ispass09, chatterjee.sc14}.
In a GPU, the memory controller coalesces (i.e., combines) multiple memory requests
that target nearby locations in memory into a single memory request.
This is particularly useful for GPU and GPGPU applications, where a large number of
threads operate in lockstep, and often operate on neighboring pieces of data.
Memory coalescing exploits the spatial locality between multiple threads,
in order to reduce pressure on the memory system and reduce queuing delays.
The coalesced memory requests take significant advantage of the
high bandwidth available in GDDR5, and the additional bank parallelism 
available in DDR4.
Coalescing is particularly helpful for \emph{sc}, where the memory requests are highly bound
by the available memory bandwidth~\cite{kloosterman.micro15}.
This leads to very high speedups on GDDR5 (253.6\%) for \changesvii{\emph{sc}} over DDR3.

Unlike the other memory-intensive applications, memory requests from \emph{sp} 
are typically not coalesced~\cite{chatterjee.sc14} (i.e., requests from multiple threads cannot be combined easily to exploit locality).
Without coalescing, the application issues many requests at once \changesvi{to DRAM,
and, thus, performs best when it is run on a DRAM type \chii{that} can
provide both high bandwidth \emph{and} high bank parallelism
to service many requests concurrently, \chii{such as HBM or HMC}.
As a result, for \emph{sp},}
HBM outperforms GDDR5 by 8.3\%, \changesvi{and} HMC outperforms
GDDR5 by 11.3\%.
HMC and HMC-Alt \changesvi{perform} within 0.2\% of each
\changesvi{other,
as} \emph{sp} does not have significant locality for
HMC-Alt to exploit for additional performance benefits.

\changesvi{\emph{We conclude that for memory-intensive GPGPU applications, GDDR5
provides significant performance improvements as long as memory requests
can be coalesced, while HBM and HMC \ch{improve performance even more when 
memory requests are not coalesced because \chii{both of these DRAM types} 
provide high bank parallelism}.}}

\subsection{DRAM Energy Consumption}
\label{sec:mobile:energy}

\obs{obs:mobileenergy}{For \changesvi{the accelerators\\ that have}
high memory throughput requirements,\\ GDDR5 provides much 
\changesvi{greater energy efficiency\\ \chiv{(i.e., large performance gains
with a\\ small energy increase)} than DDR3}.}

Figure~\ref{fig:mobile:energy} shows the normalized energy consumption
for the \changesvi{three major types of heterogeneous system workloads that
we study: (1)~multimedia acceleration, (2)~network acceleration, and
(3)~GPGPU applications, averaged across all applications \changesvii{of} each type.
Overall, for multimedia acceleration and GPGPU applications, we observe
that GDDR5 consumes more than double the energy of the other memory types,
while for network acceleration, GDDR5 consumes only 24.8\% more energy
than DDR3.}

\begin{figure}[h]
  \centering
  \includegraphics[width=0.9\columnwidth, trim=70 120 60 270, clip]{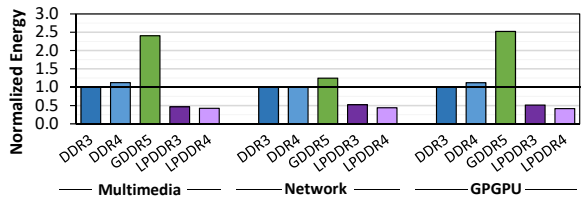}%
  \vspace{-7pt}
  \caption{Mean DRAM energy consumption for multimedia (left) and network (right) acceleration, normalized to DDR3.}
  \label{fig:mobile:energy}
\end{figure}

\changesvi{Upon closer inspection, \changesvii{however}, we find that for \changesvii{the set of} heterogeneous system
applications that require high memory throughput (H.264 decoding, all of our
network acceleration traces, \emph{sc}, \emph{bfs}, and \emph{sp}),
GDDR5's energy consumption \changesvii{comes with large performance benefits}.
\ch{Figure~\ref{fig:mobile:energyhighthroughput} shows the energy consumption of
the high-throughput multimedia and GPGPU applications (see Figure~\ref{fig:mobile:energy}
for the network \chii{accelerator} energy).}
Averaged across these high-throughput applications, GDDR5 consumes \changesvii{only}
31.4\% more energy than DDR3, 
while delivering \changesvii{a performance
improvement of 65.6\%} \ch{(not shown)}. 
In the extreme case, for \emph{sc}, GDDR5 consumes
\changesvii{only} 20.2\% more energy than DDR3 to provide a 253.6\% speedup.
For such accelerator applications, where high memory throughput
is combined with high spatial locality, \changesvii{we conclude that} GDDR5 \ch{can be} significantly more
energy efficient than the other DRAM types.}

\begin{figure}[h]
  \centering
  \includegraphics[width=0.9\columnwidth, trim=70 120 60 270, clip]{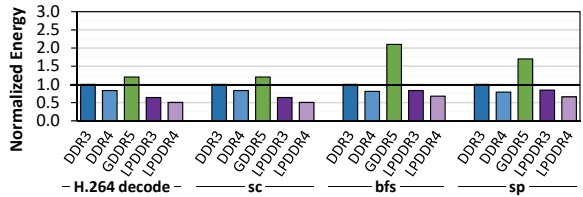}%
  \vspace{-7pt}
  \caption{\changesvii{DRAM energy consumption for high-memory-throughput multimedia and GPGPU applications, normalized to DDR3.}}
  \label{fig:mobile:energyhighthroughput}
\end{figure}

\changesvi{\emph{We conclude that certain types of accelerators can achieve 
higher energy efficiency \chiii{(i.e., large performance increases with a
small energy increase) than with DDR3, by using} aggressive DRAM types such as GDDR5,
when the accelerators perform tasks that require high memory throughput.}}

%% file: sections/os.tex
% !TEX root=../dramcharacterization.tex

\section{Common OS Routines}
\label{sec:os}

We collect several traces capturing common \changesvii{kernel-mode activities} for different benchmarks:
\begin{itemize}
    \item \emph{IOzone}~\cite{IOZone}, a file system benchmark suite that tests a number of I/O performance tasks
        (Tests 0--12);
    \item \sloppypar \emph{Netperf}~\cite{Netperf}, which tests TCP/UDP network calls
        (\emph{UDP\_RR}, \emph{UDP\_STREAM}, \emph{TCP\_RR}, \emph{TCP\_STREAM});
    \item \emph{bootup}~\cite{seshadri.micro13}, a representative phase of the boot operation in the
        Debian operating system;
    \item \emph{forkbench}~\cite{seshadri.micro13}, a microbenchmark trace that creates a 64MB array
        of random values, forks itself, and has its child process update 1K random pages; and
    \item \emph{shell}~\cite{seshadri.micro13}, a microbenchmark trace of a Unix shell script that runs \texttt{find}
        on a directory tree and executes \texttt{ls} on each subdirectory.
\end{itemize}

\subsection{Workload Characteristics}
\label{sec:os:workload}

\begin{figure}[h]
  \centering
  \includegraphics[width=0.9\columnwidth, trim=65 147 60 185, clip]{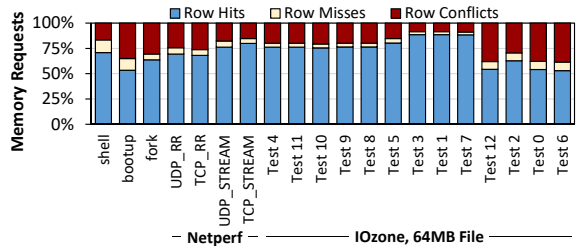}%
  \vspace{-7pt}%
  \caption{DDR3 row buffer locality for OS routines.}
  \label{fig:os:rbl}
\end{figure}

While the OS routines that we study perform a variety of different tasks, 
we find that they exhibit very similar behavior.  We \changesvii{depict} the row
buffer locality of the routines with DDR3 DRAM \changesvii{in} Figure~\ref{fig:os:rbl}. 
\changesvi{From the figure, we} find that most of the routines have exceptionally high row buffer locality,
with row buffer hit rates greater than 75\%.
This behavior occurs because many of the OS routines are based on files and
file-like structures, \changesvii{and} these files \changesvi{are often} read or written in large
sequential blocks.  This causes the routines to access most, if not all, of the
data mapped to an OS page (and therefore to the open DRAM row
\changesvi{that houses the page}).
We also observe that \changesvi{memory requests from these routines}
reach the DRAM at regular time intervals, \changesvii{as opposed to in bursts.
The regularly-timed memory requests reduce} \changesvi{the peak throughput demand} on DRAM.

\subsection{Performance}
\label{sec:os:perf}

\begin{figure}[h]
  \centering
  \includegraphics[width=0.9\columnwidth, trim=65 112 60 135, clip]{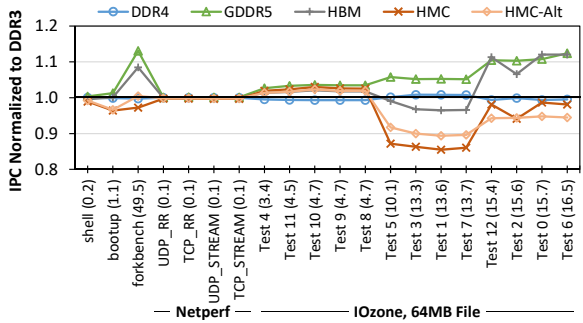}%
  \vspace{-10pt}%
  \caption{Performance of common OS routines for standard-power 
  DRAM types, normalized to performance with DDR3.
  MPKI listed in parentheses.}
  \label{fig:os:perf}
\end{figure}

Figure~\ref{fig:os:perf} shows the performance of the OS routines on 
standard-power DRAM types, normalized to their performance under DDR3. 
We find that the overall performance of the routines is similar to the
performance \changesvi{trends} observed for server and cloud workloads (see Section~\ref{sec:server:perf}):
\changesvii{only} GDDR5 memory outperforms DDR3 for the majority of routines.
The other high-throughput memories are \changesvi{generally} unable to significantly improve 
\changesvi{performance (except HBM, for some workloads)}, and in many cases actually hurt performance.

\obs{obs:os}{OS routines benefit most from\\DRAM types with 
low access latencies\\that exploit spatial locality (e.g., GDDR5).}

\chii{Lower-latency} \changesvii{DRAM types such as GDDR5 are best for OS routines due to}
\changesvi{(1)~}the serialized access \chii{patterns} of most of the
routines \ch{(e.g., dependent memory requests)} \changesvi{and
(2)~}the regular time intervals between DRAM requests (see Section~\ref{sec:os:workload}).
\changesvi{The regular, serialized accesses 
\ch{require lower latency to be served faster and do not benefit from
high memory throughput.}
As a result, DRAM types that \changesvii{increase access latency to provide}
higher \ch{throughput often} hurt the performance of OS routines.
This is particularly true of HMC, which \chii{greatly} sacrifices row buffer locality with its narrow
rows to \ch{provide high bank parallelism (which OS routines typically cannot
take advantage of due to their low BPUs).}}
As we show in Figure~\ref{fig:os:perf},
if we employ our locality-aware addressing mode for HMC (HMC-Alt),
the performance of HMC improves for some (but not all) of the routines,
\changesvi{as HMC-Alt can exploit more of the high spatial locality
\changesvii{present} in OS routines than HMC}.  
GDDR5 provides the highest performance across all OS routines because it
\ch{\chii{reduces} latency over DDR3 while also \chii{increasing bandwidth}}.

\changesiii{Figure~\ref{fig:os:perflp} shows the performance of the common OS routines
on low-power DRAM types.
We \changesvii{draw out} four \changesvi{findings} from the figure.
First, we observe that \ch{the average performance loss} from using
low-power DRAM types, compared to DDR3, \ch{can be} relatively small.
For example, LPDDR3 \ch{leads to} an average slowdown of only 6.6\% over DDR3.
Second, we observe that due to the high sensitivity of OS routines to 
DRAM access latency,
LPDDR4 \changesvi{causes} a larger slowdown (9.6\% on average over DDR3) than LPDDR3 due to 
its higher access latency.
Third, we observe that there are four routines where Wide I/O 2 provides
significant performance \emph{improvements} over DDR3: \emph{Test~12},
\emph{Test~2}, \emph{Test~0}, and \emph{Test~6}.
This is because Wide I/O 2 significantly increases the row hit rate.
As Figure~\ref{fig:os:rbl} shows,
these four routines have much lower row hit rates (an average of 56.1\%)
than the other routines in DDR3.
\changesvi{This is because the four routines have access patterns that
\changesvii{lead to a large} number of row conflicts.
Specifically, \emph{Test~12} reads data from a file and scatters it into
multiple buffers using the \texttt{preadv()} system call, while \emph{Test~2}, \emph{Test~0}, and \emph{Test~6} are dominated by write system calls that update both the data and any
associated metadata.
\changesvii{The data and associated \ch{metadata} often reside in different parts of
the memory address space, which leads to row conflicts in many
DRAM types.}
Wide I/O 2 \ch{reduces} these row conflicts, and} the average row hit rate for these four routines
\changesiv{increases} to 91.0\% (not shown).
Fourth, we observe that the \changesiv{performance reduction \changesvi{that
\emph{forkbench} experiences on low-power DRAM types versus DDR3
is much larger than the reduction other routines experience.}
\changesiv{This is} due to the fact that \emph{forkbench} is
significantly more memory intensive (with an MPKI of 49.5) than the
other OS routines.}

\begin{figure}[h]
  \centering
  \includegraphics[width=0.9\columnwidth, trim=65 148 60 137, clip]{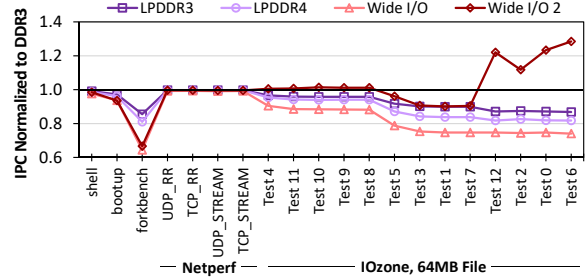}%
  \vspace{-10pt}%
  \caption{Performance of common OS routines for low-power
  DRAM types, normalized to performance with DDR3.}
  \label{fig:os:perflp}
\end{figure}

\changesvi{\emph{We conclude that OS routines \changesvii{perform better on}
memories \changesvii{that \ch{(1)~}provide} low-latency access \changesvii{and
\ch{(2)~}exploit} the high amount of spatial locality that exists in the 
memory access patterns of \changesvii{these} routines.}}

\subsection{DRAM Energy Consumption}
\label{sec:os:energy}

\changesiii{Figure~\ref{fig:os:energy} shows the energy consumed by each of the DRAM types
that we have accurate power models for, normalized to DDR3 energy consumption, and
averaged across all of the OS routines.
\changesvii{We} find 
that the DRAM energy consumption \changesvi{trends} of the \changesiv{OS} routines is very similar to
the trends that we observed for desktop workloads in Section~\ref{sec:desktop:energy}.  Without a
large \changesiv{improvement in average} performance, GDDR5 consumes 2.1x \changesvi{\emph{more}} energy than DDR3,
while \ch{LPDDR3/LPDDR4 consume 52.6\%/58.0\%} \changesvi{\emph{less}} energy \changesvii{than} DDR3.}

\begin{figure}[h]
  \includegraphics[width=\columnwidth, trim=70 170 60 270, clip]{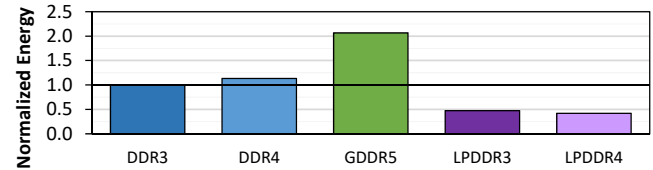}%
  \vspace{-10pt}
  \caption{Mean DRAM energy consumption for common OS routines, normalized to DDR3.}
  \label{fig:os:energy}
\end{figure}

\changesvi{\ch{LPDDR3/LPDDR4 incur} a much smaller average performance loss (\ch{6.6\%/9.6\%}; see Section~\ref{sec:os:perf})
over DDR3 for OS routines than for desktop \changesvi{and scientific} applications.
While both the OS routines and our desktop and scientific applications exhibit high
row buffer locality and low memory intensity, the average row buffer locality is higher for
the OS routines, while the average memory intensity is lower.}
As a result, LPDDR3 and LPDDR4 strike a better compromise \changesvii{between}
performance and energy consumption for OS routines than they do for 
desktop \ch{and scientific} applications.

\changesvi{\emph{We conclude that OS routines can attain high energy efficiency
\chii{(i.e., large energy reductions with a small performance impact), \chiii{compared to DDR3,}}
when run on low-power DRAM types that provide large row sizes to exploit
spatial locality \changesvii{(e.g., LPDDRx)}.}}

%% file: sections/lessons.tex
% !TEX root=../dramcharacterization.tex

\section{Key Takeaways}
\label{sec:lessons}

\shep{From our detailed characterization, and from the twelve key observations
that we make, we find that there are a number of high-level \shepii{lessons} and 
takeaways that can be of use for future DRAM architects, system architects, and
programmers.  We discuss the four \shepii{most important (as we deem)}
takeaways here.}

\begin{enumerate}[itemsep=6pt, topsep=6pt]
    \item \shep{\textbf{DRAM latency remains a critical bottleneck for many
        applications.}
        \changesvii{For} many of our applications, \changesvii{we observe that} the overall
        application performance \chv{degrades} \changesvii{when} the DRAM access
        latency \chv{increases}.  These applications cannot easily take advantage
        of greater memory-level parallelism or \shepii{higher memory} throughput,
        often because the memory intensity of the applications is not
        high enough to \changesvii{take advantage of} the maximum bandwidth offered by DDR3.
        As a result, \changesvii{even though many new DRAM types offer higher
        bandwidth and higher bank parallelism than DDR3,
        they do not significantly improve performance}.
        \changesvii{In fact, in many cases, new DRAM types
        \emph{reduce} performance, because the increased bandwidth and bank parallelism
        \ch{they provide come at the cost of higher latency}.}
	\vspace{3pt}\newline%
        \emph{For DRAM architects}, this means that newer \changesvi{DRAM} types
        \shepiii{require} ways to bring down the latency of a single
        access, \ch{which goes against} the recent trend of \changesvii{increasing latency 
        \ch{in order to provide other benefits}}.}
        \changesvi{Several recent works propose ways of reducing DRAM latency
        at low cost\chii{~\cite{lee.hpca13, kim.isca12, lee.hpca15, chang.sigmetrics16,
        hassan.hpca16, hassan.isca19, lee.sigmetrics17, kim.iccd18, chang.hpca16,
        zhang.hpca16, wang.micro18, choi.isca15, seshadri.micro13, chandrasekar.date14}}.
        We believe such approaches are very promising \chii{and critical} \changesvii{for modern and
        future applications, and we} encourage the
        development of more such novel latency reduction mechanisms for DRAM.}
	\vspace{3pt}\newline%
        \shepii{\emph{For system architects}, there is a need to reconsider
        whether systems should be built with older, lower-latency \changesvi{DRAM} types
        such as DDR3 and GDDR5 instead of with \shepiii{newer, higher-throughput}
        \changesvi{DRAM} types such as DDR4 and HMC.}
        \changesvi{The use of memory controller mechanisms to
        reduce DRAM latency is also promising\chii{~\cite{lee.hpca15, chang.sigmetrics16,
        hassan.hpca16, lee.sigmetrics17, kim.iccd18, zhang.hpca16, wang.micro18, 
        chandrasekar.date14}}.}

    \item \shep{\textbf{Bank parallelism is not fully utilized \shepii{by \changesvi{a
        wide variety} of our applications}.}
        As we show in our characterization, \changesvi{bank} parallelism utilization
        (BPU; see Section~\ref{sec:metrics}) is a key indicator of whether applications
        benefit from high-throughput and highly-parallel \changesvi{DRAM} types such as 
        HBM and HMC.  BPU expresses a combination of the memory intensity and
        the memory-level parallelism \shepii{of} an application.
        \shepii{While there are some applications (e.g., multiprogrammed workloads, 
        high-memory-intensity GPGPU applications) that have high BPU and benefit from
        using \changesvi{DRAM} types such as HBM and HMC,} \shepiii{\shepiv{a variety} of our
        applications have low BPU (e.g., single-thread desktop/scientific applications,
        irregular memory-sensitive multithreaded applications, and server/cloud
        applications)}, and \shepii{thus} do not experience appreciable
        performance gains with HBM and HMC.
	\vspace{3pt}\newline%
        \emph{For DRAM architects}, this indicates that providing a greater
        number of banks in newer \changesvi{DRAM} types may not provide significant
        gains \shepiii{for \changesvi{especially} single-thread performance}.  
        For example, while DDR4 doubles the number of banks over DDR3,
        the increased bank count requires architectural changes (e.g., the
        introduction of bank groups; see Section~\ref{sec:bkgd:arch} and
        Appendix~\ref{sec:dramarch}) that increase access latency.
        \shepiv{An important (critical) thread} with low BPU \changesvi{may not
        be able to} overcome the latency increase,
        \changesvii{and thus the additional banks would \ch{not benefit performance}.
        We see this behavior in several of our workloads, and thus their performance
        degrades when DDR4 is used instead of DDR3.}
	\vspace{3pt}\newline%
        \emph{For system architects}, \changesvi{it may be useful to consider using
        cheaper DRAM types with fewer banks \changesvii{(and also low latencies)} 
        in systems that run} \shepii{applications
        with} low BPU.
	\vspace{3pt}\newline%
        We do note that there are several cases where row conflicts remain
        high even when BPU is low (e.g., our example \shepiii{YCSB} server workload in
        Figure~\ref{fig:server:bpu}, \shepiii{irregular memory-intensive multithreaded
        workloads, \emph{cactusADM}, \emph{omnetpp}, \changesvii{and} \emph{GemsFDTD}}).
        \shepiii{This can occur if memory requests are unevenly distributed
        across the memory banks, \shepiv{causing some of the banks to be highly
        contended for} while other banks remain idle.}
        \emph{For both DRAM architects and 
        programmers}, this indicates that there are opportunities to change
        address interleavings, memory schedulers, memory allocation, and
        program access patterns to make better use of the available bank-level
        parallelism.}

    \item \shep{\textbf{Spatial locality continues to provide significant performance
        benefits if it is exploited by \shepii{the memory subsystem}.}
        One of the more significant changes made in HMC \changesvii{versus} other \changesvi{DRAM} types
        is the \shepii{reduction of the row buffer size}.  A row in HMC (\SI{256}{\byte}) is \ch{97\%}
        smaller than a row in DDR3/DDR4 (\SI{8}{\kilo\byte}).  HMC uses the shorter
        row size to significantly increase bank-level parallelism and memory
        throughput.  \shepii{Due} to the limited BPU of \shepii{many} applications, 
        \shepii{the increased parallelism and throughput \ch{only occasionally} provide}
        benefits when using HMC.  
        \changesvii{In contrast to bank parallelism,} applications and virtual memory managers still \changesvi{try} to
        maximize spatial locality as much as they can.  
        \shepii{While most \changesvi{DRAM} types exploit this spatial locality (by using large
        rows to amortize the high penalty of a row conflict), HMC's small rows are
        unable to effectively capture much of this locality.  As a result,}
        HMC provides notable performance improvements \shepiii{only}
        in cases where spatial locality is \shepiii{low (or is destroyed)}, such as 
        \chii{for highly-memory-intensive multiprogrammed workloads where 
        multiple applications significantly} interfere with
        each other.
        \changesvi{Otherwise, HMC can lead to large performance losses compared to
        other high-throughput memories (e.g., GDDR5, HBM),
        \changesvii{and HMC often performs worse than most other DRAM types}.}
	\vspace{3pt}\newline%
        \emph{For DRAM architects}, \changesvi{our observations indicate that}
        new \changesvi{DRAM} types that activate
        at a row granularity should not reduce the row width.  A reduced row 
        width typically requires more row activation operations for the same 
        amount of data, which introduces a significant overhead that cannot be \shepiii{(easily)}
        amortized by other benefits (such as higher memory-level parallelism in
        HMC).
	\vspace{3pt}\newline%
        \emph{For programmers}, applications should maximize the amount of
        spatial locality that they exploit, as most \changesvi{DRAM} types are designed to
        perform better with higher locality.  This \shepii{could require} \changesvi{(a)~}redesigning data
        structures to maximize spatial locality based on the application's
        memory access patterns, and \changesvi{(b)~}issuing fewer, larger memory allocation
        requests that the virtual memory manager can attempt to allocate
        contiguously in the physical address space.}
        \shepiii{Alternatively, for programs that will execute on systems that make
        use of low-spatial-locality memories such as HMC, programmers should
        take the poor \shepiv{hardware} locality into account and \changesvi{perhaps} rewrite their applications to
        take better advantage of the large available bank-level parallelism.}

    \item \shep{\textbf{For some classes of applications, low-power memory
        can provide \ch{large} energy savings without sacrificing \shepii{significant}
        performance.}
        In particular, if we compare DDR3 with LPDDR4, we find that there are
        two types of memory \shepii{behavior} where LPDDR4 significantly reduces
        DRAM energy consumption without a \shepii{significant} performance overhead.
        First, applications with low memory intensity do not perform a large
        number of memory accesses.  As a result, despite the \shepii{much higher} access latency of
        LPDDR4 (\SI{45.0}{\nano\second} for a row miss) \shepii{over}
        that of DDR3 (\SI{26.3}{\nano\second}), such applications
        do not \shepii{experience} a significant impact on \chii{their} overall execution time.
        Second, applications with high BPU can take advantage of the \shepii{larger
        number of banks in} LPDDR4.
        \shepii{The} greater number of
        banks in LPDDR4 (16 per rank) than in DDR3 (8 per rank) actually \changesvi{helps}
        LPDDR4 to \changesvi{(partially)} overcome the overhead of additional latency.
	\vspace{3pt}\newline%
        \emph{For system architects}, this means that there are \shepiii{a number of} cases where
        \changesvi{they can deploy systems that use LPDDR4} to reduce the system energy consumption with
        \ch{a small} impact \shepiii{on system} performance, \ch{thereby 
        improving energy efficiency}.}
	\vspace{3pt}\newline%
        \shepiii{\emph{For DRAM architects}, there is a need to develop new \changesvi{DRAM} types
        and subsystems that consume low energy without impacting system performance
        across a broad range of applications.}
\end{enumerate}

%% file: sections/related_work.tex
% !TEX root=../dramcharacterization.tex

\section{Related Work}
\label{sec:related}

To our knowledge, this is \ch{(1)~}the first work to uncover new trends about
and interactions between different DRAM types and the performance
and energy consumption of modern workloads, 
\ch{(2)~the most extensive study performed to date on the combined 
DRAM--workload behavior, and
(3)~the most comprehensive study to date \chii{that compares} the performance of
monolithic (i.e., 2D) and 3D-stacked DRAM}.
No prior work presents a comprehensive study across such a wide variety 
of workloads \ch{(115 of them)} and DRAM types \ch{(nine of them)}.
We briefly discuss the most closely related works.

Cuppu et al.\chii{~\cite{cuppu.isca99, cuppu.tc01}} present a \changesvi{study} of \ch{seven}
DRAM types \changesvi{and their interaction with a suite of
desktop and scientific applications}.  Their work, \chii{more than} two decades old now, noted
several characteristics emerging from then-con\-tem\-po\-ra\-ry DRAM designs
\ch{(many of which do not exist in the field today)},
and made recommendations based on these insights. 
\ch{Similar to some of our findings, Cuppu et al.\ recommend that
memory latency needs to be reduced, and spatial locality needs to
be further exploited by the memory subsystem.}
\chii{Other recommendations from Cuppu et al.\ are more relevant
for the older DRAM types that they study, and in some cases do not
apply to the modern DRAM types that we study in this work.}
\chiii{Later work by Cuppu and Jacob~\cite{cuppu.isca01} studies the impact of
different memory channel configurations on application performance,
which is orthogonal to the characterizations that we perform.}

Zhu and Zhang~\cite{zhu.hpca05} study how various
DRAM types can be optimized to work with \ch{simultaneous multithreading (SMT)}
processors, but
do not perform a broad characterization \changesvi{of applications}.
Zheng and Zhu~\cite{zheng.tc10}
compare the performance of DDR3 DRAM to DDR2\ch{~\cite{ddr2}} and FB-DIMM\ch{~\cite{fbdimm}}
\ch{for 26 desktop and scientific applications}.  
Gomony et al.~\cite{gomony.date12} characterize \ch{three low-power} DRAM types for
mobile systems, and propose a tool to select the right type for
real-time mobile systems.
All of these
studies predate the emergence of \ch{most} of the DRAM types that we characterize
\ch{(DDR4, LPDDR3, LPDDR4, HBM, HMC, Wide I/O 2)},
do not evaluate energy consumption, and focus only on a limited set of applications.

Li et al.~\cite{li.memsys18} evaluate the performance and power of
\ch{nine} modern DRAM types, including \ch{two versions each of}
HMC and HBM.
\ch{However, unlike our wide range of applications,
their evaluation studies} only 10 desktop and scientific applications.
Furthermore, their memory configuration uses row interleaving, 
which reduces the \chii{bank}-level parallelism compared to \ch{many} modern systems
that use \ch{cache line interleaving~\cite{kim.isca12, jeong.hpca12, kim.hpca10,
coregen7.datasheet, xeone5.datasheet, lenovo.xeon.memconfig, power9.datasheet, 
zhang.micro00, rokicki.tr96}}.

Several works study the impact of \ch{new and existing} memory controller policies on
performance
(e.g., \chii{\cite{rixner.isca00, kim.micro10, ghose.isca13,
ipek.isca08, mutlu.isca08, kim.hpca10, kaseridis.micro2011, subramanian.iccd14,
mutlu.micro07, usui.taco16, hur.micro04, nesbit.micro06, mukundan.hpca12,
subramanian.tpds16, moscibroda.podc08, ausavarungnirun.isca12, 
muralidhara.micro11, rixner.micro04, ebrahimi.micro11, yuan.micro09, 
jeong.dac12, liu.isca12, venkatesan.hpca06, isen.micro09, lee.tr10, 
zhang.micro00, rokicki.tr96, lee.pact15, chang.hpca14, lee.hpca15, chang.sigmetrics16,
hassan.hpca16, lee.sigmetrics17, kim.iccd18, zhang.hpca16, wang.micro18,
lee.micro09, lee.micro08, ebrahimi.isca11, seshadri.isca14, zhao.micro14,
jog.sigmetrics16, das.dac18, stuecheli.isca10, stuecheli.micro10, 
chandrasekar.date14}}).
These} works are orthogonal to our study, which
keeps controller policies constant and explores the effect of the underlying DRAM type.
Other works \chiii{profile} the low-level behavior of DRAM types by
characterizing real \chii{DRAM chips} (e.g., \chii{\cite{lee.hpca15, chang.sigmetrics16,
lee.sigmetrics17, ghose.sigmetrics18, chang.sigmetrics17, kim.isca14, kim.hpca19,
liu.isca13, patel.dsn19, kim.hpca18, kim.iccd18, khan.dsn16, qureshi.dsn15,
patel.isca17, khan.sigmetrics14, khan.micro17, khan.cal16, hassan.hpca17,
chandrasekar.date14}}).  \ch{These}
works focus on a single DRAM type (DDR3 \changesvi{or LPDDR4}), and do not use real-world
applications to perform their characterization, \ch{since their goal is to understand
device behavior independently of workloads}.
\chii{A few works~\cite{meza.dsn15, schroeder.sigmetrics09, hwang.asplos12,
sridharan.sc13, sridharan.asplos15} perform large-scale characterization studies of DRAM errors
in the field.  These works examine reliability in a specific setting (e.g.,
datacenters, supercomputers), and as a result they do not focus on metrics such as
performance or energy, and do not consider a broad range of application domains.}

A number of works study the memory access behavior of
benchmark suites (e.g., \ch{\cite{parsec, henning.computer00, he.pact08,
singh.icpe19, murphy.tc07, agaram.ismm06, charney.ibmjrd97, mccalpin.tcca95}}).
These works focus on only a \emph{single} DRAM type.
Conversely, several works \ch{propose DRAM simulators and use
the simulators to} study the
memory access behavior of a limited set of workloads on several
memory types~\cite{kim.cal15, rosenfeld-cal2011, radulovic.memsys15,
suresh.cluster14, giridhar.sc13, ahn.sc09, li.sc17}.  
None of these studies (1)~take a \changesvi{comprehensive
look at as wide a} range of workloads and DRAM types as
we do, or (2)~evaluate energy consumption.

%% file: sections/conclusions.tex
% !TEX root=../dramcharacterization.tex

\section{Conclusion}
\label{sec:conclusion}

\changesvi{It has become very difficult to intuitively understand
how modern applications interact with different DRAM types.
This is due to the emergence of (1)~many new DRAM types, each catering to
different \ch{needs (e.g., high \chii{bandwidth}, low power,
high memory density)}; and (2)~new applications that are often
data intensive.  The combined behavior of each pair of workload and
DRAM type 
is impacted by the complex interaction between memory latency,
\chii{bandwidth}, \ch{bank parallelism, row buffer locality, memory access
patterns, and} energy consumption.}
In this work, we perform a comprehensive experimental study to
analyze these interactions, by characterizing the
behavior of 115~applications and workloads \chii{across} nine DRAM types.
With the help of new metrics that capture the interaction between
\changesvi{memory access patterns and the underlying hardware},
we make 12~key observations \chii{and draw out many new findings}
about the combined DRAM--workload behavior.
\ch{We then provide a number of recommendations for
DRAM architects, system architects, and programmers.}
We hope that our \chii{observations inspire} \changesvi{the development
of} many memory optimizations in both hardware and software.
\changesvi{\chiii{To this end,} we have released our toolchain with the hope that
the tools can assist with future studies and research on memory
optimization} \chii{in both hardware and software}.

%% file: sections/dramarch.tex
% !TEX root=../dramcharacterization.tex

\section{Background on Modern DRAM Types}
\label{sec:dramarch}

\paratitle{DDR3}
Double Data Rate (DDR3)~\cite{ddr3} memory is the third generation of
DDR DRAM.
Each rank in DDR3 consists of eight banks, which \changesvii{ideally} allows eight
memory requests to be performed in parallel in a rank.
All of the banks share a single memory channel, and the memory controller
must \changesvii{schedule resources to ensure that request responses do not 
conflict with each other on}
the channel when \changesvii{each} response is being sent from
DRAM to the processor.
In order to reduce memory channel contention and increase memory throughput,
DDR3 transmits data on both the positive and negative edges of the bus clock, which
doubles the data rate by allowing a \emph{data burst} (i.e., a piece of data) to
be transmitted in only half a clock cycle.  In DDR3, eight 64-bit data bursts are
required for each 64-byte read request~\cite{ddr3}.
DDR3 was first released in 2007~\cite{ddr3}, but continues to be one of the
most popular types of DRAM available on the market today due to its low cost.
However, with the limited number of banks per rank \changesvii{and the
\ch{difficulties of increasing DDR3} bus clock frequencies,}
manufacturers no longer aggressively increase the density of DDR3 memories.

\paratitle{DDR4}
DDR4~\cite{ddr4} has evolved from the DDR3 DRAM type
as a response to solving some of the issues of earlier DDR designs.  
A major barrier to DRAM scalability is the eight-bank design 
used in DDR3 memories, as it is becoming more difficult to increase the size of 
the DRAM array within each bank.  In response to this, DDR4 employs
\emph{bank groups}~\cite{ddr4}, which enable DDR4 to double the number of banks in a
cost-effective manner.  A bank group represents a new
level of hierarchy, where it is faster to access two banks in two different bank
groups than it is to access two banks within the same group.  This is a
result of the additional I/O sharing that takes place within a bank group,
\changesvii{which reduces \chii{hardware cost} but leads to conflicts when two requests
access \chii{different banks in} the same bank group}.
One drawback of the DDR4 implementation of bank groups is that the average
memory access takes \emph{longer} in DDR4 than it did in DDR3.  
DRAM vendors
make the trade-off of having additional bank-level parallelism and higher bus
throughput in DDR4, which can potentially offset the latency increase when
an application effectively exploits bank-level parallelism.

\paratitle{GDDR5}
Like DDR4, Graphics DDR5 (GDDR5)~\cite{gddr5} memory uses bank groups to double the number of banks.
However, GDDR5 does so \emph{without} increasing memory latency, 
instead increasing the die area and energy.
Due to these additional costs, GDDR5 is currently unable to support the memory densities available in DDR4. 
GDDR5 \changesvii{increases} memory throughput significantly over DDR3 by \emph{quad pumping}
its data (i.e., it effectively sends four pieces of data in a
single clock cycle, as opposed to the two pieces sent by DDR3)~\cite{gddr5}.  In addition,
GDDR5 memories are clocked at a faster frequency.  
This aggressive 
throughput is especially helpful for \ch{GPUs,}
as they often perform many data-parallel operations that \chii{require high} memory
throughput.
\ch{As a result, many GPUs use GDDR5 memory.}

\paratitle{3D-Stacked DRAM}
Thanks to recent \changesvii{innovations,} manufacturers are now able
to build 3D-stacked memories, where multiple layers of DRAM are stacked on top
of one another.  A major advantage of 3D stacking is the availability of
\emph{through-silicon vias}~\cite{loh2008stacked, lee.taco16}, vertical interconnects that provide a high-bandwidth
interface across the layers.  The High Bandwidth Memory (HBM), Wide I/O, and
Wide I/O 2 DRAM types exploit 3D stacking for different purposes.
HBM~\cite{hbm, AMD.hbm} is a response to the need for improved memory bandwidth for GPUs without 
the high power costs associated with GDDR5.  HBM DRAM is clocked much slower
than GDDR5, but 
connects four to eight memory
channels to a \emph{single} DRAM \ch{module}.  The large number of memory channels allows
each HBM \ch{module} to service a large number of requests in parallel without I/O contention.  
Wide I/O~\cite{wideio} and Wide I/O 2~\cite{wideio2} apply the same principle
while targeting low-power devices (e.g., mobile phones)~\cite{kim.isscc2011}.
As mobile devices are
\emph{not} expected to require as much throughput as GPUs, Wide I/O and Wide I/O 2
have \emph{fewer} memory channels connected to each stack, and use \emph{fewer} banks than
HBM and GDDR5.

The Hybrid Memory Cube (HMC)~\cite{jeddeloh2012hybrid, pawlowski.hc11, rosenfeld.tr12, hmc.2.1}
makes more \emph{radical} changes to the memory design.  HMC is a 3D-stacked memory designed
to maximize the amount of parallelism that DRAM can deliver. 
\changesvii{\ch{It has increased access latencies} in order to provide a
significant increase in the number of banks \chii{(256 in HMC v2.1~\cite{hmc.2.1})}.}
Instead of employing a traditional on-chip
memory controller, a processor using an HMC \ch{chip} simply sends requests in
FIFO order to the memory, over a high-speed serial link.
Unlike other DRAM
types, all scheduling constraints in HMC are handled within the memory
itself,
\changesvii{as the HMC memory controller in the logic layer of the memory chip
performs scheduling.}
To keep this scheduling logic manageable, HMC partitions its DRAM into
multiple \emph{vaults}, 
\changesvii{each of which consists} of a small, multi-bank vertical slice of memory.  
\changesvii{To facilitate the partitioning of memory into vaults}, HMC reduces the size of each
row in memory from the typical \SIrange{4}{8}{\kilo\byte} down to 256~bytes.

\sloppypar
\paratitle{LPDDR3 and LPDDR4}
In order to decrease the power consumed by \chii{DDRx} DRAM, manufacturers have
created low-power (LP) variants, known as LPDDR3 and LPDDR4.
LPDDR3~\cite{lpddr3} reduces power over DDR3 by using a lower core voltage, employing
deep power-down modes, and reducing the number of chips used in each
DRAM module.  One drawback of the lower core voltage and the deep
power-down mode is that memory accesses take significantly \emph{longer} on
low-power memories (see Table~\ref{tbl:dram}).  LPDDR4~\cite{lpddr4} achieves even
greater power savings by cutting the width of each chip in half with respect
to LPDDR3.  A smaller chip \changesvii{width leads to lower power consumption,
but} requires LPDDR4
to perform \emph{double} the number of data bursts \ch{(i.e., have higher
latency)} for each request \changesvii{to keep the throughput intact}.

%% file: sections/artifactdesc.tex
% !TEX root=../dramcharacterization.tex

\section{Ramulator Modifications}
\label{sec:artdesc:sim}

We characterize the different DRAM architectures using a heavily-modified 
version of Ramulator~\cite{kim.cal15}.  Ramulator is a detailed and extensible
\changesvii{open-source DRAM simulator~\cite{ramulator.github}}.  
We make several modifications to Ramulator to improve the
fidelity of our experiments.  First, we implement a shared last-level cache, to
ensure that the initial contention between memory requests from different
cores takes place before the requests reach memory, just as they would in a real
computer.  Second, we add support for virtual-to-physical address translation.
Third, we implement a faithful model of HMC version 
2.1~\cite{hmc.2.1}.  Our model accurately replicates the high-speed
serial link in HMC, and includes a logic layer where DRAM commands
are scheduled.

Our modifications allow us to use \shep{application traces to drive}
the simple core model built into Ramulator, as
opposed to using a detailed CPU timing simulator \shep{to
execute the application}, without losing accuracy.  
\shep{As a result, we can significantly reduce the total simulation
time required \chv{(by an average of 9.8$\times$, with a range of 1.4\%--24.7\%, for our
SPEC CPU 2006 benchmarks)}, and can \changes{simulate applications with
much larger memory footprints without the need for
large computing resources}.}
We simulate a 4~GHz, 4-issue processor with a 128-entry reorder buffer, and 
\chii{an 8-way} set associative shared last-level cache
\changesvii{(see Table~\ref{tbl:config} in Section~\ref{sec:meth})}.
We \shep{have open-sourced our modified} version of Ramulator\shepii{~\cite{ramulator.github}}.

\paratitle{\shep{Validation}}
\shep{We validate our trace-based approach by comparing 
\shepiii{(a)~}the simple core model results 
from our modified version of Ramulator with
\shepiii{(b)~}results generated when we execute 
applications using \changesvii{gem5~\cite{gem5},}
a detailed, full-system, cycle-accurate CPU timing simulator.
\changesvii{We integrate gem5~\cite{gem5} with the unmodified
version of Ramulator to accurately} model the memory system.
Prior work~\cite{kim.cal15} has already validated the memory model in the unmodified
Ramulator with a Verilog memory model provided by Micron~\cite{ddr3.verilogmodel}.}

\shep{To perform our validation, we run all of our SPEC CPU2006~\cite{spec2006}
applications using both our trace-driven modified Ramulator and the full-system
gem5 with Ramulator.  We configure both simulators to use the system configuration
in Table~\ref{tbl:config}.
\shepiii{As we are interested in comparing trends across applications and across
memory types, we normalize the performance \changesvii{(i.e., execution time)} 
of each application to one benchmark
(\emph{gamess}).}
We find that \shepii{normalized \changesvii{performance} results from our trace-driven modified Ramulator differ by an average of
only 6.1\% from the \changesvii{performance} results when using full-system gem5 and Ramulator.}
As other works have shown, much larger differences \shepii{between a simulation
platform and the \changesvii{system being modeled by the simulator}}
are still representative of \changesvii{the behavior of the modeled system}.
For example, other popular and publicly-available simulators that have been validated
report average errors of 4.9\%~\cite{endo.samos14}, 12--19\%~\cite{alves.hpcc15}, 
20\%~\cite{ubal.pact12}, and 19.5\%~\cite{desikan.isca01}.
We believe that our average validation error, which at 6.1\% is on the lower end 
of this \changesvii{error} range,
represents that the quantitative values generated by our simulator can be trusted, 
and that the general observations that we make are \ch{accurate.}}

\section{Workload Details}
\label{sec:artdesc:workloads}

We study 87~different applications, spread over a diverse range of uses.
In our characterization, we categorize our applications into one of six families:
desktop/scientific~\cite{spec2006, coral, coral-2, parsec}, server/cloud~\cite{hadoop, ycsb, redis, Apache, 
memcached, difallah.vldb04}, multimedia \changesvii{acceleration}~\cite{mediabench}, 
\changesiii{network \changesvii{acceleration}}~\cite{nxp.networkaccel}, 
GPGPU~\cite{he.pact08, rodinia, lonestar}, and OS routines~\cite{IOZone, Netperf, seshadri.micro13}.  
These applications have been collected from a
wide variety of sources.
\shepiii{The 87~evaluated applications are listed across four tables:
Table~\ref{tbl:workloads:desktop} lists desktop/scientific applications
(characterized in Sections~\ref{sec:desktop} and \ref{sec:mt});
Table~\ref{tbl:workloads:server} lists server/cloud applications
(characterized in Section~\ref{sec:server});
Table~\ref{tbl:workloads:hetero} lists multimedia, network accelerator, and
GPGPU applications
(characterized in Section~\ref{sec:hetero}); and
Table~\ref{tbl:workloads:os} lists OS routines
(characterized in Section~\ref{sec:os}).
In each table, we list the input size, and the total \ch{DRAM} footprint of
memory accesses performed in DRAM \ch{(i.e., the number of
unique byte addresses that are read from or written to DRAM).
Note that the DRAM footprints consider only last-level cache misses to
DRAM, and do not include any memory \chii{accesses} that hit in the
caches.  As a result, the DRAM footprints that we report may be
smaller than the working set sizes reported in other works.}}

\begin{table}[t]
\centering
\caption{Evaluated desktop/scientific applications.\vspace{-8pt}}
\label{tbl:workloads:desktop}
{\footnotesize
\begin{tabular}{cccr}
\firsthline
\textbf{Application} & \textbf{Benchmark} & \textbf{Input Set/} & \multicolumn{1}{c}{\textbf{DRAM}} \\
\textbf{Suite} & \textbf{Name} & \textbf{Problem Size} & \multicolumn{1}{c}{\textbf{Footprint}} \\ \hline
\multirow{22}{*}{SPEC CPU2006}
& \emph{gamess} & ref & \SI{0.8}{\mega\byte}  \\
& \emph{povray} & ref & \SI{1.0}{\mega\byte} \\
& \emph{calculix} & ref & \SI{1.1}{\mega\byte} \\
& \emph{h264ref} & ref & \SI{9.4}{\mega\byte} \\
& \emph{perlbench} & ref & \SI{20.4}{\mega\byte} \\
& \emph{hmmer} & ref & \SI{7.5}{\mega\byte} \\
& \emph{bzip2} & ref & \SI{9.0}{\mega\byte} \\
& \emph{sjeng} & ref & \SI{166.0}{\mega\byte} \\
& \emph{sphinx3} & ref & \SI{17.1}{\mega\byte} \\
& \emph{namd} & ref & \SI{39.9}{\mega\byte} \\
& \emph{astar} & ref & \SI{25.1}{\mega\byte} \\
& \emph{gobmk} & ref & \SI{25.6}{\mega\byte} \\
& \emph{zeusmp} & ref & \SI{128.0}{\mega\byte} \\
& \emph{cactusADM} & ref & \SI{166.5}{\mega\byte} \\
& \emph{gcc} & ref & \SI{91.2}{\mega\byte} \\
& \emph{omnetpp} & ref & \SI{145.6}{\mega\byte} \\
& \emph{soplex} & ref & \SI{58.0}{\mega\byte} \\
& \emph{bwaves} & ref & \SI{559.8}{\mega\byte} \\
& \emph{GemsFDTD} & ref & \SI{718.5}{\mega\byte} \\
& \emph{milc} & ref & \SI{362.0}{\mega\byte} \\
& \emph{libquantum} & ref & \SI{32.0}{\mega\byte} \\
& \emph{mcf} & ref & \SI{1673.0}{\mega\byte} \\ \cline{1-4}
\multirow{9}{*}{PARSEC}
& \emph{blackscholes} & simmedium & \SI{3.8}{\mega\byte} \\
& \emph{canneal} & simmedium & \SI{2268.7}{\mega\byte} \\
& \emph{fluidanimate} & simmedium & \SI{350.5}{\mega\byte} \\
& \emph{raytrace} & simmedium & \SI{323.0}{\mega\byte} \\
& \emph{bodytrack} & simmedium & \SI{65.3}{\mega\byte} \\
& \emph{facesim} & simmedium & \SI{374.3}{\mega\byte} \\
& \emph{freqmine} & simmedium& \SI{503.3}{\mega\byte} \\
& \emph{streamcluster} & simmedium & \SI{72.1}{\mega\byte} \\
& \emph{swaptions} & simmedium & \SI{31.1}{\mega\byte} \\ \cline{1-4}
\multirow{2}{*}{CORAL}
& \multirow{2}{*}{\emph{miniFE}} & 32 x 32 x 32 & \SI{52.5}{\mega\byte} \\ \cline{3-4}
& & 64 x 64 x 64 & \SI{288.1}{\mega\byte} \\ \cline{1-4}
\multirow{2}{*}{CORAL-2}
& \emph{quicksilver} & Coral2\_P1.inp, 4 x 4 x 4 & \SI{56.6}{\mega\byte} \\
& \emph{pennant} & leblanc.pnt & \SI{8.6}{\mega\byte} \\
\lasthline
\end{tabular}
}
\end{table}

\begin{table}[t]
\centering
\caption{Evaluated server/cloud applications.\vspace{-8pt}}
\label{tbl:workloads:server}
{\footnotesize
\begin{tabular}{cccr}
\firsthline
\textbf{Application} & \textbf{Benchmark} & \textbf{Input Set/} & \multicolumn{1}{c}{\textbf{DRAM}} \\
\textbf{Suite} & \textbf{Name} & \textbf{Problem Size} & \multicolumn{1}{c}{\textbf{Footprint}} \\ \hline
\multirow{14}{*}{Hadoop}
& \multirow{5}{*}{\emph{grep}} & \multirow{5}{*}{\SI{1}{\giga\byte}}
& map~0: \SI{147.5}{\mega\byte}  \\
& & & map~1: \SI{149.2}{\mega\byte}  \\
& & & map~2: \SI{147.1}{\mega\byte}  \\
& & & map~3: \SI{145.9}{\mega\byte}  \\
& & & reduce: \SI{26.9}{\mega\byte}  \\ \cline{2-4}
& \multirow{5}{*}{\emph{wordcount}} & \multirow{5}{*}{\SI{1}{\giga\byte}}
& map~0: \SI{1332.2}{\mega\byte}  \\
& & & map~1: \SI{1307.3}{\mega\byte}  \\
& & & map~2: \SI{1308.2}{\mega\byte}  \\
& & & map~3: \SI{1360.3}{\mega\byte}  \\
& & & reduce: \SI{81.2}{\mega\byte}  \\ \cline{2-4}
& \multirow{4}{*}{\emph{sort}} & \multirow{4}{*}{\SI{1}{\giga\byte}}
& map~0: \SI{19.1}{\mega\byte}  \\
& & & map~1: \SI{19.9}{\mega\byte}  \\
& & & map~2: \SI{19.5}{\mega\byte}  \\
& & & map~3: \SI{21.0}{\mega\byte}  \\ \hline
\multirow{6}{*}{YCSB + Redis}
& \multirow{2}{*}{\emph{workload~A}} & \multirow{2}{*}{---} & server: \SI{217.9}{\mega\byte}  \\
& & & bgsave: \SI{195.0}{\mega\byte}  \\ \cline{2-4}
& \emph{workload~B} & --- & server: \SI{219.5}{\mega\byte}  \\ \cline{2-4}
& \emph{workload~C} & --- & server: \SI{218.6}{\mega\byte}  \\ \cline{2-4}
& \emph{workload~D} & --- & server: \SI{193.2}{\mega\byte}  \\ \cline{2-4}
& \emph{workload~E} & --- & server: \SI{27.0}{\mega\byte}  \\ \hline
\multirow{3}{*}{---}
& \emph{MySQL} & employeedb & \SI{65.1}{\mega\byte}  \\
& \emph{Memcached} & continuous insertions & \SI{177.4}{\mega\byte}  \\
& \emph{Apache2} & continuous \texttt{wget()} calls & \SI{200.0}{\mega\byte}  \\
\lasthline
\end{tabular}
}
\end{table}

\begin{table}[h]
\centering
\caption{Evaluated heterogeneous system applications.\vspace{-8pt}}
\label{tbl:workloads:hetero}
{\footnotesize
\begin{tabular}{cccr}
\firsthline
\textbf{Application} & \textbf{Benchmark} & \textbf{Input Set/} & \multicolumn{1}{c}{\textbf{DRAM}} \\
\textbf{Suite} & \textbf{Name} & \textbf{Problem Size} & \multicolumn{1}{c}{\textbf{Footprint}} \\ \hline
\multirow{4}{*}{MediaBench~II}
& \emph{H.264 encode} & base\_4CIF & \SI{10.2}{\mega\byte} \\
& \emph{H.264 decode} & base\_4CIF & \SI{8.3}{\mega\byte} \\
& \emph{JPEG-2000 encode} & base\_4CIF & \SI{24.4}{\mega\byte} \\
& \emph{JPEG-2000 decode} & base\_4CIF & \SI{21.5}{\mega\byte} \\ \hline
NXP Network
& \emph{Trace~A} & --- & \SI{0.7}{\mega\byte} \\
Accelerator & \emph{Trace~B} & --- & \SI{0.8}{\mega\byte} \\ \hline
\multirow{5}{*}{LonestarGPU}
& \emph{dmr} & r1M & \SI{0.1}{\mega\byte} \\
& \emph{bh} & 50K~bodies & \SI{0.5}{\kilo\byte} \\
& \emph{mst} & USA-road-d.FLA & \SI{4.0}{\mega\byte} \\
& \emph{bfs} & rmat20 & \SI{4.0}{\mega\byte} \\
& \emph{sp} & 4.2M~literals, 1M~clauses & \SI{34.3}{\mega\byte} \\ \hline
\multirow{3}{*}{Rodinia}
& \emph{hs} & 512 & \SI{3.2}{\mega\byte} \\
& \emph{bp} & 64K~nodes & \SI{4.9}{\mega\byte} \\
& \emph{sc} & 64K~points & \SI{40.1}{\mega\byte} \\ \hline
\multirow{2}{*}{Mars}
& \emph{iix} & 3~web pages & \SI{31.3}{\kilo\byte} \\
& \emph{ss} & 1024 x 256 & \SI{6.0}{\mega\byte} \\
\lasthline
\end{tabular}
}
\end{table}

\begin{table}[h]
\centering
\caption{Evaluated OS routines.\vspace{-8pt}}
\label{tbl:workloads:os}
{\footnotesize
\begin{tabular}{cccr}
\firsthline
\textbf{Application} & \textbf{Benchmark} & \textbf{Input Set/} & \multicolumn{1}{c}{\textbf{DRAM}} \\
\textbf{Suite} & \textbf{Name} & \textbf{Problem Size} & \multicolumn{1}{c}{\textbf{Footprint}} \\ \hline
\multirow{4}{*}{Netperf}
& \emph{UDP\_RR} & --- & \SI{6.2}{\mega\byte} \\
& \emph{UDP\_STREAM} & --- & \SI{7.9}{\mega\byte} \\
& \emph{TCP\_RR} & --- & \SI{6.7}{\mega\byte} \\
& \emph{TCP\_STREAM} & --- & \SI{7.8}{\mega\byte} \\ \hline
\multirow{13}{*}{IOzone}
& \emph{Test~0} (write/re-write) & \SI{64}{\mega\byte}~file & \SI{110.7}{\mega\byte} \\
& \emph{Test~1} (read/re-read) & \SI{64}{\mega\byte}~file & \SI{107.2}{\mega\byte} \\
& \emph{Test~2} (random-read/write) & \SI{64}{\mega\byte}~file & \SI{111.8}{\mega\byte} \\
& \emph{Test~3} (read backwards)& \SI{64}{\mega\byte}~file & \SI{107.0}{\mega\byte} \\
& \emph{Test~4} (record re-write) & \SI{64}{\mega\byte}~file & \SI{41.9}{\mega\byte} \\
& \emph{Test~5} (strided read) & \SI{64}{\mega\byte}~file & \SI{108.1}{\mega\byte} \\
& \emph{Test~6} (fwrite/re-fwrite) & \SI{64}{\mega\byte}~file & \SI{112.3}{\mega\byte} \\
& \emph{Test~7} (fread/re-fread) & \SI{64}{\mega\byte}~file & \SI{109.2}{\mega\byte} \\
& \emph{Test~8} (random mix) & \SI{64}{\mega\byte}~file & \SI{42.8}{\mega\byte} \\
& \emph{Test~9} (pwrite/re-pwrite) & \SI{64}{\mega\byte}~file & \SI{42.7}{\mega\byte} \\
& \emph{Test~10} (pread/re-pread) & \SI{64}{\mega\byte}~file & \SI{44.1}{\mega\byte} \\
& \emph{Test~11} (pwritev/re-pwritev) & \SI{64}{\mega\byte}~file & \SI{42.0}{\mega\byte} \\
& \emph{Test~12} (preadv/re-preadv) & \SI{64}{\mega\byte}~file & \SI{112.7}{\mega\byte} \\ \hline
\multirow{4}{*}{---}
& \emph{shell} & --- & \SI{4.3}{\mega\byte} \\ \cline{2-4}
& \emph{bootup} & --- & \SI{21.0}{\mega\byte} \\ \cline{2-4}
& \multirow{2}{*}{\emph{fork}} & \SI{64}{\mega\byte} shared data, & \multirow{2}{*}{\SI{22.8}{\mega\byte}} \\
& & 1K~updates & \\
\lasthline
\end{tabular}
}
\end{table}

We run our workloads to completion, with \shep{three} exceptions.
For our desktop benchmarks, we identify a representative phase of
execution using Simpoint~\cite{hamerly2005simpoint}.  During simulation,
we warm up the caches for 100~million instructions, and then run a 1-billion 
instruction representative phase.
\shep{We execute each GPGPU application until the application completes, or until the 
GPU executes 100~million instructions, \chii{whichever occurs first}.}
For Netperf, we
emulate 10 real-world seconds of execution time for each benchmark.

In addition to our 87~applications listed in \shepiii{Tables~\ref{tbl:workloads:desktop}--\ref{tbl:workloads:os}},
we assemble 28 multiprogrammed workloads for our desktop (Table~\ref{tbl:mwdesktop})
and server/cloud (Table~\ref{tbl:mwcloud})
applications by selecting bundles of four applications to represent varying
levels of memory intensity.  To ensure
that we accurately capture system-level contention, we restart any applications
that finish until \emph{all} applications in the bundle complete.  
Note that we
stop collecting statistics for an application once it has restarted.

\begin{table}[h]
\centering
\caption{Multiprogrammed workloads of desktop and scientific applications.
For each application, we indicate what fraction of the applications in the
workload are memory intensive (i.e., MPKI $>$ 15.0).\vspace{-8pt}}
\label{tbl:mwdesktop}
{\footnotesize
\begin{tabular}{cp{4.15cm}cr}
\firsthline
\textbf{Bundle} & \multicolumn{1}{c}{\bf Applications} & \textbf{\% Mem} & \multicolumn{1}{c}{\textbf{Memory}} \\
\textbf{Name} & \multicolumn{1}{c}{\bf in Workload} & \textbf{Intensive} & \multicolumn{1}{c}{\textbf{Footprint}} \\ \hline
\textbf{D0} & \emph{milc}, \emph{GemsFDTD}, \emph{mcf}, \emph{libquantum} & 100\% & \SI{5673.4}{\mega\byte} \\
\textbf{D1} & \emph{bwaves}, \emph{omnetpp}, \emph{mcf}, \emph{libquantum} & 100\% & \SI{3304.3}{\mega\byte} \\
\textbf{D2} & \emph{libquantum}, \emph{bwaves}, \emph{soplex}, \emph{GemsFDTD} & 100\% & \SI{3698.6}{\mega\byte} \\
\textbf{D3} & \emph{soplex}, \emph{mcf}, \emph{omnetpp}, \emph{milc} & 100\% & \SI{3512.2}{\mega\byte} \\ \hline
\textbf{D4} & \emph{milc}, \emph{mcf}, \emph{GemsFDTD}, \emph{h264ref} & 75\% & \SI{5539.9}{\mega\byte} \\
\textbf{D5} & \emph{soplex}, \emph{omnetpp}, \emph{milc}, \emph{namd} & 75\% & \SI{1801.9}{\mega\byte} \\
\textbf{D6} & \emph{libquantum}, \emph{omnetpp}, \emph{bwaves}, \emph{povray} & 75\% & \SI{1377.3}{\mega\byte} \\
\textbf{D7} & \emph{libquantum}, \emph{mcf}, \emph{milc}, \emph{zeusmp} & 75\% & \SI{3503.4}{\mega\byte} \\ \hline
\textbf{D8} & \emph{omnetpp}, \emph{GemsFDTD}, \emph{cactusADM}, \emph{hmmer} & 50\% & \SI{3141.0}{\mega\byte} \\
\textbf{D9} & \emph{GemsFDTD}, \emph{mcf}, \emph{gamess}, \emph{zeusmp} & 50\% & \SI{4006.5}{\mega\byte} \\
\textbf{D10} & \emph{milc}, \emph{mcf}, \emph{bzip2}, \emph{h264ref} & 50\% & \SI{3196.6}{\mega\byte} \\
\textbf{D11} & \emph{bwaves}, \emph{soplex}, \emph{gamess}, \emph{namd} & 50\% & \SI{1095.1}{\mega\byte} \\ \hline
\textbf{D12} & \emph{omnetpp}, \emph{sjeng}, \emph{namd}, \emph{gcc} & 25\% & \SI{1137.5}{\mega\byte} \\
\textbf{D13} & \emph{GemsFDTD}, \emph{hmmer}, \emph{zeusmp}, \emph{astar} & 25\% & \SI{2643.0}{\mega\byte} \\
\textbf{D14} & \emph{GemsFDTD}, \emph{povray}, \emph{sphinx3}, \emph{calculix} & 25\% & \SI{1341.7}{\mega\byte} \\
\textbf{D15} & \emph{soplex}, \emph{zeusmp}, \emph{sphinx3}, \emph{gcc} & 25\% & \SI{502.4}{\mega\byte} \\ \hline
\textbf{D16} & \emph{povray}, \emph{astar}, \emph{gobmk}, \emph{perlbench} & 0\% & \SI{188.4}{\mega\byte} \\
\textbf{D17} & \emph{povray}, \emph{bzip2}, \emph{sphinx3}, \emph{cactusADM} & 0\% & \SI{345.9}{\mega\byte} \\
\textbf{D18} & \emph{astar}, \emph{sjeng}, \emph{gcc}, \emph{cactusADM} & 0\% & \SI{1132.4}{\mega\byte} \\
\textbf{D19} & \emph{calculix}, \emph{namd}, \emph{perlbench}, \emph{gamess} & 0\% & \SI{117.2}{\mega\byte} \\ 
\lasthline
\end{tabular}
}
\end{table}

\begin{table}[h]
\centering
\caption{Multiprogrammed workloads of server and cloud applications.\vspace{-8pt}}
\label{tbl:mwcloud}
{\footnotesize
\begin{tabular}{ccp{3.5cm}r}
\firsthline
\textbf{Application} & \textbf{Bundle} & \multicolumn{1}{c}{\bf Applications} & \multicolumn{1}{c}{\textbf{Memory}} \\
\textbf{Suite} & \textbf{Name} & \multicolumn{1}{c}{\bf in Workload} & \multicolumn{1}{c}{\textbf{Footprint}} \\ \hline
\multirow{5}{*}[-5.5em]{YCSB + Redis} 
& \textbf{Y0} & \emph{workload A: server}, \emph{workload B: server}, \emph{workload C: server}, \emph{workload D: server} & \SI{1492.0}{\mega\byte} \\ \cline{2-4}
& \textbf{Y1} & \emph{workload A: server}, \emph{workload B: server}, \emph{workload C: server}, \emph{workload E: server} & \SI{1262.2}{\mega\byte} \\ \cline{2-4}
& \textbf{Y2} & \emph{workload A: server}, \emph{workload B: server}, \emph{workload D: server}, \emph{workload E: server} & \SI{1274.1}{\mega\byte} \\ \cline{2-4}
& \textbf{Y3} & \emph{workload A: server}, \emph{workload C: server}, \emph{workload D: server}, \emph{workload E: server} & \SI{1212.5}{\mega\byte} \\ \cline{2-4}
& \textbf{Y4} & \emph{workload B: server}, \emph{workload C: server}, \emph{workload D: server}, \emph{workload E: server} & \SI{889.9}{\mega\byte} \\ \hline
\multirow{3}{*}[-1.75em]{Hadoop} 
 & \textbf{H0} & four \emph{grep: map} processes with different inputs & \SI{1726.9}{\mega\byte} \\ \cline{2-4}
& \textbf{H1} & four \emph{wordcount: map} processes with different inputs & \SI{2240.3}{\mega\byte} \\ \cline{2-4}
& \textbf{H2} & four \emph{sort:map} processes with different inputs & \SI{94.9}{\mega\byte} \\ 
\lasthline
\end{tabular}
}
\end{table}

%% file: sections/detail.tex
% !TEX root=../dramcharacterization.tex

\section{Detailed Workload Characterization Results}
\label{sec:char}

\subsection{Single-Threaded Desktop/Scientific Applications}
\label{sec:char:desktop}
\label{sec:detail:desktopchar}

\begin{figure}[h]
  \centering
  \includegraphics[width=0.9\columnwidth, trim=68 128 60 130, clip]{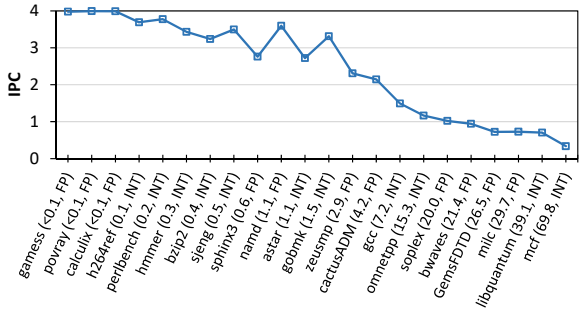}%
  \vspace{-10pt}%
  \caption{IPC for desktop and scientific applications executing on a system with DDR3-2133 memory.
  In parentheses, \changesvii{we show each benchmark's} MPKI, and whether the benchmark consists of
  predominantly integer (INT) or floating point (FP) operations.}
  \label{fig:desktop:singleipc}
\end{figure}

Figure~\ref{fig:desktop:singleipc} shows the instructions per cycle (IPC) for
each of the desktop applications when run on a system with DDR3 memory.  
The benchmarks along the x-axis are sorted in ascending order of MPKI (i.e.,
memory intensity).
As we discuss in Section~\ref{sec:desktop:workload}, our desktop applications
consist of both applications with predominantly integer computations and
applications with predominantly floating point computations.
Prior work \changes{shows} that within the CPU, there is a notable difference in the behavior of 
integer applications (typically
desktop and/or business applications) from floating point applications
(typically scientific 
applications)\mbox{~\cite{henning.computer00}}.  
From Figure~\ref{fig:desktop:singleipc}, we observe that the performance of the two
groups is interspersed throughout the range of MPKIs and IPCs.
Thus, we conclude that there is no discernible difference between 
integer and floating point applications, from the
perspective of \changesvii{main} memory.

\begin{figure}[h]
  \centering
  \includegraphics[width=0.9\columnwidth, trim=65 190 60 173, clip]{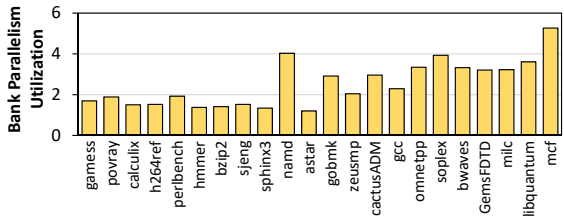}%
  \vspace{-10pt}%
  \caption{DDR3 BPU for \ch{single-threaded} desktop/scientific applications.}
  \label{fig:desktop:bpu-ddr3}
\end{figure}

We observe from Figure~\ref{fig:desktop:singleipc} that, \changesvii{in general,}
the overall IPC of desktop and scientific applications decreases as the
MPKI increases.
\changesvii{However,} there are two notable exceptions: \emph{namd}
and \emph{gobmk}.
We discuss how these exceptions are the result of \ch{differences in}
bank parallelism utilization (BPU) in Section~\ref{sec:desktop:workload}.
Figure~\ref{fig:desktop:bpu-ddr3} shows the BPU of each application
when run with the DDR3 DRAM type.  Note that our DDR3 configuration,
with four channels, and eight banks per channel, has a total of 32~banks
available.  Thus, the theoretical maximum BPU is 32, though this
does not account for
(1)~request serialization for banks that share a \changesvii{memory channel
or that are part of the same bank group}, or
(2)~maintenance operations such as refresh \changesvii{that reduce
the bank parallelism of requests}.
As we observe from the figure, none of our desktop and scientific
applications come close to the maximum BPU.
We find that \emph{namd} and \emph{gobmk} exhibit much higher
BPU values than other applications with similar MPKI values.
This indicates that these two applications often issue their memory requests
in clusters, \changesvii{i.e., they have} \emph{bursty} memory access patterns.
As a result, these two applications \changesvii{exploit \emph{memory-level parallelism} (MLP)},
where the latencies of multiple memory requests are overlapped with
each other, \changesvii{better than other applications.  This increased
MLP} reduces the application stall time\chii{~\cite{ghose.isca13, 
qureshi.isca06, mutlu.hpca03, mutlu.micro07, mutlu.ieeemicro06, kirman.hpca05,
das.micro09}}, which in turn increases the IPC of the application.

\begin{figure}[h]
  \centering
  \includegraphics[width=0.9\columnwidth, trim=65 190 60 187, clip]{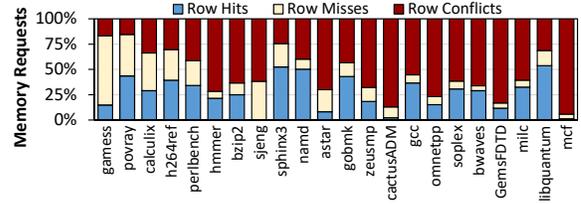}%
  \vspace{-10pt}%
  \caption{DDR3 row buffer locality for \ch{single-threaded} desktop/scientific applications.}
  \label{fig:desktop:rbl-ddr3}
\end{figure}

Figure~\ref{fig:desktop:rbl-ddr3} shows the row buffer locality of each
\ch{single-threaded} desktop/scientific application when run on DDR3.
We do not see a correlation between the memory intensity of an
application and its row buffer locality.  This suggests that the locality
is predominantly a function of the application's memory access patterns
\changesvii{and row buffer size, and is not related to memory intensity
(as expected).}
We \changesvii{compare} the row buffer locality under DDR3 to
the row buffer locality with our other DRAM types (not shown for brevity).
We find that, with the exception of HMC (which reduces the row width by 97\%),
row buffer locality \ch{characteristics} remain similar across different DRAM types.

\subsection{Multithreaded Desktop/Scientific Applications}
\label{sec:char:mt}

\begin{figure}[h]
  \centering
  \vspace{-5pt}
  \includegraphics[width=0.9\columnwidth, trim=68 122 60 104, clip]{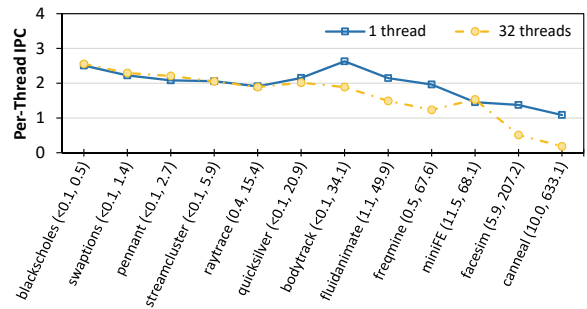}%
  \vspace{-10pt}%
  \caption{\ch{Per-thread} IPC for multithreaded applications executing on a system with DDR3-2133 memory.
  In parentheses, \changesvii{we show each benchmark's} \ch{1-thread} MPKI followed by its 32-thread MPKI.}
  \label{fig:mt:ipc}
\end{figure}

To gain insight on limiting factors on the scalability
of our multithreaded applications (see Section~\ref{sec:mt}), we study the
MPKI and IPC of each application when run using DDR3-2133, \changes{and}
when the application runs with 1, 2, 4, 8, 16, and 32~threads.
Figure~\ref{fig:mt:ipc} shows the \ch{per-thread} IPC for all
12 applications, \ch{when the applications are run with one thread and
with 32~threads,} and lists both the \ch{1-thread} and 32-thread MPKI
(which quantifies the memory intensity of the application).
We observe from the figure that 
\ch{unlike our single-threaded desktop/scientific applications,
many of our multithreaded applications maintain a relatively high IPC even at
32~threads, despite the high memory intensity.}
This is often because multithreaded applications are designed to strike a
careful balance between computation and memory usage, which is
necessary to scale the algorithms to large numbers of threads.
\ch{As a result, several memory-intensive multithreaded applications
have significantly higher IPCs when compared to single-threaded
desktop/scientific applications with similar MPKI values.  We note}
that as a general trend, multithreaded applications
with a higher MPKI tend to have a lower IPC relative to multithreaded
applications with a lower MPKI.

As an example, we see that \emph{miniFE} becomes more memory-intensive
as the number of threads increases, with its MPKI increasing from 11.5 with only
one thread to 68.1 with 32~threads.  Despite this increase in memory intensity,
its \ch{per-thread} IPC remains around 1.5, indicating that the application is not completely
memory-bound.  Prior work~\cite{balaprakash.ispass13} corroborates this behavior,
with an analysis of \emph{miniFE} showing that in its two hotspot functions,
the application spends about 40\% of its time on load instructions, but also 
spends about 40\% of its time on integer or floating-point instructions.
This exemplifies the balanced approach between computation and memory
that most of our multithreaded applications take, regardless of their
memory intensity.

\subsection{Server and Cloud Workloads}
\label{sec:char:server}

\begin{figure}[h]
  \centering
  \vspace{6pt}
  \includegraphics[width=0.9\columnwidth, trim=68 122 60 103, clip]{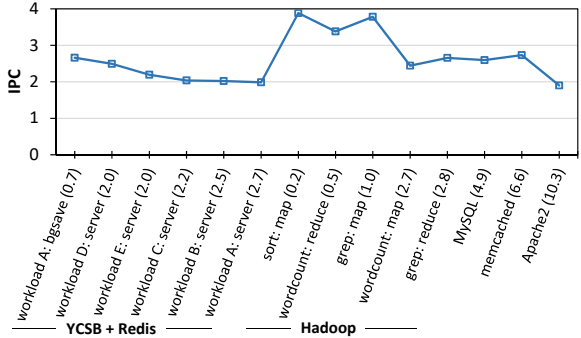}%
  \vspace{-7pt}%
  \caption{IPC for server/cloud applications executing on a system with DDR3-2133 memory.
  In parentheses, \changesvii{we show each benchmark's} MPKI.}
  \label{fig:server:ipc}
\end{figure}

To characterize our server and cloud workloads (see Section~\ref{sec:server}),
we study their performance and memory intensity using the DDR3 DRAM type.
Figure~\ref{fig:server:ipc} shows the 
performance of each application (IPC; see Section~\ref{sec:metrics}), and lists \changesvii{its} MPKI.
As we see from the figure, the IPC of all of the applications is very high, 
with the lowest-performing application (\emph{Apache2}) having an IPC of \chii{1.9}
\ch{(out of a maximum possible IPC of 4.0)}.
The high performance is a result of the low memory utilization of \changesvii{our} server and cloud
workloads, \changesvii{which 
are} highly optimized 
to take advantage of on-chip caches.

\section{DRAM Power Breakdown}
\label{sec:power:desktop}
\label{sec:detail:desktoppwr}

\begin{figure}[h]
  \includegraphics[width=\columnwidth, trim=57 120 60 210, clip]{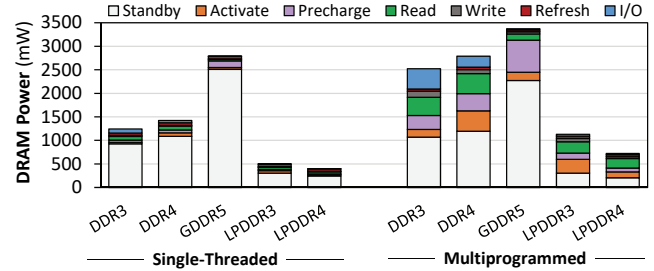}%
  \vspace{-10pt}
  \caption{Breakdown of mean DRAM power consumption when executing \ch{single-threaded} (left) and multiprogrammed (right) desktop
and scientific applications.}
  \label{fig:desktop:power}
\end{figure}

Figure~\ref{fig:desktop:power}
shows the breakdown of power consumed by \ch{the five DRAM types
for which we have accurate power models},
averaged across all of our \ch{single-threaded} applications and across our
multiprogrammed workloads from Section~\ref{sec:desktop}.
We observe that all of \ch{the evaluated} DRAM types consume a large amount of standby power
\changesvii{(we simulate a DRAM capacity of \SI{4}{\giga\byte})}, with
DDR3's standby power representing 77.8\% of its total power consumption.  As
the density \changesvii{and capacity} of DRAM continues to increase, the standby power consumption is
expected to grow as well.
However, the total power consumed varies widely between DRAM types.
For example, \changesvii{for our \ch{single-threaded} desktop and scientific
workloads, GDDR5 consumes 2.25x the power of DDR3, while LPDDR4 consumes
67.7\% less power than DDR3}.

Across all of our workloads (including other workload categories, not shown for brevity), 
we observe three \ch{major} trends:
(1)~standby power is the single biggest source of average power consumption
in standard-power DRAM types \changesvii{(because most of the DRAM capacity
is idle at any given time)};
(2)~LPDDR3 and LPDDR4 cut down standby power consumption significantly, 
due to a number of design optimizations that specifically target standby power
(see Appendix~\ref{sec:dramarch}); and
\ch{(3)~}workloads with a high rate of row conflicts and/or row misses spend more power on
activate and precharge commands, as a new row must be opened for each conflict or miss.

%% file: ms.bbl
% Generated by IEEEtranS.bst, version: 1.13 (2008/09/30)
\begin{thebibliography}{100}
\providecommand{\url}[1]{#1}
\csname url@samestyle\endcsname
\providecommand{\newblock}{\relax}
\providecommand{\bibinfo}[2]{#2}
\providecommand{\BIBentrySTDinterwordspacing}{\spaceskip=0pt\relax}
\providecommand{\BIBentryALTinterwordstretchfactor}{4}
\providecommand{\BIBentryALTinterwordspacing}{\spaceskip=\fontdimen2\font plus
\BIBentryALTinterwordstretchfactor\fontdimen3\font minus
  \fontdimen4\font\relax}
\providecommand{\BIBforeignlanguage}[2]{{%
\expandafter\ifx\csname l@#1\endcsname\relax
\typeout{** WARNING: IEEEtranS.bst: No hyphenation pattern has been}%
\typeout{** loaded for the language `#1'. Using the pattern for}%
\typeout{** the default language instead.}%
\else
\language=\csname l@#1\endcsname
\fi
#2}}
\providecommand{\BIBdecl}{\relax}
\BIBdecl

\bibitem{AMD.hbm}
{Advanced Micro Devices, Inc.}, ``{High Bandwidth Memory (HBM) DRAM},'' 2013.

\bibitem{agaram.ismm06}
K.~K. Agaram, S.~W. Keckler, C.~Lin, and K.~S. McKinley, ``{Decomposing Memory
  Performance: Data Structures and Phases},'' in \emph{ISMM}, 2006.

\bibitem{ahn.sc09}
J.~Ahn, N.~Jouppi, C.~Kozyrakis, J.~Leverich, and R.~Schreiber, ``{Future
  Scaling of Processor-Memory Interfaces},'' in \emph{SC}, 2009.

\bibitem{alves.hpcc15}
M.~A.~Z. Alves, C.~Villavieja, M.~Diener, F.~B. Moreira, and P.~O.~A. Navaux,
  ``{SiNUCA: A Validated Micro-Architecture Simulator},'' in
  \emph{HPCC/CSS/ICESS}, 2015.

\bibitem{hadoop}
{Apache Foundation}, ``{Apache Hadoop},'' \url{http://hadoop.apache.org/}.

\bibitem{Apache}
{Apache Foundation}, ``{Apache HTTP Server Project},''
  \url{http://www.apache.org/}.

\bibitem{ausavarungnirun.isca12}
R.~Ausavarungnirun, K.~K. Chang, L.~Subramanian, G.~H. Loh, and O.~Mutlu,
  ``{Staged Memory Scheduling: Achieving High Performance and Scalability in
  Heterogeneous Systems},'' in \emph{ISCA}, 2012.

\bibitem{bakhoda.ispass09}
A.~Bakhoda, G.~Yuan, W.~W.~L. Fung, H.~Wong, and T.~M. Aamodt, ``{Analyzing
  CUDA Workloads Using a Detailed GPU Simulator},'' in \emph{ISPASS}, 2009.

\bibitem{balaprakash.ispass13}
P.~Balaprakash, D.~Buntinas, A.~Chan, A.~Guha, R.~Gupta, S.~H.~K. Narayanan,
  A.~A. Chien, P.~Hovland, and B.~Norris, ``{Exascale Workload Characterization
  and Architecture Implications},'' in \emph{ISPASS}, 2013.

\bibitem{parsec}
C.~Bienia, S.~Kumar, J.~P. Singh, and K.~Li, ``{The PARSEC Benchmark Suite:
  Characterization and Architectural Implications},'' Princeton Univ.\ Dept.\
  of Computer Science, Tech. Rep. TR-811-08, 2008.

\bibitem{gem5}
N.~Binkert, B.~Beckmann, G.~Black, S.~K. Reinhardt, A.~Saidi, A.~Basu,
  J.~Hestness, D.~R. Hower, T.~Krishna, S.~Sardashti, R.~Sen, K.~Sewell,
  M.~Shoaib, N.~Vaish, M.~D. Hill, and D.~A. Wood, ``{gem5: A Multiple-ISA Full
  System Simulator with Detailed Memory Model},'' \emph{CAN}, 2011.

\bibitem{lonestar}
M.~Burtscher, R.~Nasre, and K.~Pingali, ``{A Quantitative Study of Irregular
  Programs on {GPUs}},'' in \emph{IISWC}, 2012.

\bibitem{ubuntu.14.04}
{Canonical Ltd.}, ``{Ubuntu 14.04 LTS (Trusty Tahr)},''
  \url{http://releases.ubuntu.com/14.04/}, 2014.

\bibitem{ubuntu.16.04}
{Canonical Ltd.}, ``{Ubuntu 16.04 LTS (Xenial Xerus)},''
  \url{http://releases.ubuntu.com/16.04/}, 2016.

\bibitem{chandrasekar.date14}
K.~Chandrasekar, S.~Goossens, C.~Weis, M.~Koedam, B.~Akesson, N.~Wehn, and
  K.~Goossens, ``{Exploiting Expendable Process-Margins in DRAMs for Run-Time
  Performance Optimization},'' in \emph{DATE}, 2014.

\bibitem{drampower}
K.~Chandrasekar, C.~Weis, Y.~Li, S.~Goossens, M.~Jung, O.~Naji, B.~Akesson,
  N.~Wehn, and K.~Goossens, ``{DRAMPower: Open-Source DRAM Power \& Energy
  Estimation Tool},'' \url{http://www.drampower.info}.

\bibitem{chang.hpca14}
K.~K. Chang, D.~Lee, Z.~Chishti, A.~Alameldeen, C.~Wilkerson, Y.~Kim, and
  O.~Mutlu, ``{Improving DRAM Performance by Parallelizing Refreshes With
  Accesses},'' in \emph{HPCA}, 2014.

\bibitem{chang.hpca16}
K.~K. Chang, P.~J. Nair, S.~Ghose, D.~Lee, M.~K. Qureshi, and O.~Mutlu,
  ``{Low-Cost Inter-Linked Subarrays (LISA): Enabling Fast Inter-Subarray Data
  Movement in DRAM},'' in \emph{HPCA}, 2016.

\bibitem{chang.sigmetrics17}
K.~K. Chang, A.~G. Ya\u{g}l{\i}k\c{c}{\i}, S.~Ghose, A.~Agrawal, N.~Chatterjee,
  A.~Kashyap, D.~Lee, M.~O'Connor, H.~Hassan, and O.~Mutlu, ``{Understanding
  Reduced-Voltage Operation in Modern DRAM Devices: Experimental
  Characterization, Analysis, and Mechanisms},'' in \emph{SIGMETRICS}, 2017.

\bibitem{chang.thesis17}
K.~K. Chang, ``{Understanding and Improving the Latency of DRAM-Based Memory
  Systems},'' Ph.D. dissertation, Carnegie Mellon Univ., 2017.

\bibitem{chang.sigmetrics16}
K.~K. Chang, A.~Kashyap, H.~Hassan, S.~Ghose, K.~Hsieh, D.~Lee, T.~Li,
  G.~Pekhimenko, S.~Khan, and O.~Mutlu, ``{Understanding Latency Variation in
  Modern DRAM Chips: Experimental Characterization, Analysis, and
  Optimization},'' in \emph{SIGMETRICS}, 2016.

\bibitem{charney.ibmjrd97}
M.~J. Charney and T.~R. Puzak, ``{Prefetching and Memory System Behavior of the
  SPEC95 Benchmark Suite},'' \emph{IBM JRD}, 1997.

\bibitem{chatterjee.sc14}
N.~Chatterjee, M.~O'Connor, G.~H. Loh, N.~Jayasena, and R.~Balasubramonian,
  ``{Managing DRAM Latency Divergence in Irregular GPGPU Applications},'' in
  \emph{SC}, 2014.

\bibitem{rodinia}
S.~Che, M.~Boyer, J.~Meng, D.~Tarjan, J.~Sheaffer, S.-H. Lee, and K.~Skadron,
  ``{Rodinia: A Benchmark Suite for Heterogeneous Computing},'' in
  \emph{IISWC}, 2009.

\bibitem{choi.isca15}
J.~Choi, W.~Shin, J.~Jang, J.~Suh, Y.~Kwon, Y.~Moon, and L.-S. Kim, ``{Multiple
  Clone Row DRAM: A Low Latency and Area Optimized DRAM},'' in \emph{ISCA},
  2015.

\bibitem{chou.isca04}
Y.~Chou, B.~Fahs, and S.~Abraham, ``{Microarchitecture Optimizations for
  Exploiting Memory-Level Parallelism},'' in \emph{ISCA}, 2004.

\bibitem{ycsb}
B.~F. Cooper, A.~Silberstein, E.~Tam, R.~Ramakrishnan, and R.~Sears,
  ``{Benchmarking Cloud Serving Systems with YCSB},'' in \emph{SoCC}, 2010.

\bibitem{cuppu.isca99}
V.~Cuppu, B.~Jacob, B.~Davis, and T.~Mudge, ``{A Performance Comparison of
  Contemporary DRAM Architectures},'' in \emph{ISCA}, 1999.

\bibitem{cuppu.tc01}
V.~Cuppu, B.~Jacob, B.~Davis, and T.~Mudge, ``{High-Performance DRAMs in
  Workstation Environments},'' in \emph{IEEE Transactions on Computers}, 2001.

\bibitem{cuppu.isca01}
V.~Cuppu and B.~Jacob, ``{Concurrency, Latency, or System Overhead: Which Has
  the Largest Impact on Uniprocessor DRAM-System Performance?}'' in
  \emph{ISCA}, 2001.

\bibitem{das.dac18}
A.~Das, H.~Hassan, and O.~Mutlu, ``{VRL-DRAM: Improving DRAM Performance via
  Variable Refresh Latency},'' in \emph{DAC}, 2018.

\bibitem{das.micro09}
R.~Das, O.~Mutlu, T.~Moscibroda, and C.~Das, ``{Application-Aware
  Prioritization Mechanisms for On-Chip Networks},'' in \emph{MICRO}, 2009.

\bibitem{david.icac11}
H.~David, C.~Fallin, E.~Gorbatov, U.~R. Hanebutte, and O.~Mutlu, ``{Memory
  Power Management via Dynamic Voltage/Frequency Scaling},'' in \emph{ICAC},
  2011.

\bibitem{mapreduce}
J.~Dean and S.~Ghemawat, ``{MapReduce: Simplified Data Processing on Large
  Clusters},'' in \emph{OSDI}, 2004.

\bibitem{desikan.isca01}
R.~Desikan, D.~Burger, and S.~W. Keckler, ``{Measuring Experimental Error in
  Microprocessor Simulation},'' in \emph{ISCA}, 2001.

\bibitem{difallah.vldb04}
D.~E. Difallah, A.~Pavlo, C.~Curino, and P.~Cudre-Mauroux, ``{OLTP-Bench: An
  Extensible Testbed for Benchmarking Relational Databases},'' in \emph{VLDB},
  2004.

\bibitem{dong.tcad12}
X.~Dong, C.~Xu, Y.~Xie, and N.~P. Jouppi, ``{NVSim: A Circuit-Level
  Performance, Energy, and Area Model for Emerging Nonvolatile Memory},''
  \emph{TCAD}, 2012.

\bibitem{memcached}
Dormando, ``{Memcached: High-Performance Distributed Memory Object Caching
  System},'' \url{http://memcached.org/}.

\bibitem{ebrahimi.isca11}
E.~Ebrahimi, C.~J. Lee, O.~Mutlu, and Y.~N. Patt, ``{Prefetch-Aware Shared
  Resource Management for Multi-Core Systems},'' in \emph{ISCA}, 2011.

\bibitem{ebrahimi.micro11}
E.~Ebrahimi, R.~Miftakhutdinov, C.~Fallin, C.~J. Lee, J.~A. Joao, O.~Mutlu, and
  Y.~N. Patt, ``Parallel application memory scheduling,'' in \emph{MICRO},
  2011.

\bibitem{endo.samos14}
F.~A. Endo, D.~Corouss{\'{e}}, and H.-P. Charles, ``{Micro-Architectural
  Simulation of In-Order and Out-of-Order ARM Microprocessors with gem5},'' in
  \emph{SAMOS}, 2014.

\bibitem{eyerman.ieeemicro08}
S.~Eyerman and L.~Eeckhout, ``{{System-Level Performance Metrics for
  Multiprogram Workloads}},'' \emph{IEEE Micro}, 2008.

\bibitem{mediabench}
J.~Fritts and B.~Mangione-Smith, ``{MediaBench II - Technology, Status, and
  Cooperation},'' in \emph{The Workshop on Media and Stream Processors}, 2002.

\bibitem{ghose.isca13}
S.~Ghose, H.~Lee, and J.~F. Mart{\'\i}nez, ``{Improving Memory Scheduling via
  Processor-Side Load Criticality Information},'' in \emph{ISCA}, 2013.

\bibitem{ghose.sigmetrics18}
S.~Ghose, A.~G. Ya\u{g}l{\i}k\c{c}{\i}, R.~Gupta, D.~Lee, K.~Kudrolli, W.~X.
  Liu, H.~Hassan, K.~K. Chang, N.~Chatterjee, A.~Agrawal, M.~O'Connor, and
  O.~Mutlu, ``{What Your DRAM Power Models Are Not Telling You: Lessons from a
  Detailed Experimental Study},'' in \emph{SIGMETRICS}, 2018.

\bibitem{giridhar.sc13}
B.~Giridhar, M.~Cieslak, D.~Duggal, R.~Dreslinski, H.~Chen, R.~Patti, B.~Hold,
  C.~Chakrabarti, T.~Mudge, and D.~Blaauw, ``{Exploring DRAM Organizations for
  Energy-Efficient and Resilient Exascale Memories},'' in \emph{SC}, 2013.

\bibitem{glew.waci98}
A.~Glew, ``{MLP Yes! ILP No! Memory Level Parallelism, or Why I No Longer Care
  About Instruction Level Parallelism},'' in \emph{ASPLOS WACI}, 1998.

\bibitem{gomony.date12}
M.~D. Gomony, C.~Weis, B.~Akesson, N.~Wehn, and K.~Goossens, ``{DRAM Selection
  and Configuration for Real-Time Mobile Systems},'' in \emph{DATE}, 2012.

\bibitem{hamerly2005simpoint}
G.~Hamerly, E.~Perelman, J.~Lau, and B.~Calder, ``{Simpoint 3.0: Faster and
  More Flexible Program Phase Analysis},'' \emph{JILP}, 2005.

\bibitem{hassan.isca19}
H.~Hassan, M.~Patel, J.~S. Kim, A.~G. Ya\u{g}l{\i}k\c{c}{\i}, N.~Vijaykumar,
  N.~M. Ghiasi, S.~Ghose, and O.~Mutlu, ``{CROW: A Low-Cost Substrate for
  Improving DRAM Performance, Energy Efficiency, and Reliability},'' in
  \emph{ISCA}, 2019.

\bibitem{hassan.hpca17}
H.~Hassan, N.~Vijaykumar, S.~Khan, S.~Ghose, K.~Chang, G.~Pekhimenko, D.~Lee,
  O.~Ergin, and O.~Mutlu, ``{SoftMC: A Flexible and Practical Open-Source
  Infrastructure for Enabling Experimental DRAM Studies},'' in \emph{HPCA},
  2017.

\bibitem{hassan.hpca16}
H.~Hassan, G.~Pekhimenko, N.~Vijaykumar, V.~Seshadri, D.~Lee, O.~Ergin, and
  O.~Mutlu, ``{ChargeCache: Reducing DRAM Latency by Exploiting Row Access
  Locality},'' in \emph{HPCA}, 2016.

\bibitem{he.pact08}
B.~He, W.~Fang, Q.~Luo, N.~Govindaraju, and T.~Wang, ``{Mars: A MapReduce
  Framework on Graphics Processors},'' in \emph{PACT}, 2008.

\bibitem{henning.computer00}
J.~L. Henning, ``{SPEC CPU2000: Measuring {CPU} Performance in the New
  Millennium},'' \emph{IEEE Computer}, 2000.

\bibitem{Netperf}
Hewlett-Packard, ``{Netperf: A Network Performance Benchmark (Rev. 2.1)},''
  1996.

\bibitem{holzle.book09}
U.~Holzle and L.~A. Barroso, \emph{{The Datacenter as a Computer: An
  Introduction to the Design of Warehouse-Scale Machines}}.\hskip 1em plus
  0.5em minus 0.4em\relax {Morgan \& Claypool}, 2009.

\bibitem{hur.micro04}
I.~Hur and C.~Lin, ``{Adaptive History-Based Memory Schedulers},'' in
  \emph{MICRO}, 2004.

\bibitem{hwang.asplos12}
A.~Hwang, I.~Stefanovici, and B.~Schroeder, ``{Cosmic Rays Don’t Strike
  Twice: Understanding the Nature of DRAM Errors and the Implications for
  System Design},'' in \emph{ASPLOS}, 2012.

\bibitem{hmc.2.1}
{Hybrid Memory Cube Consortium}, ``{Hybrid Memory Cube Specification 2.1},''
  2015.

\bibitem{power9.datasheet}
{IBM Corp.}, \emph{{POWER9 Processor RegistersSpecification, Vol. 3}}, May
  2017.

\bibitem{corei7-2600k.spec}
{Intel Corp.}, ``{Product Specification:
  Intel{\textsuperscript{\textregistered}} Core{\texttrademark} i7-2600K},''
  \url{https://ark.intel.com/products/52214/}.

\bibitem{corei7-975.spec}
{Intel Corp.}, ``{Product Specification:
  Intel{\textsuperscript{\textregistered}} Core{\texttrademark} i7-975
  Processor Extreme Edition},'' \url{https://ark.intel.com/products/37153/}.

\bibitem{xeone5-2630v4.spec}
{Intel Corp.}, ``{Product Specification:
  Intel{\textsuperscript{\textregistered}}
  Xeon{\textsuperscript{\textregistered}} Processor E5-2630 v4},''
  \url{https://ark.intel.com/products/92981/}.

\bibitem{coregen7.datasheet}
{Intel Corp.}, \emph{{7th Generation Intel{\textsuperscript{\textregistered}}
  Processor Families for S Platforms and
  Intel{\textsuperscript{\textregistered}} Core{\texttrademark} X-Series
  Processor Family Datasheet, Vol. 1}}, December 2018.

\bibitem{xeone5.datasheet}
{Intel Corp.}, \emph{{Intel{\textsuperscript{\textregistered}}
  Xeon{\textsuperscript{\textregistered}} Processor E5-1600/2400/2600/4600
  (E5-Product Family) Product Families Datasheet Vol. 2}}, May 2018.

\bibitem{IOZone}
{IOzone Lab}, ``{IOzone Filesystem Benchmark},'' \url{http://www.iozone.org/},
  2016.

\bibitem{ipek.isca08}
E.~\.{I}pek, O.~Mutlu, J.~F. Mart{\'\i}nez, and R.~Caruana, ``{Self-Optimizing
  Memory Controllers: A Reinforcement Learning Approach},'' in \emph{ISCA},
  2008.

\bibitem{isen.micro09}
C.~Isen and L.~John, ``{ESKIMO --- Energy Savings Using Semantic Knowledge of
  Inconsequential Memory Occupancy for DRAM Subsystem},'' in \emph{MICRO},
  2009.

\bibitem{jeddeloh2012hybrid}
J.~Jeddeloh and B.~Keeth, ``{Hybrid Memory Cube New DRAM Architecture Increases
  Density and Performance},'' in \emph{VLSIT}, 2012.

\bibitem{fbdimm}
{JEDEC Solid State Technology Assn.}, \emph{{{JESD}206: FBDIMM Architecture and
  Protocol}}, January 2007.

\bibitem{ddr2}
{JEDEC Solid State Technology Assn.}, \emph{{{JESD}79-2F: DDR2 SDRAM
  Standard}}, November 2009.

\bibitem{wideio}
{JEDEC Solid State Technology Assn.}, \emph{{JESD229: Wide I/O Single Data Rate
  (Wide I/O SDR) Standard}}, December 2011.

\bibitem{ddr3}
{JEDEC Solid State Technology Assn.}, \emph{{{JESD}79-3F: DDR3 SDRAM
  Standard}}, July 2012.

\bibitem{hbm}
{JEDEC Solid State Technology Assn.}, \emph{{JESD235: High Bandwidth Memory
  (HBM) DRAM}}, October 2013.

\bibitem{wideio2}
{JEDEC Solid State Technology Assn.}, \emph{{JESD229-2: Wide I/O 2 (WideIO2)
  Standard}}, August 2014.

\bibitem{lpddr3}
{JEDEC Solid State Technology Assn.}, \emph{{JESD209-3C: Low Power Double Data
  Rate 3 (LPDDR3) Standard}}, August 2015.

\bibitem{gddr5}
{JEDEC Solid State Technology Assn.}, \emph{{JESD}212C: {Graphics Double Data
  Rate (GDDR5) SGRAM Standard}}, February 2016.

\bibitem{lpddr4}
{JEDEC Solid State Technology Assn.}, \emph{{JESD209-4B: Low Power Double Data
  Rate 4 (LPDDR4) Standard}}, March 2017.

\bibitem{ddr4}
{JEDEC Solid State Technology Assn.}, \emph{{{JESD}79-4B: {DDR4} {SDRAM}
  Standard}}, June 2017.

\bibitem{jeong.hpca12}
M.~K. Jeong, D.~H. Yoon, D.~Sunwoo, M.~Sullivan, I.~Lee, and M.~Erez,
  ``{Balancing DRAM Locality and Parallelism in Shared Memory CMP Systems},''
  in \emph{HPCA}, 2012.

\bibitem{jeong.dac12}
M.~K. Jeong, M.~Erez, C.~Sudanthi, and N.~Paver, ``{A QoS-Aware Memory
  Controller for Dynamically Balancing GPU and CPU Bandwidth Use in an
  MPSoC},'' in \emph{DAC}, 2012.

\bibitem{jog.sigmetrics16}
A.~Jog, O.~Kayiran, A.~Pattnaik, M.~T. Kandemir, O.~Mutlu, R.~Iyer, and C.~R.
  Das, ``{Exploiting Core Criticality for Enhanced GPU Performance},'' in
  \emph{SIGMETRICS}, 2016.

\bibitem{kang.memoryforum14}
U.~Kang, H.-S. Yu, C.~Park, H.~Zheng, J.~Halbert, K.~Bains, S.~Jang, and
  J.~Choi, ``{Co-Architecting Controllers and {DRAM} to Enhance {DRAM} Process
  Scaling},'' in \emph{The Memory Forum}, 2014.

\bibitem{kaseridis.micro2011}
D.~Kaseridis, J.~Stuecheli, and L.~K. John, ``{Minimalist Open-Page: A DRAM
  Page-Mode Scheduling Policy for the Many-Core Era},'' in \emph{{MICRO}},
  2011.

\bibitem{khan.dsn16}
S.~Khan, D.~Lee, and O.~Mutlu, ``{PARBOR: An Efficient System-Level Technique
  to Detect Data Dependent Failures in DRAM},'' in \emph{{DSN}}, 2016.

\bibitem{khan.sigmetrics14}
S.~Khan, D.~Lee, Y.~Kim, A.~R. Alameldeen, C.~Wilkerson, and O.~Mutlu, ``{The
  Efficacy of Error Mitigation Techniques for DRAM Retention Failures: A
  Comparative Experimental Study},'' in \emph{SIGMETRICS}, 2014.

\bibitem{khan.cal16}
S.~Khan, C.~Wilkerson, D.~Lee, A.~R. Alameldeen, and O.~Mutlu, ``{A Case for
  Memory Content-Based Detection and Mitigation of Data-Dependent Failures in
  {DRAM}},'' \emph{CAL}, 2016.

\bibitem{khan.micro17}
S.~Khan, C.~Wilkerson, Z.~Wang, A.~Alameldeen, D.~Lee, and O.~Mutlu,
  ``{Detecting and Mitigating Data-Dependent DRAM Failures by Exploiting
  Current Memory Content},'' in \emph{MICRO}, 2017.

\bibitem{kim2013memory}
G.~Kim, J.~Kim, J.~H. Ahn, and J.~Kim, ``{Memory-Centric System Interconnect
  Design with Hybrid Memory Cubes},'' in \emph{PACT}, 2013.

\bibitem{kim.isscc2011}
J.~S. Kim, C.~Oh, H.~Lee, D.~Lee, H.~R. Hwang, S.~Hwang, B.~Na, J.~Moon, J.~G.
  Kim, H.~Park, J.~W. Ryu, K.~Park, S.~K. Kang, S.~Y. Kim, H.~Kim, J.~M. Bang,
  H.~Cho, M.~Jang, C.~Han, J.~B. Lee, K.~Kyung, J.~S. Choi, and Y.~H. Jun, ``{A
  1.2V 12.8GB/s 2Gb Mobile Wide-I/O DRAM with 4x128 I/Os Using TSV-Based
  Stacking},'' in \emph{ISSCC}, 2011.

\bibitem{kim.iccd18}
J.~S. Kim, M.~Patel, H.~Hassan, and O.~Mutlu, ``{Solar-DRAM: Reducing DRAM
  Access Latency by Exploiting the Variation in Local Bitlines},'' in
  \emph{ICCD}, 2018.

\bibitem{kim.hpca18}
J.~S. Kim, M.~Patel, H.~Hassan, and O.~Mutlu, ``{The {DRAM} Latency {PUF}:
  Quickly Evaluating Physical Unclonable Functions by Exploiting the
  Latency--Reliability Tradeoff in Modern {DRAM} Devices},'' in \emph{HPCA},
  2018.

\bibitem{kim.hpca19}
J.~S. Kim, M.~Patel, H.~Hassan, L.~Orosa, and O.~Mutlu, ``{D-RaNGe: Using
  Commodity DRAM Devices to Generate True Random Numbers with Low Latency and
  High Throughput},'' in \emph{HPCA}, 2019.

\bibitem{kim.hpca10}
Y.~Kim, D.~Han, O.~Mutlu, and M.~Harchol-Balter, ``{ATLAS: A Scalable and
  High-Performance Scheduling Algorithm for Multiple Memory Controllers},'' in
  \emph{HPCA}, 2010.

\bibitem{kim.micro10}
Y.~Kim, M.~Papamichael, O.~Mutlu, and M.~Harchol-Balter, ``{Thread Cluster
  Memory Scheduling: Exploiting Differences in Memory Access Behavior},'' in
  \emph{MICRO}, 2010.

\bibitem{kim.cal15}
Y.~Kim, W.~Yang, and O.~Mutlu, ``{Ramulator: A Fast and Extensible DRAM
  Simulator},'' \emph{CAL}, 2015.

\bibitem{kim.isca14}
Y.~Kim, R.~Daly, J.~Kim, C.~Fallin, J.~H. Lee, D.~Lee, C.~Wilkerson, K.~Lai,
  and O.~Mutlu, ``{Flipping Bits in Memory Without Accessing Them: An
  Experimental Study of DRAM Disturbance Errors},'' in \emph{ISCA}, 2014.

\bibitem{kim.isca12}
Y.~Kim, V.~Seshadri, D.~Lee, J.~Liu, and O.~Mutlu, ``{A Case for Exploiting
  Subarray-Level Parallelism (SALP) in DRAM},'' in \emph{ISCA}, 2012.

\bibitem{kirman.hpca05}
N.~K{\i}rman, M.~K{\i}rman, M.~Chaudhuri, and J.~F. Mart{\'i}nez,
  ``{Checkpointed Early Load Retirement},'' in \emph{HPCA}, 2005.

\bibitem{kloosterman.micro15}
J.~Kloosterman, J.~Beaumont, M.~Wollman, A.~Sethia, R.~Dreslinski, T.~Mudge,
  and S.~Mahlke, ``{WarpPool: Sharing Requests with Inter-Warp Coalescing for
  Throughput Processors},'' in \emph{MICRO}, 2015.

\bibitem{lawton2006bochs}
K.~Lawton, B.~Denney, and C.~Bothamy, ``{The Bochs IA-32 emulator project},''
  \url{http://bochs.sourceforge.net}, 2006.

\bibitem{lee.micro08}
C.~J. Lee, O.~Mutlu, V.~Narasiman, and Y.~N. Patt, ``{Prefetch-Aware DRAM
  Controllers},'' in \emph{MICRO}, 2008.

\bibitem{lee.micro09}
C.~J. Lee, V.~Narasiman, O.~Mutlu, and Y.~N. Patt, ``{Improving Memory
  Bank-Level Parallelism in the Presence of Prefetching},'' in \emph{MICRO},
  2009.

\bibitem{lee.tr10}
C.~J. Lee, E.~Ebrahimi, V.~Narasiman, O.~Mutlu, and Y.~N. Patt, ``{DRAM-Aware
  Last-Level Cache Writeback: Reducing Write-Caused Interference in Memory
  Systems},'' Univ.\ of Texas at Austin, High Performance Systems Group, Tech.
  Rep. TR-HPS-2010-002, 2010.

\bibitem{lee.sigmetrics17}
D.~Lee, S.~Khan, L.~Subramanian, S.~Ghose, R.~Ausavarungnirun, G.~Pekhimenko,
  V.~Seshadri, and O.~Mutlu, ``{Design-Induced Latency Variation in Modern DRAM
  Chips: Characterization, Analysis, and Latency Reduction Mechanisms},'' in
  \emph{{SIGMETRICS}}, 2017.

\bibitem{lee.pact15}
D.~Lee, L.~Subramanian, R.~Ausavarungnirun, J.~Choi, and O.~Mutlu, ``{Decoupled
  Direct Memory Access: Isolating CPU and IO Traffic by Leveraging a
  Dual-Data-Port DRAM},'' in \emph{PACT}, 2015.

\bibitem{lee.thesis16}
D.~Lee, ``{Reducing DRAM Latency at Low Cost by Exploiting Heterogeneity},''
  Ph.D. dissertation, Carnegie Mellon Univ., 2016.

\bibitem{lee.taco16}
D.~Lee, S.~Ghose, G.~Pekhimenko, S.~Khan, and O.~Mutlu, ``{Simultaneous
  Multi-Layer Access: Improving 3D-Stacked Memory Bandwidth at Low Cost},''
  \emph{TACO}, 2016.

\bibitem{lee.hpca15}
D.~Lee, Y.~Kim, G.~Pekhimenko, S.~Khan, V.~Seshadri, K.~Chang, and O.~Mutlu,
  ``{Adaptive-Latency DRAM: Optimizing DRAM Timing for the Common-Case},'' in
  \emph{HPCA}, 2015.

\bibitem{lee.hpca13}
D.~Lee, Y.~Kim, V.~Seshadri, J.~Liu, L.~Subramanian, and O.~Mutlu,
  ``{Tiered-Latency DRAM: A Low Latency and Low Cost DRAM Architecture},'' in
  \emph{HPCA}, 2013.

\bibitem{lefurgy.computer03}
C.~Lefurgy, K.~Rajamani, F.~Rawson, W.~Felter, M.~Kistler, and T.~Keller,
  ``{Energy Management for Commercial Servers},'' \emph{{Computer}}, 2003.

\bibitem{lenovo.xeon.memconfig}
{Lenovo Group Ltd.}, ``{Intel Xeon Scalable Family Balanced Memory
  Configurations},'' \url{https://lenovopress.com/lp0742.pdf}, 2017.

\bibitem{li.sc17}
A.~Li, W.~Liu, M.~R.~B. Kistensen, B.~Vinter, H.~Wang, K.~Hou, A.~Marquez, and
  S.~L. Song, ``{Exploring and Analyzing the Real Impact of Modern On-Package
  Memory on HPC Scientific Kernels},'' in \emph{SC}, 2017.

\bibitem{li.memsys18}
S.~Li, D.~Reddy, and B.~Jacob, ``{A Performance \& Power Comparison of Modern
  High-Speed DRAM Architectures},'' in \emph{MEMSYS}, 2018.

\bibitem{liu.isca12}
J.~Liu, B.~Jaiyen, R.~Veras, and O.~Mutlu, ``{RAIDR: Retention-Aware
  Intelligent DRAM Refresh},'' in \emph{ISCA}, 2012.

\bibitem{liu.isca13}
J.~Liu, B.~Jaiyen, Y.~Kim, C.~Wilkerson, and O.~Mutlu, ``{An Experimental Study
  of Data Retention Behavior in Modern {DRAM} Devices: Implications for
  Retention Time Profiling Mechanisms},'' in \emph{ISCA}, 2013.

\bibitem{loh2008stacked}
G.~H. Loh, ``{3D-Stacked Memory Architectures for Multi-Core Processors},'' in
  \emph{ISCA}, 2008.

\bibitem{luk2005pin}
C.-K. Luk, R.~Cohn, R.~Muth, H.~Patil, A.~Klauser, G.~Lowney, S.~Wallace, V.~J.
  Reddi, and K.~Hazelwood, ``{Pin: Building Customized Program Analysis Tools
  with Dynamic Instrumentation},'' in \emph{PLDI}, 2005.

\bibitem{luo.ispass01}
K.~Luo, J.~Gummaraju, and M.~Franklin, ``{Balancing Throughput and Fairness in
  SMT Processors},'' in \emph{ISPASS}, 2001.

\bibitem{malladi.isca12}
K.~T. Malladi, F.~A. Nothaft, K.~Periyathambi, B.~C. Lee, C.~Kozyrakis, and
  M.~Horowitz, ``{Towards Energy-Proportional Datacenter Memory with Mobile
  DRAM},'' in \emph{ISCA}, 2012.

\bibitem{mandelman.ibmjrd02}
J.~A. Mandelman, R.~H. Dennard, G.~B. Bronner, J.~K. DeBrosse, R.~Divakaruni,
  Y.~Li, and C.~J. Radens, ``{Challenges and Future Directions for the Scaling
  of Dynamic Random-Access Memory ({DRAM})},'' \emph{IBM JRD}, 2002.

\bibitem{mccalpin.tcca95}
J.~D. McCalpin, ``{Memory Bandwidth and Machine Balance in Current High
  Performance Computers},'' \emph{TCCA Newsletter}, 1995.

\bibitem{meza.dsn15}
J.~Meza, Q.~Wu, S.~Kumar, and O.~Mutlu, ``{Revisiting Memory Errors in
  Large-Scale Production Data Centers: Analysis and Modeling of New Trends from
  the Field},'' in \emph{DSN}, 2015.

\bibitem{micronlp}
{Micron Technology, Inc.}, \emph{{Technical Note TN-46-12: Mobile DRAM
  Power-Saving Features and Calculations}}, May 2009,
  \url{https://www.micron.com/~/media/documents/products/technical-note/dram/tn4612.pdf}.

\bibitem{ddr3.verilogmodel}
{Micron Technology, Inc.}, ``{DDR3 SDRAM Verilog Model, v.\ 1.74},''
  \url{https://www.micron.com/-/media/client/global/documents/products/sim-model/dram/ddr3/ddr3-sdram-verilog-model.zip},
  2015.

\bibitem{micron.lpddr3.8gb.datasheet}
{Micron Technology, Inc.}, \emph{{178-Ball 2E0F Mobile LPDDR3 SDRAM Data
  Sheet}}, April 2016.

\bibitem{micron.ddr3.2gb.datasheet}
{Micron Technology, Inc.}, \emph{{2Gb: x4, x8, x16 DDR3 SDRAM Data Sheet}},
  February 2016.

\bibitem{micron.lpddr4.8gb.datasheet}
{Micron Technology, Inc.}, \emph{{200-Ball Z01M LPDDR4 SDRAM Automotive Data
  Sheet}}, May 2018.

\bibitem{micron.ddr4.4gb.datasheet}
{Micron Technology, Inc.}, \emph{{4Gb: x4, x8, x16 DDR4 SDRAM Data Sheet}},
  June 2018.

\bibitem{moscibroda.podc08}
T.~Moscibroda and O.~Mutlu, ``{Distributed Order Scheduling and Its Application
  to Multi-Core DRAM Controllers},'' in \emph{PODC}, 2008.

\bibitem{mukundan.hpca12}
J.~Mukundan and J.~F. Mart{\'\i}nez, ``{MORSE: Multi-objective Reconfigurable
  Self-Optimizing Memory Scheduler},'' in \emph{HPCA}, 2012.

\bibitem{muralidhara.micro11}
S.~P. Muralidhara, L.~Subramanian, O.~Mutlu, M.~Kandemir, and T.~Moscibroda,
  ``{Reducing Memory Interference in Multicore Systems via Application-Aware
  Memory Channel Partitioning},'' in \emph{MICRO}, 2011.

\bibitem{murphy.tc07}
R.~C. Murphy and P.~M. Kogge, ``{On the Memory Access Patterns of Supercomputer
  Applications: Benchmark Selection and Its Implications},'' \emph{TC}, 2007.

\bibitem{mutlu.date17}
O.~Mutlu, ``{The RowHammer Problem and Other Issues We May Face as Memory
  Becomes Denser},'' in \emph{DATE}, 2017.

\bibitem{mutlu.isca08}
O.~Mutlu and T.~Moscibroda, ``{Parallelism-Aware Batch Scheduling: Enhancing
  Both Performance and Fairness of Shared {DRAM} Systems},'' in \emph{ISCA},
  2008.

\bibitem{scale.imw13}
O.~Mutlu, ``{Memory Scaling: A Systems Architecture Perspective},'' in
  \emph{IMW}, 2013.

\bibitem{mutlu.isca05}
O.~Mutlu, H.~Kim, and Y.~N. Patt, ``{Techniques for Efficient Processing in
  Runahead Execution Engines},'' in \emph{ISCA}, 2005.

\bibitem{mutlu.ieeemicro06}
O.~Mutlu, H.~Kim, and Y.~N. Patt, ``{Efficient Runahead Execution:
  Power-Efficient Memory Latency Tolerance},'' \emph{IEEE Micro}, 2006.

\bibitem{mutlu.tcad19}
O.~Mutlu and J.~S. Kim, ``{RowHammer: A Retrospective},'' \emph{TCAD}, 2019.

\bibitem{mutlu.micro07}
O.~Mutlu and T.~Moscibroda, ``{Stall-Time Fair Memory Access Scheduling for
  Chip Multiprocessors},'' in \emph{MICRO}, 2007.

\bibitem{mutlu.hpca03}
O.~Mutlu, J.~Stark, C.~Wilkerson, and Y.~N. Patt, ``{Runahead Execution: An
  Alternative to Very Large Instruction Windows for Out-of-Order Processors},''
  in \emph{HPCA}, 2003.

\bibitem{nesbit.micro06}
K.~J. Nesbit, N.~Aggarwal, J.~Laudon, and J.~E. Smith, ``{Fair Queuing Memory
  Systems},'' in \emph{MICRO}, 2006.

\bibitem{gtx480.spec}
{NVIDIA Corp.}, ``{GeForce GTX 480: Specifications},''
  \url{https://www.geforce.com/hardware/desktop-gpus/geforce-gtx-480/specifications}.

\bibitem{nxp.networkaccel}
{NXP Semiconductors}, ``{QorIQ Processing Platforms: 64-Bit Multicore SoCs},''
  \url{https://www.nxp.com/products/processors-and-microcontrollers/applications-processors/qoriq-platforms:QORIQ_HOME}.

\bibitem{patel.dsn19}
M.~Patel, J.~S. Kim, H.~Hassan, and O.~Mutlu, ``{Understanding and Modeling
  On-Die Error Correction in Modern DRAM: An Experimental Study Using Real
  Devices},'' in \emph{DSN}, 2019.

\bibitem{patel.isca17}
M.~Patel, J.~Kim, and O.~Mutlu, ``{The Reach Profiler (REAPER): Enabling the
  Mitigation of DRAM Retention Failures via Profiling at Aggressive
  Conditions},'' in \emph{ISCA}, 2017.

\bibitem{paul.isca15}
I.~Paul, W.~Huang, M.~Arora, and S.~Yalamanchili, ``{Harmonia: Balancing
  Compute and Memory Power in High-Performance GPUs},'' in \emph{ISCA}, 2015.

\bibitem{pawlowski.hc11}
J.~T. Pawlowski, ``{Hybrid Memory Cube (HMC)},'' in \emph{HC}, 2011.

\bibitem{multithreadedpin.github}
S.~Pelley, ``{atomic-memory-trace},''
  \url{https://github.com/stevenpelley/atomic-memory-trace}, 2013.

\bibitem{peter2016arrakis}
S.~Peter, J.~Li, I.~Zhang, D.~R. Ports, D.~Woos, A.~Krishnamurthy, T.~Anderson,
  and T.~Roscoe, ``{Arrakis: The Operating System Is the Control Plane},''
  \emph{TOCS}, 2016.

\bibitem{qureshi.dsn15}
M.~K. Qureshi, D.~H. Kim, S.~Khan, P.~J. Nair, and O.~Mutlu, ``{AVATAR: A
  Variable-Retention-Time (VRT) Aware Refresh for DRAM Systems},'' in
  \emph{{DSN}}, 2015.

\bibitem{qureshi.isca06}
M.~K. Qureshi, D.~N. Lynch, O.~Mutlu, and Y.~N. Patt, ``{A Case for MLP-Aware
  Cache Replacement},'' in \emph{ISCA}, 2006.

\bibitem{radulovic.memsys15}
M.~Radulovic, D.~Zivanovic, D.~Ruiz, B.~R. {de Supinski}, S.~A. McKee,
  P.~Radojkovi{\'c}, and E.~Ayaguad{\'e}, ``{Another Trip to the Wall: How Much
  Will Stacked DRAM Benefit HPC?}'' in \emph{MEMSYS}, 2015.

\bibitem{rixner.micro04}
S.~Rixner, ``{Memory Controller Optimizations for Web Servers},'' in
  \emph{MICRO}, 2004.

\bibitem{rixner.isca00}
S.~Rixner, W.~J. Dally, U.~J. Kapasi, P.~Mattson, and J.~D. Owens, ``{Memory
  Access Scheduling},'' in \emph{ISCA}, 2000.

\bibitem{rokicki.tr96}
T.~Rokicki, ``{Indexing Memory Banks to Maximize Page Mode Hit Percentage and
  Minimize Memory Latency},'' HP Laboratories Palo Alto, Tech. Rep. HPL-96-95,
  1996.

\bibitem{rosenfeld.tr12}
P.~Rosenfeld, E.~Cooper-Balis, T.~Farrell, D.~Resnick, and B.~Jacob, ``{Peering
  Over the Memory Wall: Design Space and Performance Analysis of the Hybrid
  Memory Cube},'' Univ.\ of Maryland Systems and Computer Architecture Group,
  Tech. Rep. UMD-SCA-2012-10-01, 2012.

\bibitem{rosenfeld-cal2011}
P.~Rosenfeld, E.~Cooper-Balis, and B.~Jacob, ``{DRAMSim2: A Cycle Accurate
  Memory System Simulator},'' \emph{CAL}, 2011.

\bibitem{gpgpusimramulator.github}
{SAFARI Research Group}, ``{GPGPUSim+Ramulator --- GitHub Repository},''
  \url{https://github.com/CMU-SAFARI/GPGPUSim-Ramulator}.

\bibitem{memben.github}
{SAFARI Research Group}, ``{MemBen: A Memory Benchmark Suite for Ramulator ---
  GitHub Repository},'' \url{https://github.com/CMU-SAFARI/MemBen}.

\bibitem{ramulator.github}
{SAFARI Research Group}, ``{Ramulator: A DRAM Simulator --- GitHub
  Repository},'' \url{https://github.com/CMU-SAFARI/ramulator}.

\bibitem{schroeder.sigmetrics09}
B.~Schroeder, E.~Pinheiro, and W.-D. Weber, ``{DRAM Errors in the Wild: A
  Large-Scale Field Study},'' in \emph{SIGMETRICS}, 2009.

\bibitem{seshadri.isca14}
V.~Seshadri, A.~Bhowmick, O.~Mutlu, P.~B. Gibbons, M.~A. Kozuch, and T.~C.
  Mowry, ``{The Dirty-Block Index},'' in \emph{ISCA}, 2014.

\bibitem{seshadri.micro13}
V.~Seshadri, Y.~Kim, C.~Fallin, D.~Lee, R.~Ausavarungnirun, G.~Pekhimenko,
  Y.~Luo, O.~Mutlu, P.~B. Gibbons, M.~A. Kozuch, and T.~C. Mowry, ``{RowClone:
  Fast and Energy-Efficient In-{DRAM} Bulk Data Copy and Initialization},'' in
  \emph{MICRO}, 2013.

\bibitem{seshadri.micro17}
V.~Seshadri, D.~Lee, T.~Mullins, H.~Hassan, A.~Boroumand, J.~Kim, M.~A. Kozuch,
  O.~Mutlu, P.~B. Gibbons, and T.~C. Mowry, ``{Ambit: In-Memory Accelerator for
  Bulk Bitwise Operations Using Commodity DRAM Technology},'' in \emph{MICRO},
  2017.

\bibitem{seshadri.bookchapter20}
V.~Seshadri and O.~Mutlu, ``{In-DRAM Bulk Bitwise Execution Engine},'' in
  \emph{Advances in Computers}, 2020, available as arXiv:1905.09822 [cs.AR].

\bibitem{singh.icpe19}
S.~Singh and M.~Awasthi, ``{Memory Centric Characterization and Analysis of
  SPEC CPU2017 Suite},'' in \emph{ICPE}, 2019.

\bibitem{hynix.gddr5.2gb.datasheet}
{SK Hynix Inc.}, \emph{{2Gb (64Mx32) GDDR5 SGRAM Data Sheet}}, November 2011.

\bibitem{snavely.asplos00}
A.~Snavely and D.~M. Tullsen, ``{Symbiotic Jobscheduling for a Simultaneous
  Multithreading Processor},'' in \emph{ASPLOS}, 2000.

\bibitem{son.isca13}
Y.~H. Son, S.~O, Y.~Ro, J.~W. Lee, and J.~H. Ahn, ``{Reducing Memory Access
  Latency with Asymmetric DRAM Bank Organizations},'' in \emph{ISCA}, 2013.

\bibitem{sridharan.sc13}
V.~Sridharan, J.~Stearley, N.~DeBardeleben, S.~Blanchard, and S.~Gurumurthi,
  ``{Feng Shui of Supercomputer Memory: Positional Effects in DRAM and SRAM
  Faults},'' in \emph{SC}, 2013.

\bibitem{sridharan.asplos15}
V.~Sridharan, N.~DeBardeleben, S.~Blanchard, K.~B. Ferreira, J.~Stearley,
  J.~Shalf, and S.~Gurumurthi, ``{Memory Errors in Modern Systems: The Good,
  The Bad, and the Ugly},'' in \emph{ASPLOS}, 2015.

\bibitem{spec2006}
{Standard Performance Evaluation Corp.}, ``{SPEC CPU2006 Benchmarks},''
  \url{http://www.spec.org/cpu2006/}.

\bibitem{stuecheli.micro10}
J.~Stuecheli, D.~Kaseridis, H.~C. Hunter, and L.~K. John, ``{Elastic Refresh:
  Techniques to Mitigate Refresh Penalties in High Density Memory},'' in
  \emph{MICRO}, 2010.

\bibitem{stuecheli.isca10}
J.~Stuecheli, D.~Kaseridis, D.~Daly, H.~C. Hunter, and L.~K. John, ``{The
  Virtual Write Queue: Coordinating DRAM and Last-Level Cache Policies},'' in
  \emph{ISCA}, 2010.

\bibitem{subramanian.tpds16}
L.~Subramanian, D.~Lee, V.~Seshadri, H.~Rastogi, and O.~Mutlu, ``{BLISS:
  Balancing Performance, Fairness and Complexity in Memory Access
  Scheduling},'' \emph{TPDS}, 2016.

\bibitem{subramanian.iccd14}
L.~Subramanian, D.~Lee, V.~Seshadri, H.~Rastogi, and O.~Mutlu, ``{The
  Blacklisting Memory Scheduler: Achieving High Performance and Fairness at Low
  Cost},'' in \emph{ICCD}, 2014.

\bibitem{sun2014behavior}
B.~Sun, X.~Li, Z.~Zhu, and X.~Zhou, ``{Behavior Gaps and Relations between
  Operating System and Applications on Accessing DRAM},'' in \emph{ICECCS},
  2014.

\bibitem{suresh.cluster14}
A.~Suresh, P.~Cicotti, and L.~Carrington, ``{Evaluation of Emerging Memory
  Technologies for HPC, Data Intensive Applications},'' in \emph{CLUSTER},
  2014.

\bibitem{tang.micro16}
X.~Tang, M.~Kandemir, P.~Yedlapalli, and J.~Kotra, ``{Improving Bank-Level
  Parallelism for Irregular Applications},'' in \emph{MICRO}, 2016.

\bibitem{tuck.micro06}
J.~Tuck, L.~Ceze, and J.~Torrellas, ``{Scalable Cache Miss Handling for High
  Memory-Level Parallelism},'' in \emph{MICRO}, 2006.

\bibitem{ubal.pact12}
R.~Ubal, B.~Jand, P.~Mistry, D.~Schaa, and D.~Kaeli, ``{Multi2Sim: A Simulation
  Framework for CPU--GPU Computing},'' in \emph{PACT}, 2012.

\bibitem{coral}
{United States Department of Energy}, ``{CORAL Benchmark Codes},''
  \url{https://asc.llnl.gov/CORAL-benchmarks/}, 2014.

\bibitem{coral-2}
{United States Department of Energy}, ``{CORAL-2 Benchmarks},''
  \url{https://asc.llnl.gov/coral-2-benchmarks/}, 2017.

\bibitem{usui.taco16}
H.~Usui, L.~Subramanian, K.~K. Chang, and O.~Mutlu, ``{DASH: Deadline-Aware
  High-Performance Memory Scheduler for Heterogeneous Systems with Hardware
  Accelerators},'' \emph{TACO}, 2016.

\bibitem{venkatesan.hpca06}
R.~K. Venkatesan, S.~Herr, and E.~Rotenberg, ``{Retention-Aware Placement in
  DRAM (RAPID): Software Methods for Quasi-Non-Volatile DRAM},'' in
  \emph{HPCA}, 2006.

\bibitem{wang.micro18}
Y.~Wang, A.~Tavakkol, L.~Orosa, S.~Ghose, N.~{Mansouri Ghiasi}, M.~Patel, J.~S.
  Kim, H.~Hassan, M.~Sadrosadati, and O.~Mutlu, ``{Reducing DRAM Latency via
  Charge-Level-Aware Look-Ahead Partial Restoration},'' in \emph{MICRO}, 2018.

\bibitem{ware.hpca10}
M.~Ware, K.~Rajamani, M.~Floyd, B.~Brock, J.~C. Rubio, F.~Rawson, and J.~B.
  Carter, ``{Architecting for Power Management: The IBM POWER7 Approach},'' in
  \emph{{HPCA}}, 2010.

\bibitem{yoon.isca12}
D.~H. Yoon, J.~Chang, N.~Muralimanohar, and P.~Ranganathan, ``{BOOM: Enabling
  Mobile Memory Based Low-Power Server DIMMs},'' in \emph{ISCA}, 2012.

\bibitem{yuan.micro09}
G.~L. Yuan, A.~Bakhoda, and T.~M. Aamodt, ``{Complexity Effective Memory Access
  Scheduling for Many-Core Accelerator Architectures},'' in \emph{MICRO}, 2009.

\bibitem{redis}
J.~Zawodny, ``{Redis: Lightweight Key/Value Store That Goes the Extra Mile},''
  in \emph{Linux Magazine}, 2009.

\bibitem{zhang.hpca16}
X.~Zhang, Y.~Zhang, B.~R. Childers, and J.~Yang, ``{Restore Truncation for
  Performance Improvement in Future DRAM Systems},'' in \emph{HPCA}, 2016.

\bibitem{zhang.micro00}
Z.~Zhang, Z.~Zhu, and X.~Zhang, ``{A Permutation-Based Page Interleaving Scheme
  to Reduce Row-Buffer Conflicts and Exploit Data Locality},'' in \emph{MICRO},
  2000.

\bibitem{zhao.micro14}
J.~Zhao, O.~Mutlu, and Y.~Xie, ``{FIRM: Fair and High-Performance Memory
  Control for Persistent Memory Systems},'' in \emph{MICRO}, 2014.

\bibitem{zheng.tc10}
H.~Zheng and Z.~Zhu, ``{Power and Performance Trade-Offs in Contemporary DRAM
  System Designs for Multicore Processors},'' \emph{TC}, 2010.

\bibitem{zhu.hpca05}
Z.~Zhu and Z.~Zhang, ``{A Performance Comparison of DRAM Memory System
  Optimizations for SMT Processors},'' in \emph{HPCA}, 2005.

\bibitem{zuravleff.patent97}
W.~Zuravleff and T.~Robinson, ``{Controller for a Synchronous DRAM That
  Maximizes Throughput by Allowing Memory Requests and Commands to Be Issued
  Out of Order},'' U.S. Patent No. 5,630,096, 1997.

\end{thebibliography}
